%% file: fq_log_fsigma8.tex
\documentclass[
    twocolumn,       % final version in two-column format
    aps, prd,          % APS style, Phys. Rev. D
    superscriptaddress, % author affiliations as superscripts
    secnumarabic,      % Arabic section numbers
    floatfix,        % try to resolve float placement issues
    nobibnotes,      % suppress author footnotes in bibliography
    amsmath,           % advanced math (align, eqref) 
    amssymb            % extra math symbols
]{revtex4-1}
\usepackage{graphicx}      % include figure files
\usepackage{wrapfig}       % wrap text around figures
\usepackage{dcolumn}       % align table columns on decimal point
\graphicspath{{./}{figures_model/}{figures/}{other/}{figures_paper/}{figures_paper_rsd/}}
\usepackage{amsmath}                % advanced math (align, eqref)
\usepackage{amssymb}                % extra math symbols
\usepackage{eqnarray}               % old-style eqnarray (avoid if possible)
\usepackage{bm}% bold math

\usepackage{pifont}                 % dingbat symbols
\usepackage{wasysym}                % misc. symbols
\usepackage{textcomp}               % arrows, symbols, text extras
\usepackage[dvipsnames]{xcolor}     % colors
\usepackage{soul,ulem}              % highlighting + strikethrough
\soulregister\cite7
\soulregister\ref7 
\soulregister\eqref7                % allow \hl{} with refs/cites/eqref

\usepackage{cancel}                 % strike through math
\usepackage{comment}                % block commenting
\usepackage{marginnote}             % margin notes
\usepackage[colorinlistoftodos,
  textsize=tiny]{todonotes}         % inline todos

\usepackage[colorlinks=true,linkcolor=blue]{hyperref}%
\usepackage{cleveref}               % smart cross-references
\usepackage{algorithm}              % algorithm environment

\usepackage[ddmmyyyy]{datetime}     % compile date in ddmmyyyy format
\usepackage{acro}          % modern acronym handling
\input{acros}              % load acronym definitions
\input{macross}
\begin{document}

\title{Constraints on Logarithmic Model Extensions of Symmetric Teleparallel Gravity}
% \title{Constraints on Logarithmic model extensions in F(Q) gravity: in light of DESI DR2}

\author{Purnendu Karmakar}
\email{purnendu.karmakar@ut.ee}
\affiliation{Institute of Physics, University of Tartu, W. Ostwaldi 1, 50411 Tartu, Estonia}
%\affiliation{INAF - Osservatorio Astronomico di Roma, Via Frascati 33, 00040, Monteporzio Catone, Roma, Italy}

\author{Sandeep Haridasu}
\email{sharidas@sissa.it}
\affiliation{SISSA, Via Bonomea 265, 34136 Trieste, Italy}
\affiliation{IFPU - Institute for fundamental physics of the Universe, Via Beirut 2, 34014 Trieste, Italy}
\affiliation{INFN-Sezione di Trieste, via Valerio 2, 34127 Trieste, Italy}

\author{Atsushi Nishizawa}
\email{atnishi@hiroshima-u.ac.jp}
\affiliation{Physics Program, Graduate School of Advanced Science and Engineering, Hiroshima University, Higashi-Hiroshima, Hiroshima 739-8526, Japan}
\affiliation{Astrophysical Science Center, Hiroshima University, Higashi-Hiroshima, Hiroshima 739-8526, Japan}

%%%%%%%%% Abstract %%%%%%%%%%
\begin{abstract}
We address various cosmological phenomenologies in the symmetric teleparallel framework both in background and perturbation such as cosmic expansion, gravitational coupling constant, gravitational waves propagation. Focusing on logarithmic extensions of $f(Q)$ models, we performed Bayesian analysis using the most-recent cosmological data, DESI DR2, Pantheon+. We also utilized a compilation of redshift space distortions ($f \sigma_8$) dataset to constrain the growth of structures in each of the models modulated by the effective gravitational coupling. We find that our extended Logarithmic $f(Q)$ models are well-constrained by the current cosmological data and are able to describe the late-time cosmic acceleration. The inverse Logarithmic model we introduce is also able to accommodate a phantom-like dark energy equation of state at late times, which is consistent with the recent DESI DR2 observations. We report explicitly  predictions for the effective gravitational coupling ($\mu$), and the amplitude damping parameter of gravitational wave ($\nu$) solely based on the background data, which can be tested against future observations. While the two Log-based extensions we have introduced here perform equivalently on the background level, they provide contrasting predictions for the evolution of effective Gravitational constant and propagation of gravitational waves, which should be constrained against the future perturbation data. 
\end{abstract}
% \keywords{keywords}

\date{\today \ at \currenttime \ of Rome} % Dated: November 16, 2018 at 01:27 of Japan %\currenttime
\maketitle
% \input{acros}
% \tableofcontents % Table of the contents

%%%%%%%%%%%%%%%%%%%%%%%%%%%%%%%%%%%%%%%%
%%%%% Main Content starts from here %%%%

\section{Introduction}
\label{sec:Introduction}
The standard cosmological model, $\Lambda$CDM, has been successful in explaining a wide range of cosmological observations \cite{Planck:2018nkj, Planck:2015fie, DESI:2025zgx, Scolnic:2021amr, Alam:2016hwk}. However, it faces challenges, such as the cosmological constant problem and the nature of dark energy, which keep open the possibility of exploring alternative theories of gravity at cosmological scales (see for instance \cite{CosmoVerse:2025txj,Perivolaropoulos:2024yxv,Perivolaropoulos:2021jda}). Given \ac{GR} describes gravity through spacetime geometry, it is natural to consider modifications involving alternative geometric properties, such as torsion and non-metricity \cite{BeltranJimenez:2019esp}. 
Findings of the equivalent formulation to \ac{GR}\cite{BeltranJimenez:2019tjy}, such as the \ac{TEGR} for torsion and the \ac{STEGR} for non-metricity \cite{Nester:1998mp}, which motivate the exploration of these alternative geometrical components further \cite{Bahamonde:2021gfp}. 
% \cite{Peebles:2002gy, Copeland:2006wr, Copeland:2006wr}

Due to the natural geometrical extension, one of such non-metricity extension, whiting the action as a function of non-metricity, called $f(Q)$ gravity models recently received significant attention from both theoretical and observational cosmology communities. $f(Q)$ gravity \cite{BeltranJimenez:2019tme} produces many observables similar to $f(R)$ gravity models \cite{DeFelice:2010aj}. The modifications of $f(Q)$ gravity allow for a more flexible framework to explore the dynamics of cosmic expansion and structure formation. The applications of $f(Q)$ model span a range of contexts, from late-time acceleration to gravitational wave propagation \cite{BeltranJimenez:2019tme, Ferreira:2022jcd,Ferreira:2023tat}. In this context, we aim to constrain the parameters of the $f(Q)$ gravity model using observational data.

Various functional forms of $f(Q)$ have been explored in the literature, including 
exponential \cite{Anagnostopoulos:2022gej, Yang:2024tkw, Gadbail:2024een, Su:2024avk, Gadbail:2024rpp, Mhamdi:2024xqd, Vasquez:2025hrz, Ayuso:2025vkc, Roy:2025nde, Boiza:2025xpn}, 
power-law \cite{BeltranJimenez:2019tme, Anagnostopoulos:2022gej, Frusciante:2021sio, Bhardwaj:2024mop, Gadbail:2024een, Sahlu:2024pxk, Enkhili:2024dil, Gadbail:2024rpp, Kolhatkar:2024oyy, Sakr:2024eee, Wang:2024eai, Mhamdi:2024xqd, Yang:2024tkw, Ayuso:2025vkc, Paliathanasis:2025hjw, Roy:2025nde, DAgostino:2022tdk}, 
hyperbolic tangent \cite{Anagnostopoulos:2022gej, Su:2024avk}, 
logarithmic \cite{Najera:2023wcw, Anagnostopoulos:2021ydo, Anagnostopoulos:2022gej, Goncalves:2024sem, Gadbail:2024een, Sakr:2024eee, Roy:2025nde} 
models, and a few other extensions \cite{Jarv:2023sbp, Bahamonde:2022zgj, Heisenberg:2023lru}. 
The latter of which is defined by a logarithmic dependence on the non-metricity scalar $Q$ introducing an additional parameter that can be redundant from the flatness/closure condition, offering less flexibility to explain the dynamic of the cosmic expansion \cite{Najera:2023wcw}. Also, the logarithmic extension of $f(Q)$ gravity has been studied the least among these. 

% Logarithmic extensions can emerge naturally in quantum field theory - inspired corrections, entanglement entropy, and renormalization group logarithms \cite{Donoghue:1994dn, Solodukhin:2011gn, Calabrese:2004eu}.
% Its slowly growing, with cumulative deviations wrt \ac{GR} persisting at cosmological scales, in contrast to the rapidly suppressed exponential or power-law forms \cite{BeltranJimenez:2020iyx, Harko:2018gxr}. 
% This makes it a promising candidate to capture long-lasting infrared modifications relevant for late-time cosmic acceleration in $f(Q)$ extensions.

To address the limitations of the existing simple logarithmic $f(Q)$ model, particularly its dependence on the phenomenological normalization parameter $Q_0$, we introduce two new extensions: the generalized logarithmic (gLog) and inverse logarithmic (iLog) models. The gLog model incorporates an additional parameter $\xi$ that removes the teleological issue of the Lagrangian depending on the present-day value $Q_0$ and, at the same time, allows for modulation of perturbative observables beyond the constraints imposed closure condition alone. In contrast, the iLog model is designed to yield perturbative behavior opposite to that of gLog case, thereby offering a complementarity. Therefore, while both models share nearly identical background expansion histories, they generate contrasting predictions for key perturbative quantities such as the modification of the effective gravitational couplings ($\mu$), the dark energy equation of state, ranging from phantom-like to quintessence-like behavior. This dual construction provides a phenomenologically rich setup in which two models with degenerate background dynamics can nevertheless be distinguished at the perturbative level, making them ideal targets for precision cosmological tests with future data.

% \AN{I know logarithmic corrections often appear in physics. But particularly for nonmetricity? I am not convinced yet as a reason for concentrating on only the logarithmic model.}

% The \pk{simple} logarithmic model has been shown in \cite{Najera:2023wcw} to provide a good fit to various cosmological observations, including the \ac{BAO}, and supernovae data and appropriate priors from \ac{CMBR}. \PK{please review this line. I don't understand if you refer to Najera:2023wcw paper, or our pending reconstruction paper!}

The main goal of this work is to provide constraints on the logarithmic model extension of $f(Q)$ gravity using observational data. In this work, we perform a Bayesian analysis to constrain the parameters of a logarithmic extensions of $f(Q)$ gravity using the \ac{BAO} data reported by most-recent DESI data-release 2 (DR2) \cite{DESI:2025zgx}, SNe Type-Ia compilation in Pantheon+ \cite{Scolnic:2021amr} and priors derived from \ac{CMBR} \cite{Planck:2018vyg}. Additionally, we explore the implications of these constraints for the nature of dark energy, the dynamics of cosmic expansion and evolution of the growth of structures against the redshift space distortion data \cite{Nesseris:2017vor}. Several attempts at reconstructing the dynamical connection have also been done \cite{Kar:2021juu, Capozziello:2022wgl, Gadbail:2022jco, Gadbail:2023klq, Naik:2023ykt, Mahmood:2023mac, Gadbail:2023mvu, Kaczmarek:2024yju, Gadbail:2024rpp, Kaczmarek:2024quk, Gadbail:2024lek, ElOuardi:2025okl, Gadbail:2024rpp, Yang:2024tkw, Yadav:2024vmt, Saha:2024gmk, Esposito:2021ect, Kang:2021osc}, designer approach \cite{Albuquerque:2022eac}, and also investigating the possibility to alleviate tensions \cite{Sakr:2024eee}.

Most recent DESI-DR2 \cite{DESI:2025fii} observations have provided hints/evidence of an evolving dark energy equation of state, $w_{\rm de}$. Motivated by this, we investigate whether the $f(Q)$ gravity framework, particularly logarithmic models—can account for such evolution. It has been shown on several occasions that $f(Q)$ models are evident to a phantom-like \cite{Basilakos:2025olm, Lymperis:2022oyo, Narawade:2022cgb, Arora:2022mlo} dark energy equation of state at late times.

The possible dynamical degrees of freedom (DOF) of $f(Q)$ models and Hamilton analysis have been discussed in \cite{Hu:2022anq, DAmbrosio:2023asf, Tomonari:2023wcs, Guzman:2023oyl}, and there are ambiguities regarding the counting and nature of the degrees of freedom (DOF), with some analyses suggesting up to 7 propagating modes. Recent studies have raised concerns about strong coupling issues in $f(Q)$ gravity, that the DOF are background-dependent, and at least one of 7 DOF is ghost instability in the perturbation regime, and  \cite{Gomes:2023tur}. Authors in \cite{Guzman:2024cwa} show that for the simple connection set (which we adopt here) exhibit stable behaviour throughout all evolutionary epochs in the \ac{GR} regime, and the general conditions for the physically viable f(Q) models.  
Moreover, a possible ghost-free class of symmetric teleparallel theories based on a symmetry under transverse diffeomorphisms has recently been proposed \cite{Bello-Morales:2024vqk}. Although all $f(Q)$ models are strongly coupled at the fundamental level, implying that hidden degrees of freedom reappear at higher orders in perturbation theory the cosmological branch relevant for late-time dynamics and observational tests does not exhibit explicit instabilities at background or linear order. An unified dynamical systems framework around spatially flat \ac{FLRW} cosmology has been studied for $f(Q)$ for dynamics viability \cite{Dutta:2025fqw, Khyllep:2022spx, Narawade:2022jeg}. 
Thus, it remains meaningful to constrain these models phenomenologically \cite{Tomonari:2025axd}, while recognizing that a consistent ultraviolet (UV) completion would ultimately be required to resolve the strong coupling problem. 
In this context, our observational analysis of $f(Q)$ gravity using current cosmological datasets provides a robust test of its phenomenological viability, despite these open theoretical challenges.

The organization of the paper is as follows: In Sec.~\ref{sec:fQgravity}, we introduce the $f(Q)$ gravity and its background equations and the perturbation equations, including the Poisson equation, and the propagation of gravitational waves. In Sec.~\ref{sec:data}, we describe the observational data used in this work. In Sec.~\ref{sec:results}, we present our results. Finally, in Sec.~\ref{sec:conc}, we conclude our findings.

%%%%%%%%%%%%%%%%%%%%%%%%%%%%%%%%%%%%%
%%%%%%%%%% f(Q) gravity %%%%%%%%%%%%%
\section{f(Q) gravity}
\label{sec:fQgravity}
% The new simple geometrical formulation of GR motivates a promising new framework for studies of modified gravity. To demonstrate this with an example, 

The action of $f(Q)$ gravity is given by  
 \begin{equation}   
 \label{ac:fq:grav}
\mathcal{S}  =   \int {\diff^4} x \sqrt{-g} \Big[ - \frac{1}{2\kappa^2} f(Q) + \mathcal L_M (\rm{CDM, b, rad})\Big]\,.
 \end{equation}
$\kappa^2 = 8\pi G_N$, $c=1$ and $G$ is the Newton's constant. The matter Lagrangian $\mathcal L_M (\rm{CDM, b, rad})$ includes both, relativistic matter (radiation, ignoring neutrino) and non relativistic matter (\ac{CDM}and baryons). $f(Q)$ is a function of $Q$, which is the non-metricity scalar. 
The non metricity scalar is made of the contraction of the non-metricity tensor $Q_{\alpha\mu\nu}$, and its conjugate, $P^{\alpha\mu\nu}$, 
\beq
Q = -Q_{\alpha\mu\nu}P^{\alpha\mu\nu}\,.
\eeq
The fundamental object is the non-metricity tensor is, $Q_{\alpha\mu\nu}=\nabla_\alpha g_{\mu\nu}$, which is by nature  metric incompatible. 
The non-metricity conjugate comes with the combinations of the two independent traces $Q_\alpha=g^{\mu\nu}Q_{\alpha\mu\nu}$ and $\tilde{Q}_\alpha=g^{\mu\nu}Q_{\mu\alpha\nu}$ 
\begin{equation}
P^\alpha{}_{\mu\nu} = -\frac{1}{2}L^\alpha{}_{\mu\nu} + \frac{1}{4}\left( Q^\alpha - \tilde{Q}^\alpha \right) g_{\mu\nu} - 
\frac{1}{4}\delta^\alpha_{(\mu}Q_{\nu)}\,.
\end{equation}

The non-metricity conjugate, $P^{\alpha\mu\nu}$, satisfies the variation of non-metricity scalar with respect the the non-metricity tensor, 
\beq
P^{\alpha\mu\nu}=-\frac12\frac{\partial Q}{\partial Q_{\alpha\mu\nu}}.
\eeq

At the background level we assume a homogeneous, isotropic and spatially flat \ac{FLRW} \cite{Green:2014aga} spacetime, whose line element is in the form, $\diff s^2 = -\diff t^2 + a^2(t)\delta_{ij}\diff x^i \diff x^j$ which gives the expansion rate, i.e., Hubble parameter, $H=\dot{a}/a$ with respect to the cosmic time, $t$. The changes in Hubble parameter is $\frac{\ddot a}{a} = \dot H + H^2$. 
%  \AN{No $\kappa^2$? What is the dimension of $Q$? Isn't it same as $R$?} \PK{Yes} 

There are three sets of coordinate choices for the connections. From here onwards, we shall use the simplest coordinate choice, $Q=6H^2$ \cite{Gomes:2023tur}. Note that the one should not confuse this coordinate choice with coincident gauge. A vanishing total affine connection is called as coincident gauge, i.e., $\Gamma^\alpha_{\phantom{\alpha}\mu\nu}=0$, which leads to the disformation equal to minus Levi-Civita connection in the absence of torsion. One can choose to restrict the model within coincident gauge, on the top of the coordinate choice, $Q=6H^2$.

% The general total affine connection \cite{Ortin:2015hya} 
% \begin{equation} \label{decomposition}
% \Gamma^\alpha_{\phantom{\alpha}\mu\nu}=\left\{^{\phantom{i} \alpha}_{\mu\nu}\right\} + K^\alpha_{\phantom{\alpha}\mu\nu} + L^\alpha_{\phantom{\alpha}\mu\nu}\,,
% \end{equation}

% In the coincident gauge, the total affine connection vanishes, $\Gamma^\alpha_{\phantom{\alpha}\mu\nu}=0$, which leads to the disformation equal to minus Levi-Civita connection in the absence of torsion. We shaaluse 
% In the coincident gauge, we have $Q=6H^2$. 

\subsection{Background Equations}
\label{sec:background}
Under the  coordinate choice, $Q=6H^2$, the model produces the background Friedmann equations,
\begin{eqnarray}
6H^2f_Q-\frac{f}{2}&=& \kappa^2 \rho_M 
\,, \label{eom:fQ:back:scalar:1}  %\nonumber 
\\
2 \big(12H^2f_{QQ}+f_Q\big)\dot{H}&=&-\kappa^2(\rho_M+p_M)
\,, \label{eom:fQ:back:scalar:2} %\nonumber
\end{eqnarray}
% \AN{Do we need the subscript M? The $\rho$ and $p$ are general ones and I suggest not fixing M.} \PK{General in the sense, relativistic and non-relativistic matters? what else does $\rho$ in the RHS include other than relativistic and non-relativistic matters? The alternative way to write would be $\rho_{\rm m} + \kappa^2 \rho_{\rm {r}}$ explicitly as written in \eqeqref{eom:fQ:Friedmann:F:explicit}. If we remove $M$, in order to be consistent, the action also needs explicit $\mathcal L_m (\rm{CDM, b}) + \mathcal L_r (\rm{rad}) $. We can write in that way, if you like. To keep the starting action looks shorter, I used $M$. I like to write explicitly what the black-box of matter looks like, particularly if we are using radiation or neutrino.} \AN{If you stick to the subscript M, I am ok with it but the standard terminology in cosmology is,  relativistic particles $=$ radiation and non-relativistic particles $=$ matter.} \an{I just simply fixing $M$ or not fixing $M$ in Eqs.(1), (5), and (6). Either is fine, but it would be good to note that $M$ is not {\it{matter}} in cosmology.}
% We need to include radiation, therefore, I wanted to $M$ is explicitly We can explicitly write }
where $f_Q \equiv f_Q (Q)\equiv\partial f/\partial Q$ and $f_{QQ} \equiv f_{QQ} (Q)\equiv\partial^2 f/\partial Q^2$. In general, $\rho_M$ and $p_M$ in the \ac{RHS} represent the total matter density. $\rho_M=\rho_m + \rho_r$, where the $\rho_m $\footnote{We assume that the matter, i.e., baryons and \ac{CDM}is pressureless ($p_m = 0$) dust. } includes cold dark matter and baryons. Assuming the flat universe, which is strongly supported by \ac{CMBR} observations \cite{Planck:2015fie}. These individual matter field components satisfy the continuity equation, $\dot{\rho_i}=-3H(\rho+p_i)$.

Since the action with linear $Q$ produces the \ac{GR} equivalence, called \ac{STEGR} \cite{BeltranJimenez:2017tkd}, we are considering the modifications to the \ac{GR} comes through an additional term, $F(Q)$. Therefore, we will consider the function in the form of $f(Q)=Q+F(Q)$. Note that we can always write the linear part separately by defining $F(Q) \equiv f(Q)-Q$. The Friedmann and Raychaudhuri equations now become, 
\begin{eqnarray} 
3H^2+6H^2 F_Q - \frac{F}{2}&=& \kappa^2 \rho_{\rm m} + \kappa^2 \rho_{\rm {r}}
\,, \label{eom:fQ:Friedmann:F:explicit}
\\
2\dot{H} \big(12H^2F_{QQ}+1+F_Q\big) \nonumber &&\\
+6H^2(1+F_Q)-\frac{6H^2+F}{2} &=&-\kappa^2 p_{\rm m} -\kappa^2 p_{\rm r} 
\,, \label{eq:acceleration}
\end{eqnarray}
where $F_Q \equiv F_Q (Q)\equiv\partial F/\partial Q$ and $F_{QQ} \equiv F_{QQ} (Q)\equiv\partial^2 F/\partial Q^2$. Assuming the coordinate choice, $Q=6H^2$,
\begin{eqnarray} 
3H^2&=& \kappa^2 (\rho_{\rm m} + \rho_{\rm r}+\rho_{\rm de})
\,, \label{eq:Friedmann-eq}
\\
2\dot{H} +3H^2 &=&-\kappa^2 (p_{\rm m} + p_{\rm r}+p_{\rm de})
\,, \label{eq:acceleration-eq}
\end{eqnarray}

where we have defined
\begin{eqnarray}
\rho_{\rm de} &\equiv& \frac{1}{\kappa^2}\left( \frac{F}{2} -6H^2 F_Q \right)\;, 
\label{eqn:rhode} \\
p_{\rm de} &\equiv& \frac{1}{\kappa^2}\left( 2 \dot{H} ( 12 H^2F_{QQ} + F_Q) - \frac{F}{2} + 6H^2 F_Q \right) \;.
\label{eqn:pde}
\end{eqnarray}
From Eq.~\eqref{eq:Friedmann-eq}, we can see that the critical density of the universe is defined in a standard way as (in the absence of curvature, the total energy density equals to critical density),  
\begin{equation}
\rho_{\rm c} \equiv \frac{3H^2}{\kappa^2} \;.
\end{equation}
Then the normalized/fractional energy densities are 
\begin{equation}
\Omega_{\rm m} \equiv \frac{\rho_{\rm m}}{\rho_{\rm c}} \,, \, 
\Omega_{\rm r} \equiv \frac{\rho_{\rm r}}{\rho_{\rm c}} \,,
\quad \Omega_{\rm de} \equiv \frac{\rho_{\rm de}}{\rho_{\rm c}} = \frac{F}{6H^2} - 2F_Q \;, \label{eq:Omega-parameters}
\end{equation}
where the present values of $\rho_{\rm c}$, $\Omega_{\rm m}$, $\Omega_{\rm r}$, and $\Omega_{\rm de}$ are denoted by $\rho_{\rm c0}$, $\Omega_{\rm m0}$, $\Omega_{\rm r}$, and $\Omega_{\rm de0}$.
Defining $E\equiv H/H_0$, we rewrite Eq.~\eqref{eq:Friedmann-eq} as
\bea 
E^2 
&=& \Omega_{{\rm m}0}(1+z)^3 
+ \Omega_{{\rm r}0}(1+z)^4
+ \Omega_{{\rm de}0} R_{\rm de}  \,
% &=& \Omega_{{\rm m}0}(1+z)^3 + (1-\Omega_{{\rm m}0})R_{\rm de} \,,
\label{eq:Friedmann-eq-E}
\eea
with
\beq
R_{\rm de} \equiv \frac{\rho_{\rm de}}{\rho_{\rm{de}0}} = \frac{ F - 12H^2 F_Q }{F_0 - 12H_0^2 F_{Q_0} } \,.
\;, \label{eq:Rde}
\eeq
% \begin{align}
% \rho_{\rm de 0} &\equiv \frac{1}{\kappa^2}\left( \frac{F_0}{2} -6H_0^2 F_{Q_0} \right)\;.
% \end{align}
From the flatness of the Universe, we obtain the relation
\beq
\Omega_{{\rm de}0} = 1- \Omega_{{\rm m}0} - \Omega_{{\rm r}0} = \frac{\rho_{\rm de 0}}{\rho_{\rm c 0}} =\frac{F_0}{6 H_0^2} - 2 F_{Q_0}\;. \label{eq:flatness-cond}
\eeq
% One can use the Eq. \eqref{eq:flatness-cond} to reduce one model parameter. Assuming pressureless dust, $\dot{H}$ can be solved from the Eq.~\eqref{eq:acceleration}. 
% \begin{eqnarray}
% \dot{H} &=& \frac{F-6H^2(1+2F_Q)}{4(1+F_Q+12H^2F_{QQ})} \;.\label{exp:Hdot:1}\\
% &=& \frac{-3H^2 + \rho_{\rm de} }{ 4 (1+6 H^2 F_{QQ})}\\
% &=& \frac{-3H^2 + 3H_0^2 R_{\rm de} \Omega_{\rm de 0} }{ 4 (1+6 H^2 F_{QQ})}\,. \label{exp:Hdot:Rde_Omegade}
% \end{eqnarray}
The equation of state of the dark energy component $(w_{\rm de})$ (from modification of gravity) is the ratio of pressure ($p_{\rm de}$) and density ($\rho_{\rm de}$) of the dark energy, i.e., 
% \begin{equation}
% w_{\rm de} \equiv \frac{p_{\rm de}}{\rho_{\rm de}} = -1-\frac{4\dot{H}(12 H^2F_{QQ}+F_Q)}{12H^2 F_Q - F} \;. \label{eq:DE-EOS}
% \end{equation}
% Substituting this into Eq.~\eqref{eq:DE-EOS}, we have 
\bea
w_{\rm de} &=& \frac{-F +4 \left(3
   H^2+\dot{H}\right)F_Q + 48 H^2 \dot{H} F_{QQ}}{F-12 H^2 F_Q}\,,
   % &=&\frac{-F+6 H^2 F_Q - 72 H^4 F_{QQ}}
   % {\left(F-12 H^2 F_Q\right) 
   % \left(1+F_Q+12 H^2 F_{QQ}\right)}
    \; \label{eq:wde}
\eea
and the deceleration parameter is defined as, 
\begin{eqnarray} \label{eq:q}
q &=& -\frac{\ddot{a} a}{\dot a^2} \equiv - \left(\frac{ \dot{H}}{H} +1 \right).
% &=& -1 -\frac{-6 H^2+F-12 H^2 F_Q}{4 H \left(1+F_Q+12 H^2 F_{QQ}\right)}
\end{eqnarray}

%%%%%%%%%%%%%%%%%%%%%%%%%%%%%%%%%%%%%%
%%%%%%%%% Poission Equation %%%%%%%%%%
\subsection{Perturbation equations}
\label{sec:perturb}
In this section we briefly describe the scalar and tensorial perturbations that govern the structure formation and the propagation of gradational waves. 

\subsubsection{Scalar perturbations}
{\it Effective gravitational coupling}, $\mu$,  
% and the {\it light deflection parameter}
% , $\Sigma$, enter the modified Poisson and lensing equations, respectively, which 
can be written in Fourier space \cite{Amendola:2007rr} as, 
\begin{eqnarray}
&&-\frac{k^2}{a^2} \Psi = 4 \pi G_N \,\mu(a, k)  \rho_\mathrm{m} \delta_{\rm m}\,,
\label{eq:muSigma}
% &&-\frac{k^2}{a^2} (\Psi+\Phi) = 8\pi G_N\Sigma(a,k) \rho_\mathrm{m}\delta_{\rm m}\,, \label{eq:lenseq}
\end{eqnarray}
where the density contrast w.r.t. the background is represented as $\delta_{\rm m}=\delta \rho_{\rm m}/\rho_{\rm m}$ and the wave number is represented as $k$. Here $\mu$ encodes the deviations in the strength of gravity within $f(Q)$ gravity which converges to $\mu\to1$ in the $\Lambda$CDM case. Assuming a quasi-static approximation which is valid on scales much smaller than the Horizon, one can write \cite{BeltranJimenez:2019tme}, 
\begin{eqnarray}
\mu(a)&=&\frac{1}{f_Q}\, \equiv\frac{1}{1+F_Q}\,.
\label{eq:Possion-mu}
\end{eqnarray}

% The deviation of the effective gravitational coupling from $\Lambda$CDM scenario is captured in $\mu$. 
% The deviation in the lensing gravitational potential ($\phi_{\rm len}=(\Phi+\Psi)/2$) is captured in $\Sigma$. 
% The $\Lambda$CDM model is recovered when $\mu=\Sigma=1$.

% GR : 
% In order to map the $f(Q)$ gravity within the above formalism one can employ 

% In the perturbations deep inside the Hubble radius, the quasi-static approximation. Following this, one can find that the gravitational potentials are equal as in GR (i.e. $\Phi=\Psi$) and that the two above equations match~\cite{Jimenez:2019ovq}:
% %
% \begin{equation}
%     -k^2\Psi= \frac{4 \pi G_N}{f_Q}a^2\rho_{\rm m} \delta_{\rm m}\,.\label{eq:fq:Poisson}
% \end{equation}
% %
% The effective gravitational coupling is then defined as:
% %
% \begin{eqnarray}
%          \mu(a)&=&\frac{1}{f_Q}\, \equiv\frac{1}{1+F_Q}\,.
%      \label{eq:Possion-mu}
% \end{eqnarray}

%%%%%%%%%%%%%%%%%%%%%%%%%%%%%%%%%%%%%
%%%%%%%%%% GW %%%%%%%%%%%%%

\subsubsection{Tensor perturbations }

The propagation of \acp{GW} is governed by the modified equation (in conformal time) \cite{BeltranJimenez:2019tme}
\begin{equation}\label{eom:fq:gw:conformal}
h''_{(\lambda)}+2\mathcal{H}\left(1+\frac{\diff \log f_Q}{2\mathcal{H}\diff \eta}\right) h'_{(\lambda)}+k^2h_{(\lambda)}=0.
\end{equation}

Comparing with the general propagation equation of GW \cite{Nishizawa:2017nef}
\begin{equation}\label{eom:general:gw:conformal}
h''_{ij}+\left(2+\nu\right) \mathcal{H}  h'_{ij}+(c_T^2k^2 + a^2\mu^2) h_{ij}= a^2 \Gamma\gamma_{ij}.
\end{equation}
The deviation of the amplitude damping in the propagation of GW from \ac{GR}is 
\begin{eqnarray}
\nu&=& \frac{1}{\mathcal{H}} \frac{\diff \log f_Q}{\diff \eta} \equiv \frac{1}{H}\frac{\diff \log f_Q}{\diff t} \,, \nonumber \\
&=& 12\dot{H}\frac{f_{QQ}}{f_Q} \;,\\
&=& 12\dot{H}\frac{F_{QQ}}{1+F_Q} \;, \label{exp:nu:FQ-FQQ-Hdot}
\label{eq:GW-nu}
\end{eqnarray}
where the dot is the derivative with respect to cosmic time, $t$ and prime is defined in conformal time ($\eta$). These two times are related by $dt = a d \eta$. 
Just a gentle reminder, these model produces the same speed of gravitational waves ($c_T$) as light (c). Just a reminder, we replaced $Q$ with $6H^2$ as coordinate choice assumed throughout this work.

\subsection{Interpretation of modified gravitational coupling $G_{\rm eff}$}
\label{sec:interpretation}
The effective gravitational coupling ($8\pi G_{\rm eff} \equiv \kappa_{\rm eff}$) in $f(Q)$ gravity can be defined in several equivalent ways, which appear differently in various equations and observables. 
From the covariant form of Einstein’s equations, the coupling between matter energy–momentum and the Einstein tensor yields, 
\begin{equation}
\kappa_{\rm eff}^2 = \frac{\kappa^2}{f_Q}\;,
\end{equation}
as defined in \cite{Guzman:2024cwa}. 
A similar expression could arise in the weak-field limit around a Schwarzschild background at $R \to \infty$, where the identification of the Newtonian coupling again gives $\kappa_{\rm eff}^2 = \kappa^2 / f_Q$. 
Furthermore, post-Newtonian (PPN) analyses suggest that $f(Q)$ gravity reproduces the same PPN parameters as GR, implying consistency with the standard gravitational coupling in Solar System tests \cite{Flathmann:2020zyj}. 
At the perturbative level, the effective coupling entering the Poisson equation \eqeqref{eq:muSigma} is also identical to the $G_{\rm eff}$ defined covariantly, i.e.\ $\kappa_{\rm eff}^2 = \kappa^2 / f_Q$.  

Of course, when comparing the $G_{\rm eff}$ from observations, one should directly compare $G_{\rm eff}$ evaluated from respective equations. 
When comparing other observables where $G_{\rm eff}$ enters only implicitly, its definition can be chosen for convenience. 
For instance, the background expression and solution for $H^2$ remains unchanged in \eqeqref{eom:fQ:back:scalar:1} regardless of whether one uses definition $G_{\rm eff}$ or the standard $G_{\rm N}$. 
However, the choice of gravitational constant directly affects the definition of the critical density $\rho_c$, which in turn modifies the derived densities $\rho_m$ and $\rho_{\rm DE}$. 
In practice, for a given modified gravity model, one always compares the density parameters $(\Omega_m, \Omega_{\rm DE})$ against their standard observational values in $\Lambda$CDM, typically $\Omega_m \approx 0.3$ and $\Omega_{\rm DE} \approx 0.7$ . 
Therefore, it is convenient to retain the standard Newtonian constant $G_{\rm N}$ in defining $\rho_c$, and consequently $\rho_m$ and $\rho_{\rm DE}$. 
If one instead chooses to use the modified definition $\kappa_{\rm eff}^2 = \kappa^2 / f_Q$, then the matter and dark energy densities must also be consistently rescaled.  

What is physically relevant in observational comparisons is the actual matter density $\rho_m$, not the choice of critical density. 
If one adopts a modified gravitational coupling $G_{\rm eff}$ to define $\rho_c$, then the observed density parameters must be rescaled accordingly, such that  
\begin{equation}
(\Omega_m \rho_c)_{\rm def,1} = (\Omega_m \rho_c)_{\rm def,2} = (\rho_m)_{\Lambda{\rm CDM}},
\end{equation}
ensuring a consistent comparison. In our background analysis, we adopt the standard \ac{GR} approach consistently. 

\section{Modelling}
In this section we describe the three specific $f(Q)$ models involving the Logarithmic form as a function of scalar non-metricity $Q$ as an extension to \ac{STEGR}.  
\subsection{Logarithmic model}
The logarithmic model (hereafter referred as `Log') considered in \cite{Najera:2023wcw} is
\beq
f(Q)= Q- \Gamma \ln \left( \frac{Q}{Q_0} \right) \;.
\eeq

\begin{subequations}\label{eq:log-xi-derivs}
\begin{align}
F(Q)&
= -\Gamma \ln \left( \frac{Q}{Q_0} \right) \;, \\
F_Q&
= -\frac{\Gamma}{Q} \;, \quad F_{QQ} = \frac{\Gamma}{Q^2} \;. 
\end{align}
\end{subequations}

From Eq.~\eqref{eq:Rde} the Log-model provides,
\beq
R_{\rm de} = 1 - \ln E \;,
\eeq
% where we replaced $Q$ with $6H^2$ in the coincidence gauge. 
% The Friedmann equation in Eq.~\eqref{eq:Friedmann-eq-E} is reduced to
% \beq 
% E^2 = \Omega_{{\rm m}0}(1+z)^3 + (1-\Omega_{{\rm m}0} -\Omega_{{\rm r}0})\left[ 1 - \ln E \right]
% \;.
% \eeq
The flatness/closure condition in \eqeqref{eq:flatness-cond} the Log-model gives 
\beq
\Gamma = 3H_0^2 \Omega_{{\rm de}0} \equiv Q_0 \frac{\Omega_{{\rm de}0}}{2}.
\label{eqn:gammalog}
\eeq
which is correctly tracking the dimensions of $Q$.

%------------------------------------------------------------
\subsection{Generalised logarithmic model}
\label{sec:log-model-c}

We extend the above logarithmic model to a generalised case (hereafter referred as `gLog')
\begin{equation}
  f(Q)\;=\;Q-\Gamma\,
  \ln\!\Bigl(\frac{Q}{Q_{0}}+\xi\Bigr),
  \qquad \xi = \rm{const,}
  \label{eq:fQ-log-xi}
\end{equation}
% With the coincidence gauge \(Q=6H^{2}=Q_{0}E^{2}\) and
% \(Q_{0}=6H_{0}^{2}\),
The quantities entering the field equations are then 
\begin{subequations}\label{eq:glog-xi-derivs}
\begin{align}
  F(Q)&
  =-\Gamma\ln\!\bigl(E^{2}+\xi\bigr), \\[4pt]
  F_{Q}&
  % =-\frac{\Gamma}{Q+\xi Q_{0}}    
        =-\frac{\Gamma}{Q_{0}\bigl(E^{2}+\xi\bigr)}\,,\\
    F_{QQ}& 
    =\frac{\Gamma}{Q_{0}^2\bigl(E^{2}+\xi\bigr)^2}\,.
\end{align}
\end{subequations}

From Eq.~\eqref{eq:Rde}, the gLog model provides, 
% Using the definition we obtain, 
\begin{equation}
  R_{\rm de}(E;\xi)
  \;=\;
  \frac{\displaystyle
        \dfrac{2E^{2}}{E^{2}+\xi}
        -\ln\!\bigl(E^{2}+\xi\bigr)}
       {\displaystyle
        \dfrac{2}{1+\xi}-\ln(1+\xi)}.
  \label{eq:Rde-log-xi}
\end{equation}
The flatness/closure condition in Eq.~\eqref{eq:flatness-cond} gives the $\Gamma$ for gLog-model, 
\begin{equation}
\Gamma = \frac{6 H_0^2 \, \Omega_{\mathrm{de}0}}{\frac{2}{1 + \xi} -  \log(1 + \xi)} \,,
\label{Sol:Gamma:Gen_Log}
\end{equation}
wherein it is immediate to validate that the as $\xi\to0$, it reduces to \eqeqref{eqn:gammalog}.

%------------------------------------------------------------
\subsection{Inverse–logarithmic model}
\label{sec:invlog-model}

We now introduce and generalise the logarithmic ansatz to the inverse (hereafter referred as `iLog') form
\begin{equation}
  f(Q)\;=\;Q-\frac{\Gamma}{\ln\!\Bigl(\dfrac{Q}{Q_{0}}+\xi\Bigr)}\,,
  \qquad \xi = \rm{const},
  \label{eq:fQ-invlog}
\end{equation}
so that \(\xi\!\to\!0\) recovers \(\,f(Q)=Q-\Gamma/\ln(Q/Q_{0})\). Writing the denominator of \eqeqref{eq:fQ-invlog}, 
\bea
  \mathcal{D} &\equiv& \ln\!\Bigl(\dfrac{Q}{Q_{0}}+\xi\Bigr),\qquad
  Q_{0}\equiv6H_{0}^{2}\,.
\eea

Identifying \(Q=Q_{0}E^{2}\) with \(E(z)\equiv H(z)/H_{0}\) and having once can infer, 

\bea
  \mathcal{D} &=&\ln(E^{2}+\xi)\,,\\
  \mathcal{D}_{0}\equiv \mathcal{D} \big|_{z=0} &=&\ln(1+\xi)\,.  
\eea

Then
\begin{subequations}
\begin{align}
  F(Q)&\equiv = -\frac{\Gamma}{\mathcal{D}},
  \\[2pt]
  F_{Q}&=
         \frac{\Gamma}{Q_{0} \mathcal{D}^{2} (E^2+\xi) }\,, \label{exp:FQ:inverse-log}\\
 F_{QQ}&=-
         \frac{\Gamma  (2+\mathcal{D})}{Q_0^2 \mathcal{D}^3 (E^2 + \xi)^2}\,.
         % &=-
         % \frac{ F_{Q} (2+\mathcal{D})}{Q_0 \mathcal{D} (E^2 + \xi)}\,. \label{exp:FQQ:solve_Gamma}
\end{align}
\end{subequations}

Using the Eq.~\eqref{eq:Rde}, the iLog model provides, 
\begin{equation}
    R_{\rm de}(E;\xi)=
    \frac{\displaystyle
          \dfrac{2E^{2}}{(E^{2}+\xi)\,\mathcal{D}^{2}}
          +\dfrac{1}{\mathcal{D}}}
         {\displaystyle
          \dfrac{2}{(1+\xi)\,\mathcal{D}_{0}^{2}}
          +\dfrac{1}{\mathcal{D}_{0}}}\,.
  \label{eq:Rde-invlog-final}
\end{equation}

The flatness/closure condition in \eqeqref{eq:flatness-cond} the iLog-model gives 

\begin{equation}
    \Gamma = - \frac{6 H_0^2 \, \Omega_{\mathrm{de}0} \, (1 + \xi) \mathcal{D}_0^2}{2 + (1 + \xi) \mathcal{D}_0} \,.
    \label{eqn:gammainv}
\end{equation}

As indicated in \eqeqref{eq:fQ-invlog}, we impose a condition $\xi>0$ to avoid the singularity at $Q=Q_0$, while $\xi \in (-1, 0)$ is still a valid range for the background cosmology, yields inability at various redsfhit ranges.
% \PK{Would it be good to add plots in the appendix since we have them and does not require additional efforts.} \SH{I'd say not needed to not have unnecessary discussion and questions} 
Finally, for all the three models described here, one can compute the analytical expressions of 
$w_{\rm de}, q, \mu$ and $\nu$ from Eqs.~\eqref{eq:wde}, \eqref{eq:q}, \eqref{eq:Possion-mu}, and \eqref{eq:GW-nu} respectively.  For our paper, we have numerically computed $\dot{H}, w_{\rm de}, q, \mu$ and $\nu$ from the numerical solution of the Hubble parameter, $H$. 

\subsection{On the normalization with $Q_0$} 
\label{sec:Q0norm}
In each of the logarithmic-type models we have introduced a constant $Q(z = 0) \equiv Q_0$ (the present-day value of $Q$) to make the quantity dimensionless. Clearly, the fundamental action is not expected to know about the present day value of $Q$, while for cosmological formulation it is convenient to introduce this normalization, which particularly in the case of Log model as introduced in \cite{Najera:2023wcw} implies the $f(Q) \to Q $, converging to standard \ac{GR} at the present day. In fact, it is this choice deters the model from performing better against data as we will see in the results section. 

To alleviate this we introduce the generalised logarithmic model in \eqeqref{eq:fQ-log-xi} which has a free parameter $\xi$ that can be used to relax the condition of $f(Q) \to Q$ at the present day. In fact, we can set $\xi = 0$ to recover the simple-Log model. Note that the gLog model can be rewritten as $F(Q) = -\Gamma  \log(Q/\xi Q_0 +1) - \Gamma \log(\xi) $, and for $\xi>0$ it is a constant shift in the $F(Q)$ function. Implying that the model would never converge to GR, neither at the present day nor at early-times. 

Similarly, the iLog model also does not converge to $f(Q) \to Q$ at the present day, but it can converge at early times, where $Q \gg Q_0$ and $F(Q) \to 0$ as $\log(Q/Q_0 +\xi) \gg \Gamma $. 
This in fact introduces a slowly varying $F(Q)$ function, converging asymptotically to \ac{GR}at early-times. 
In this context, writing $Q/Q_0$ in the fundamental model is only a choice of mathematical convenience however the physical rescaling is represented by $Q/(\xi Q_0)$ in the gLog and iLog models. Therefore, evading any teleological issues of the fundamental model knowing about the present day value of $Q$.

\section{Dataset and Analysis}
\label{sec:data}
%%%%%%%%%%%%%%%%%%%%%%%%%%%%%%%%%%%%%%

We describe here the datasets employed to constrain the models. We keep the description brief, as the datasets are standard and well-known in the literature. We refer to \cite{DESI:2024mwx,DESI:2025fii} for more details on the datasets and analysis.

\textit{SNe Datasets}: We use the Pantheon+ (\Panp) sample of 1550 spectroscopically confirmed SNe Ia ($0.001 \leq z \leq 2.26$), calibrated with $\Mb = -19.253 \pm 0.029$ as in \cite{Scolnic:2021amr}. We also include the DESy5 dataset \cite{DES:2024haq}, comprising 1830 SNe Ia ($0.02 \leq z \leq 1.13$) identified via ML-based light curve classification. In joint analysis, both samples are treated as uncalibrated; individually, \Panp uses the $\Mb$ prior while DESy5 employs analytical marginalization \cite{DES:2024haq}, precluding direct $\Hzero$ estimation. The \Panp dataset is pivotal for phantom-crossing evidence \cite{DESI:2024mwx}.

{The SNe likelihood is:
\begin{equation}
    \mathcal{L}_{\rm SNe} = \exp\left(-\frac{1}{2}\sum_{i=1}^{N_{\rm SNe}} \frac{\left(\mu_{\rm th}(z_i) - \mu_{\rm obs}(z_i)\right)^{2}}{\sigma_{\mu,i}^{2}}\right)\,,
\end{equation}
with $\mu_{\rm th}(z) = 5\log_{10}\left(D_{\rm L}(z)/\text{Mpc}\right) + 25$ and
\begin{equation}
    D_{\rm L}(z) = \frac{c}{H_{0}}\int_{0}^{z}\frac{dz'}{E(z')}\,, \quad E(z) = \frac{H(z)}{H_0}\,.
\end{equation}
Note $D_{\rm L}(z) = (1+z)D_{\rm M}(z)$.}

\textit{\ac{BAO} Dataset}: We use DESI-DR2 measurements \cite{DESI:2025zgx}, covering $0.1 \leq z \leq 4.2$ in six bins, providing $D_{\rm M}(z)/\rd$ and $D_{\rm H}(z)/\rd \equiv c/[H(z)\rd]$. Within-bin correlations are mild; bins are independent. The dataset is analysed using the inverse distance ladder approach \cite{Addison:2017fdm,BOSS:2014hhw,Haridasu:2018gqm}, with $\rd$ priors from Planck 2018 \cite{Planck:2018vyg} as detailed in later in this section.

{The \ac{BAO} likelihood reads:
\begin{equation}
    \mathcal{L}_{\rm BAO} = \exp\left(-\frac{1}{2}\sum_{i=1}^{6} \left(\Delta \mu \cdot C_{D_{\rm M}, D_{\rm H}}^{-1} \cdot \Delta \mu\right)^{2}\right)\,,
\end{equation}
where $\Delta \mu = \{\Delta D_{\rm M,i}, \Delta D_{\rm H,i}\}$ are residuals between observed and theoretical values.}

\textit{Redshift space Distortions (RSD)}: We use $18$ independent growth-rate measurements from galaxy surveys spanning $z\sim0.02$–$1.4$ (as implemented earlier in \cite{Lapi:2025mgr}), including correlations of subsets of different surveys, which, however, have negligible impact on the results. The fitted observable is
$f\sigma_8(z) = \sigma_8 \,\Big|\tfrac{{\rm d}\delta_+}{{\rm d}\ln a}\Big|$,
with $\sigma_8$ the rms density fluctuation on $8\,h^{-1}$Mpc scale and $\delta_+$ the linear growing mode. The linear perturbation equation in $f(Q)$ gravity is \cite{BeltranJimenez:2019tme, Barros:2020bgg, Albuquerque:2022eac},

\begin{equation}
\delta'' + \left(2 + \frac{H'}{H}\right)\delta' - \frac{3}{2}\,\mu(a,k)\,\Omega_m(a)\,\delta = 0 \,,
\label{eq:linpert}
\end{equation} 
where the gravitational coupling ($\GNeff$) is modified by the factor $\mu(a,k)$ following \eqeqref{eq:muSigma} and \eqeqref{eq:Possion-mu}. Here primes denote derivatives w.r.t.~$\ln a$. Initial conditions are set at $\ln a_{\rm ini}\approx -2$, taking $\delta_{\rm ini}=\delta’_{\rm ini}=a_{\rm ini}$, appropriate for the matter-dominated epoch where $\delta \propto a$. Note also that the data has to be rescaled for the fiducial cosmology assumed in acquiring the data through Alcock-Paczynski term \cite{Alcock:1979mp}, as implemented in \cite{Alam:2015rsa}. In the current analysis we assume no scale-dependence and set $\mu(a,k) = 1/f_Q$ as in \eqeqref{eq:muSigma}.
The RSD likelihood is: 
\begin{equation}
    \mathcal{L}_{\rm RSD} = \exp\left(-\frac{1}{2}\sum_{i=1}^{N_{\rm RSD}} \frac{\left(f\sigma_{8,{\rm th}}(z_i) - f\sigma_{8,{\rm obs}}(z_i)\right)^{2}}{\sigma_{f\sigma_8,i}^{2}}\right)\,.
\end{equation}

\textit{Inverse Distance Ladder (IDL)}: We adopt early-universe priors on $\{\rd, \Hrec\}$ from Planck 2018 \cite{Planck:2018vyg,Haridasu:2020pms}, shown to be robust against late-time assumptions \cite{Verde:2016wmz,Lemos:2023xhs}. These are similar to priors in \cite{DESI:2024mwx,DESI:2025fii,Lynch:2025ine}, except being less stringent and more robust to the late-time deviations different from $\Lambda$CDM model. 
% \label{sec:idl}
{The Gaussian prior is:
\begin{equation} 
    \chi^2_{\rm IDL} = \Delta {\mu_{\rm IDL}}^{T}\, C_{\Hrec,\, \rd}^{-1}\, \Delta {\mu_{\rm IDL}}\, ,
\label{eqn:idl}
\end{equation}
with $\Delta {\mu_{\rm IDL}} = \{\Delta \Hrec, \, \Delta \rd\}$, where $\Hrec = H(z=1089)$. We combine the likelihoods to write a total likelihood as:
\begin{equation}
     \ln{\mathcal{L}_{\rm Tot}} =  \ln{\mathcal{L}_{\rm SNe}} + \ln{\mathcal{L}_{\rm BAO}} + \ln{\mathcal{L}_{\rm RSD}} + \ln{\mathcal{L}_{\rm IDL}}\,,
\end{equation}
and sample the joint posterior to constrain model parameters.} We employ a Bayesian MCMC analysis using \texttt{emcee} \cite{Foreman-Mackey:2012any} and \texttt{Nautilus} \cite{Lange:2023ydq}, visualized via \texttt{corner}\footnote{\url{https://corner.readthedocs.io/en/latest/}} and/or \texttt{GetDist}\footnote{\url{https://getdist.readthedocs.io/}} \cite{Lewis:2019xzd}. \Cref{tab:priors} lists the priors used. For the gLog model, $\xi = 0$ fixed corresponds to the Log model implemented in \cite{Najera:2023wcw}. For the gLog and iLog models, we sample $\xi$ as a free parameter with an uniform prior in log scale to have sufficient resolution within the range, $\log_{10}(\xi) \in (-2, \log_{10}(\xi_{\rm upper})]$, with varying as $\xi_{\rm upper} = \{10, 100\}$.

{\renewcommand{\arraystretch}{1.4} 
\setlength{\tabcolsep}{6pt} 
\begin{table}[h!]
    \centering
    \caption{Priors used in the Bayesian analysis for the standard parameters.}
    \label{tab:priors}
    \begin{tabular}{ccc}\hline
        \textbf{Parameter} &  & \textbf{Prior} \\ \hline\hline  
        $\Omzero$ & & $[0.1, 1.0]$\\
        $\Hzero$ & & $[60.0, 80.0]$\\
        $M_{\rm b}$ & & $[-20.0, -18.0]$ \\ 
        $\rd$ [Mpc] & & $[130.0, 160.0]$ \\
        $\sigma_8$ & & $[0.5, 1.2]$ \\ 
        $\log_{10}\xi$ & & $[-2.0, 2.0]$ \\ \hline
    \end{tabular}
\end{table}
}

We also study mildly different variations for the priors on the $\xi$ parameter, finding consistent results with the results presented here. For the gLog model we study a scenario $\xi \in (-1, 0)$, which we discuss in the \Cref{sec:results}.

\section{Results and Discussion}
\label{sec:results}

{\renewcommand{\arraystretch}{1.6} 
\setlength{\tabcolsep}{6pt} 
\begin{table*}[t!]
    \centering
    \caption{We show 68\% confidence limits for the model parameters from the joint analysis of DESI \ac{BAO} and SNe datasets. We show the results for models: generalized Log (gLog), and inverse Log (iLog) $f(Q)$ models. The $\xi$ parameter is a free parameter in the last two models. In the last two rows, we show the results for the gLog and iLog models with the $\Hrec$ prior, which also includes the $r_d$ prior. Note the limits on the parameter $\xi \in \{0, 100\}$}
    \label{tab:results}
\begin{tabular}{l c c c c c c}
\hline
Model/ & {$H_0$} & {$\Omega_{\rm m 0}$} & {$\log_{10}\xi$} & {$M_{\rm b}$} & {$\rd$} & {$\sigma_8$} \\
\hline
\hline
{/Data} & \multicolumn{6}{c}{DESI-DR2 + \Panp + $r_d$} \\
\hline
\hline
gLog & $68.79 \pm 0.50$ & $0.3039 \pm 0.0079$ & $> 0.918$ & $-19.400 \pm 0.014$ & $147.24 \pm 0.31$ & -- \\ 
iLog & $68.63 \pm 0.49$ & $0.3043 \pm 0.0082$ & $1.09^{+0.34}_{-0.56}$ & $-19.403 \pm 0.014$ & $147.25 \pm 0.31$ & -- \\ 
\hline
{/Data} & \multicolumn{6}{c}{DESI-DR2 + \Panp + $\Hrec$} \\
\hline
gLog & $69.10 \pm 0.37$ & $0.2981 \pm 0.0046$ & $> 0.981$ & $-19.392 \pm 0.011$ & $147.13 \pm 0.30$ & -- \\ 
iLog & $68.83 \pm 0.36$ & $0.2998 \pm 0.0046$ & $1.19^{+0.32}_{-0.52}$ & $-19.398 \pm 0.011$ & $147.18 \pm 0.30$ & -- \\ 
\hline
{/Data} & \multicolumn{6}{c}{DESI-DR2 + \Panp + $\Hrec$ + RSD} \\
\hline
gLog & $69.13 \pm 0.39$ & $0.2977 \pm 0.0047$ & $> 0.932$ & $-19.391 \pm 0.012$ & $147.14 \pm 0.31$ & $0.740\pm 0.023$\\ 
iLog & $68.85 \pm 0.37$ & $0.2994 \pm 0.0047$ & $1.26^{+0.38}_{-0.50}$ & $-19.398 \pm 0.011$ & $147.19 \pm 0.31$ & $0.733\pm 0.023$\\ 
\hline

\hline
\end{tabular}
\end{table*}
}

We begin by presenting the general constraints on the model parameters from the joint analysis of the DESI \ac{BAO} and SNe datasets, using the inverse distance ladder approach. We then discuss the dark energy evolution in \Cref{sec:dark_energy_evolution} and the implications for the perturbations and gravitational wave propagation in \Cref{sec:perturbations}, that can be inferred from the current background analysis. 

We present the results of our analysis in \Cref{tab:results}, focusing on the constraints on the model parameters and the implications for cosmological evolution. The results are shown for the two models: generalized Log (gLog), and inverse Log (iLog) $F(Q)$. The $\xi$ parameter is a free parameter in the last two models. Firstly, we constrain the models in the standard inverse distance ladder approach, using only the DESI \ac{BAO} and \Panp datasets, with the $\rd$ prior. The results are shown in the first two rows of \Cref{tab:results}. 

We find that the Log models are generally consistent with the $\Lambda$CDM constraints, with a slightly larger $H_0$ and $\Omega_m$ values. When we assume $\xi = 0$  in the gLog model which coincides withe the Log model utilised in \cite{Najera:2023wcw}, we find a larger value of $\Hzero \sim 70.44 \pm 0.58$ km/s/Mpc, which is larger than the constrain obtained in the $\Lambda$CDM model ($\sim 68.5$) \cite{DESI:2025fii}, being slightly larger than the Planck 2018 value of $H_0 = 67.4 \pm 0.5$ km/s/Mpc \cite{Planck:2018vyg}. The $\Omega_m$ parameter is constrained to be $\Omega_m = 0.3265^{+0.0083}_{-0.0095}$, which is larger than the DESI-DR2 constraints on $\Lambda$CDM model, while being mildly inconsistent with the \ac{CMB} constraints. However, contrasting the {iLog} ($\xi\neq0$) and gLog ($\xi\neq0$), we find that the $\xi = 0$ case is not preferred by the data, as the gLog completely excludes this case at more than $\sim 3 \sigma$ C.L,. This in turn indicates that the generalization is in fact a necessary modification and the choice to fix $\xi = 0$ restricts the model by enforcing a particular prior assumption. This is also a important modification to assess logarithmic models for non-metrical gravity theories. Therefore, we do not extend the analysis of the Log model to include the $\Hrec$ prior.

\begin{figure}[t!]
	\centering 
    \includegraphics[width=1\linewidth]{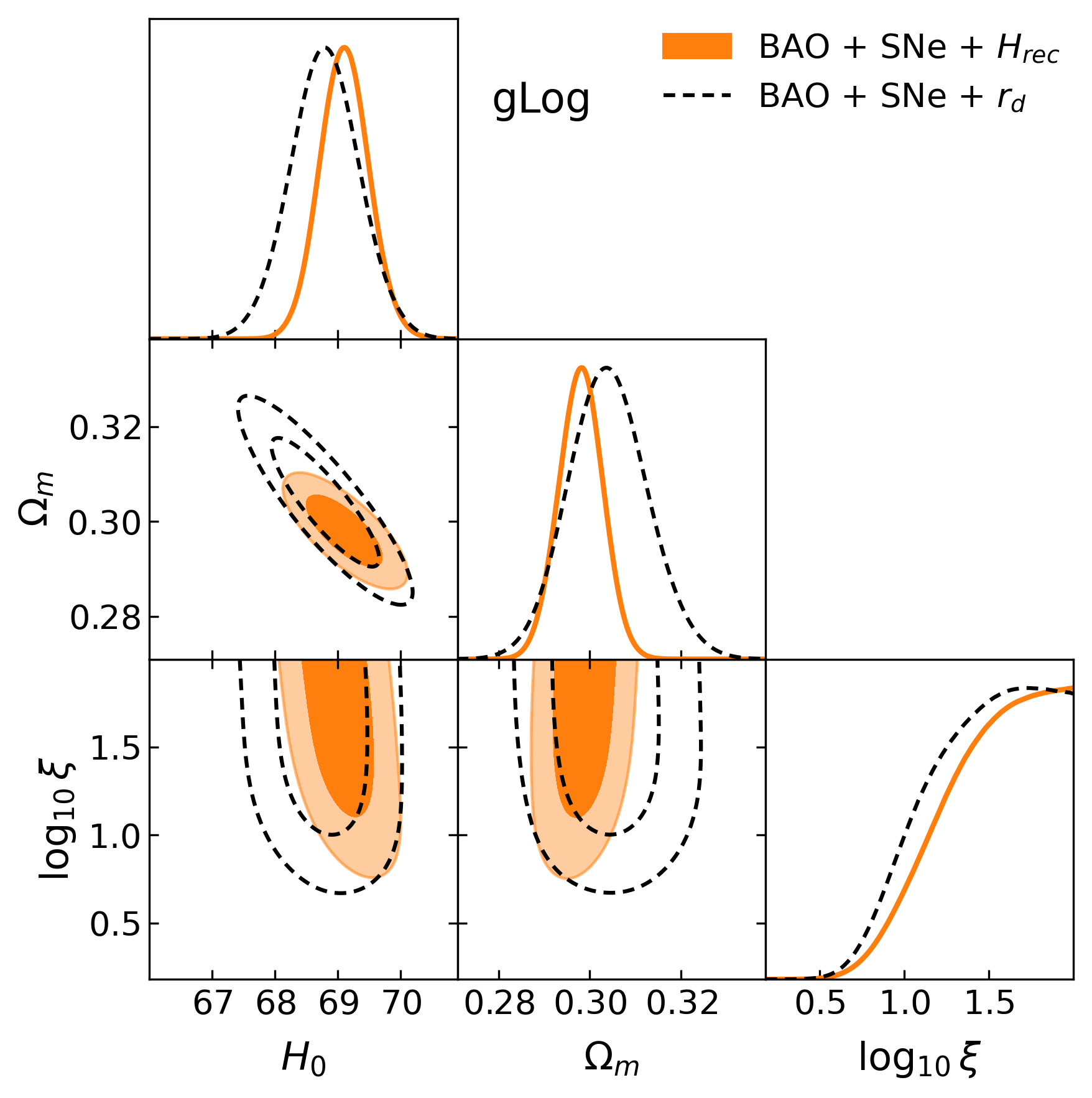}
    	\caption{Contours showing the $68\%$ and $95\%$ confidence levels for the generalized Log (gLog) model with $\xi\neq0$. The dataset combination used here is DESI-DR2 \ac{BAO} + \Panp SNe + $r_d$ prior for the filled contours and DESI-DR2 \ac{BAO} + \Panp SNe + $\Hrec$ prior for the dashed contours in black. }
\label{fig:contours_gLog}
\end{figure}

The gLog and iLog models are then constrained in the same way, with the $\xi$ parameter being a free parameter. The contours for the gLog model are shown in \Cref{fig:contours_gLog}, where we find that the $\xi$ parameter is constrained to be positive, with $\gtrsim 5$ at 95\% C.L., indicating $\xi = 0$ case is excluded at more than $3\sigma$ C.L. The $\Hzero$ and $\Omega_m$ parameters are constrained to be $\sim 69$ km/s/Mpc and $\sim 0.30$, respectively, which are consistent with the $\Lambda$CDM model, using the same dataset. 

\begin{figure}[t!]
	\centering 
    \includegraphics[width=1\linewidth]{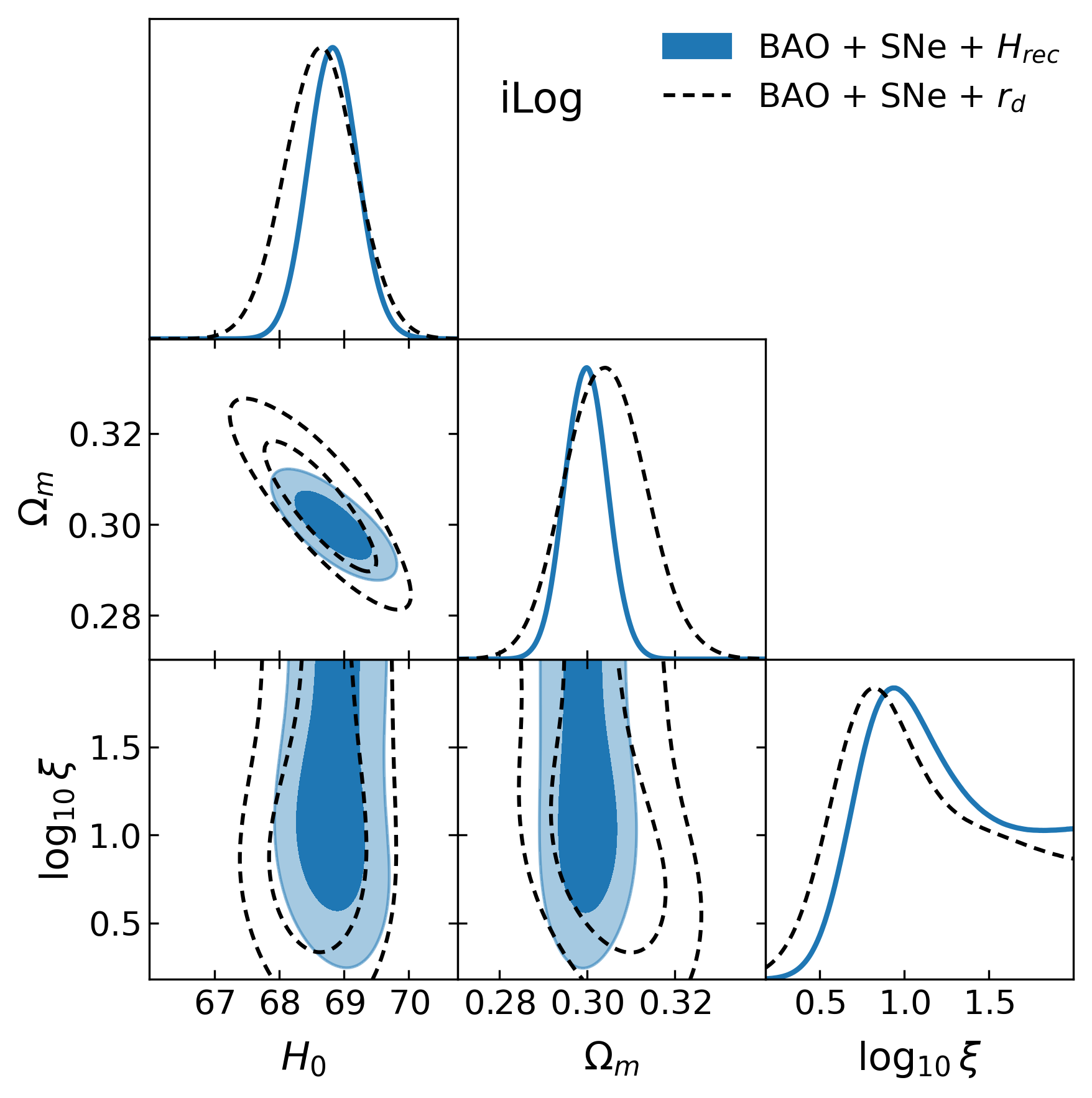}
    	\caption{Contours showing the $68\%$ and $95\%$ confidence levels for the inverse Log (iLog) model with $\xi\neq0$. The dataset combination used here is DESI-DR2 \ac{BAO} + \Panp SNe + $r_d$ prior for the filled contours and DESI-DR2 \ac{BAO} + \Panp SNe + $\Hrec$ prior for the dashed contours in black. }
\label{fig:contours_iLog}
\end{figure}

As can be noticed in both \Cref{fig:contours_gLog} and \Cref{fig:contours_iLog}, the inclusion of the $\Hrec$ priors leads to tightening of constraints on the model parameters, especially on $\Omega_m$ and $\Hzero$. While the constraints on parameter $\xi$ show no practical improvement. Clearly implying that the current Log based modelling is capable of explaining the background data with a large parameter range available for the parameter $\xi$, also implying that the information of the perturbations and gravitational wave propagation is not fully captured by the current background analysis. 

For the iLog model as well we find a lower limit on the $\xi \gtrsim 5 $ at $2\sigma$ level using both the data combinations DESI+\Panp+$\rd$ and DESI+\Panp+$\Hrec$. This is also accommodated by the the upper limit due to the peak in the posterior and flattened tails towards larger values of $\xi$. On the other hand, in the gLog model we don't find a upper limit as far as $\xi\to 100$, while the posterior flattens to have a extend tail. We find a lower limit $\xi \gtrsim 8.6$ at $2 \sigma$ C.L. using the full dataset combination as shown in the last rows of table \cref{tab:results}. These lower limits of $\xi \gtrsim $ in both the gLog and iLog model, in other words, indicate that the modeling choice of $Q/Q_0$ in the Log model is inadequate where $\xi = 0$ is imposed. 

{Restricting the analysis in the gLog model to $\xi \in (-1,0)$, {we find the posteriors} to converge towards $\xi \to 0$, with constraints similar to Log model and larger values of $\Hzero$ and lower values of $\Omzero$, consequently. }

\subsection{Dark Energy Evolution}
\label{sec:dark_energy_evolution}

\begin{figure}
    \centering
    \includegraphics[width=1.\linewidth]{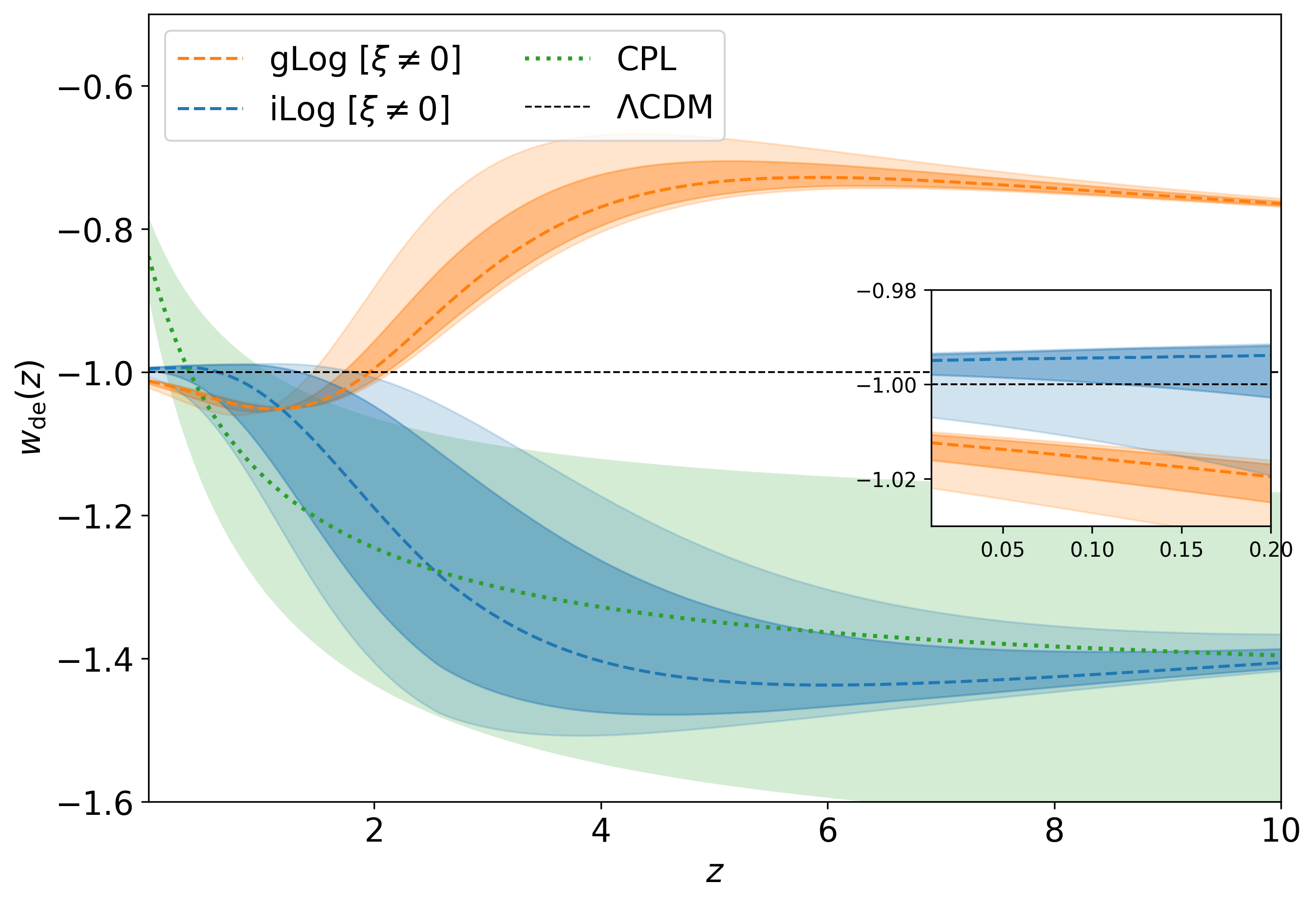}
    \caption{Evolution of the dark energy equation of state $w_{\rm de}$ for the gLog and iLog models, when using the $\Hrec$ prior along with the DESI \ac{BAO} and \Panp SNe datasets. The dashed lines show the $w_{\rm de} = -1$ line. The shaded regions show the 68\% and 95\% confidence levels for the gLog and iLog models. We show the $68\%$ C.L. limits of CPL parametrization for comparison in in green. The CPL model shows a phantom crossing around $z \sim 0.6$. Within the inset we show a zoom-in of the low-redshift range $z< 0.2$ to highlight the nature of gLog and iLog models $\wdezero$.}
    \label{fig:wde_fQ}
\end{figure}

We now discuss the dark energy evolution in the context of the $f(Q)$ models. In \Cref{fig:wde_fQ}, we show the evolution of the dark energy equation of state $w_{\rm de}$ gLog, and iLog models when utilising the $\Hrec$ prior along with the \ac{BAO} and SNe datasets. Note as already mentioned in \Cref{sec:interpretation}, it is possible to anticipate a fluid-like dark energy component in the $f(Q)$ models, which can be described by an effective equation of state. While in the modified gravity interpretation, the dark energy equation of state is usually not defined and an evolution of the gravitational constant is expected. In this draft we work, we remain with the prior interpretation to demonstrate the capability of the $f(Q)$ models to explain the background data. 

As shown in \Cref{fig:wde_fQ}, we find a evolution of the dark energy equation of state $w_{\rm de}$, within both the gLog and inverse Log models, also showing mild hints of phantom crossing in the gLog model. The gLog model shows a more complex evolution, with a phantom crossing from $w_{de}<-1$ to $w_{de}>-1$ at $z \sim 2$ and a phantom behavior at late-times. This behavior is in turn contrary to the phenomenological CPL based expectation form the same set of data \cite{DESI:2025fii}, which shows a phantom crossing at $z \sim 0.5$ in the opposite direction. On the other hand, iLog model tends to become phantom at high-redsfhit ($z \gtrsim 1.5$), indicating mild hints of phantom crossing at lower redshifts (please see the inset in \cref{fig:wde_fQ}). While this is more in line with the CPL expectation in \cite{DESI:2025fii}, does not indicate any phantom crossing. 

While we have not explicitly shown, the Log model shows singular behavior transiting from $w_{de}<-1$ to $w_{de}>-1$ at $z \sim 0.5$, as expected in several other models and at times in model-independent reconstructions\footnote{Within the context of model-independent reconstructions the singularity in the dark energy EoS is accompanied by transition of dark energy density from positive to negative.}. Note that, for the gLog model extending the Log model with $\xi > 0$, large values of the same allow for a smooth transition from $w_{de}<-1$ to $w_{de}>-1$ at $z \sim 0.5$, alleviating the singularity in the Log model. 

In the last two columns of \Cref{tab:summary}, we show the constraints on the dark energy equation of state $w_{\rm de}(z)$ and the deceleration parameter $q(z)$ for the three models today ($z\to 0$). As can be seen, we find the values of the deceleration parameter today $q_0$ to be in very good agreement with the standard values in the $\Lambda$CDM model, with $q_0 \sim -0.5$. In general we find that the gLog model provides slightly lower values of $q_0$ compared to the iLog model, with $q_0 \sim -0.56$ for the gLog model and $q_0 \sim -0.54$ for the iLog model. The Log model with $\xi = 0$ provides a larger value of $q_0 \sim -0.63$. 

The values of the dark energy equation of state $w_{\rm de,0}$, while for both the gLog and iLog models tends to $w_{\rm de,0} \sim -1.01$, the Log model with $\xi = 0$ provides a larger value of $w_{\rm de,0} \sim -1.1219^{+0.0034}_{-0.0042}$. As discussed earlier in this section, the iLog model provides a quintessence-like ($w_{\rm de,0} > -1$) dark energy equation of state, while the gLog model provides a phantom-like ($w_{\rm de,0} < -1$) dark energy equation of state. Also, the constrains in the gLog model show to be distinct from $\Lambda$CDM model ($w = -1$) at more than $\sim 3.0 \sigma$ C.L., owing to the extremely tight constrain on the same when utilizing the DESI \ac{BAO} and \Panp SNe datasets and \ac{CMB} based $\Hrec$ prior. In \cref{fig:wde_fQ}, we show for comparison with results in \cite{DESI:2025fii,DESI:2025zgx}, the EoS for the CPL parametrization \cite{Chevallier:2000qy,Linder:2002et}\footnote{Dark energy EoS in the CPL model is parametrized as $w(a) = w_0 +w_{\rm a} (1-a)$.}. The standard CPL parametrization predicts a phantom crossing around $z \sim 0.6$ using the same dataset combination. We find the iLog model generally consistent with the CPL parametrization at late times, while the gLog model shows a phantom crossing at $z \sim 2$ and a phantom behavior at late-times. The iLog models prediction of phantom behavior at early-times ($z \gtrsim 1.5$), in fact, is in an excellent agreement with the CPL model.

\subsection{Perturbations and GW Propagation}
\label{sec:perturbations}

\begin{figure}[t!]
	\centering 
    \includegraphics[width=1\linewidth]{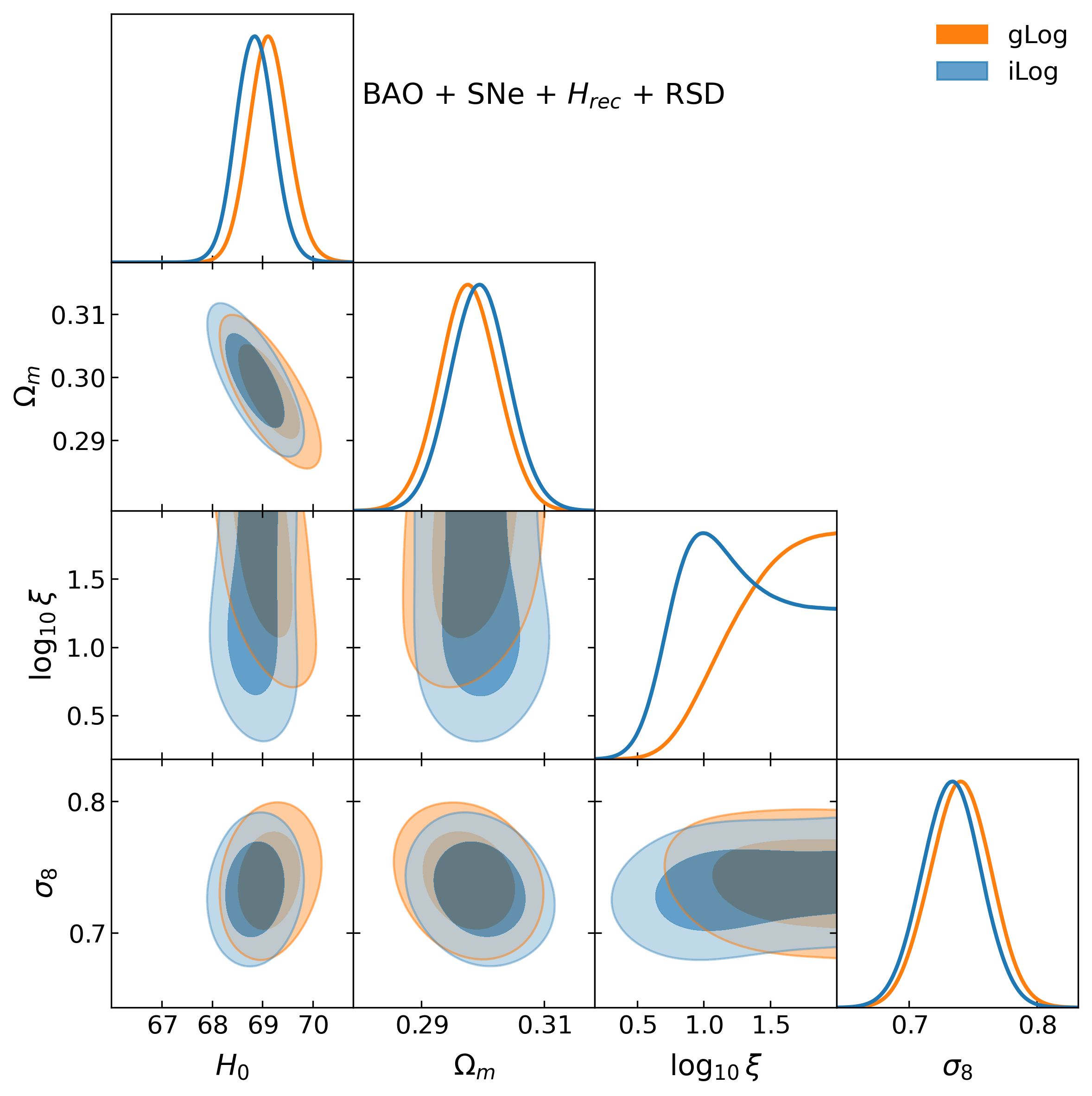}
    	\caption{Contours showing the $68\%$ and $95\%$ confidence levels for the inverse Log (iLog) and generalized Log (gLog) models with $\xi\neq0$. The dataset combination used here is DESI-DR2 \ac{BAO} + \Panp SNe + $\Hrec$ prior + RSD data. The filled contours in blue and orange are for the iLog and gLog models, respectively. }
\label{fig:contours_sigma8}
\end{figure}

In this section, we discuss the implications of the $f(Q)$ models for the perturbations and gravitational wave propagation. As discussed in \Cref{sec:data}, we use the RSD data to constrain the models at the perturbation level. We solve the linear perturbation equation \cref{eq:linpert} to compute the growth rate $f\sigma_8(z)$, which is then compared with the RSD data. In  \Cref{fig:contours_sigma8}, we show the contours for the iLog and gLog models when using the DESI {BAO}, \Panp SNe, $\Hrec$ prior, and RSD data. We find that the inclusion of the RSD data leads to no major change of constraints on the $\Omega_m$ and $H_0$ parameters, while the constraints on the $\xi$ change mildy with clearly no consequences for the background evolution. For the value of $\sigma_8$, we find $\sigma_8 = 0.740\pm 0.023$ and $\sigma_8 = 0.733\pm 0.023$ for the gLog and iLog models, respectively, which are consistent with the $\Lambda$CDM ($\sigma_8 \sim 0.74$) model constraints using the same dataset combination.

{\renewcommand{\arraystretch}{1.6} 
\setlength{\tabcolsep}{6pt} 
\begin{table*}[t!]
\centering
    \caption{We show the constraints on the $\mu$, $\nu$, $w_{\rm de,0}$ and $q_0$ parameters for the three models: Log, generalized Log (gLog), and inverse Log (iLog) $f(Q)$ models. The $\xi$ parameter is a free parameter in the last two models. The last two rows show the results for the gLog and iLog models with the $\Hrec$ prior, which also includes the $r_d$ prior.}
    \label{tab:summary}
\begin{tabular}{l c c c c}
\hline
{Model/} & {$\mu_0$} & {$\nu_0$} & {$w_\mathrm{de,0}$} & {$q_0$} \\
\hline
\hline
 /Data& \multicolumn{4}{c}{DESI-DR2 + \Panp + $r_d$}\\
\hline
iLog & $1.0328^{+0.0375}_{-0.0214}$ & $-0.0082^{+0.0070}_{-0.0249}$ & $-0.9956^{+0.0018}_{-0.0046}$ & $-0.5357^{+0.0106}_{-0.0149}$ \\
gLog & $0.9639^{+0.0061}_{-0.0165}$ & $+0.0035^{+0.0028}_{-0.0009}$  & $-1.0126^{+0.0018}_{-0.0044}$ & $-0.5567^{+0.0132}_{-0.0138}$ \\
\hline
 /Data & \multicolumn{4}{c}{DESI-DR2 + \Panp + $\Hrec$}\\
\hline
iLog & $1.0343^{+0.0298}_{-0.0208}$ & $-0.0089^{+0.0073}_{-0.0175}$ & $-0.9950^{+0.0012}_{-0.0029}$ & $-0.5438^{+0.0097}_{-0.0088}$ \\
gLog & $0.9640^{+0.0061}_{-0.0133}$ & $+0.0033^{+0.0022}_{-0.0009}$  & $-1.0121^{+0.0017}_{-0.0036}$ & $-0.5672^{+0.0069}_{-0.0083}$ \\
\hline
\end{tabular}
\end{table*}
}

\begin{figure*}[t!]
    \centering 
    \includegraphics[width=0.48\linewidth]{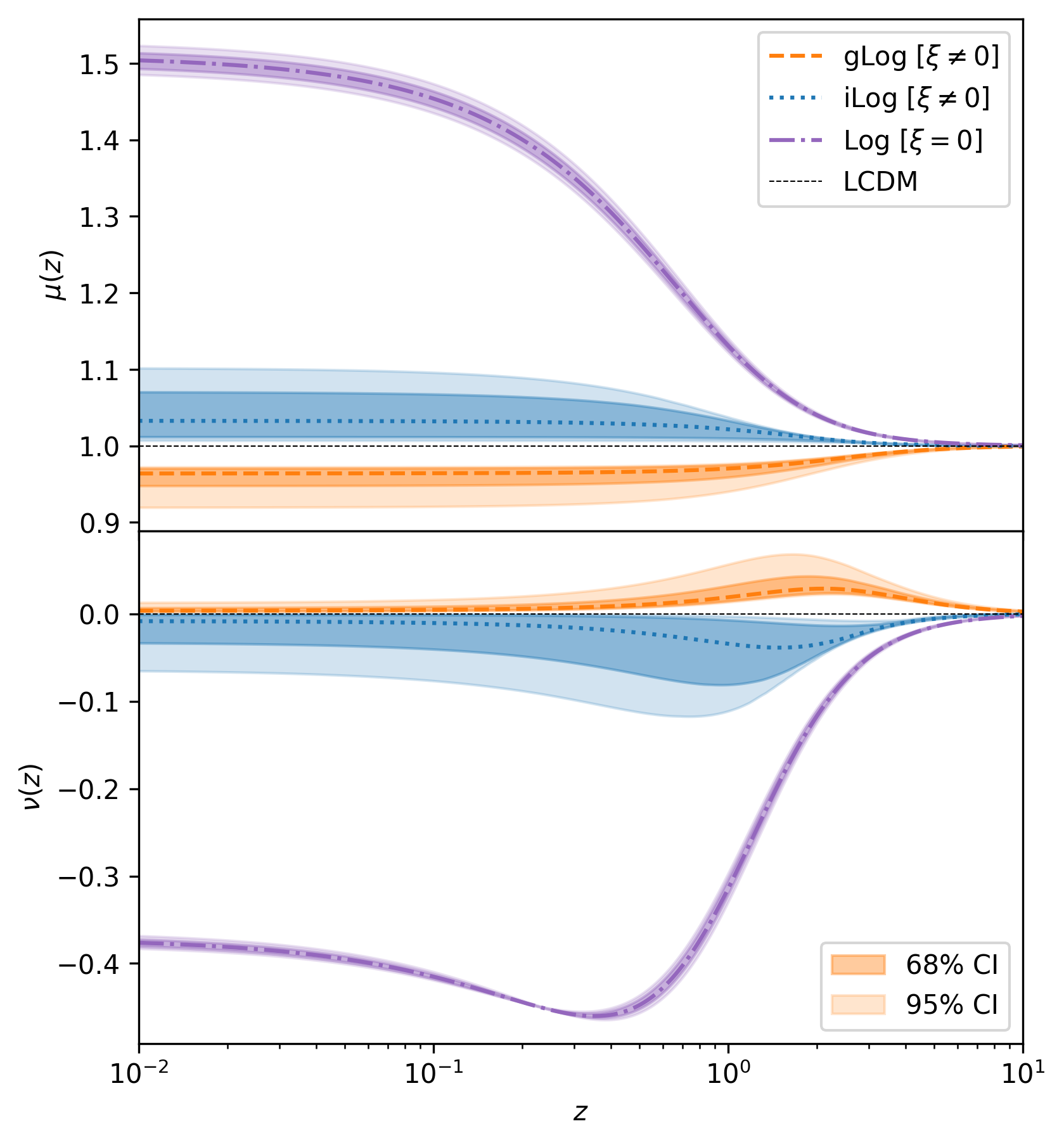}
    \includegraphics[width=0.495\linewidth]{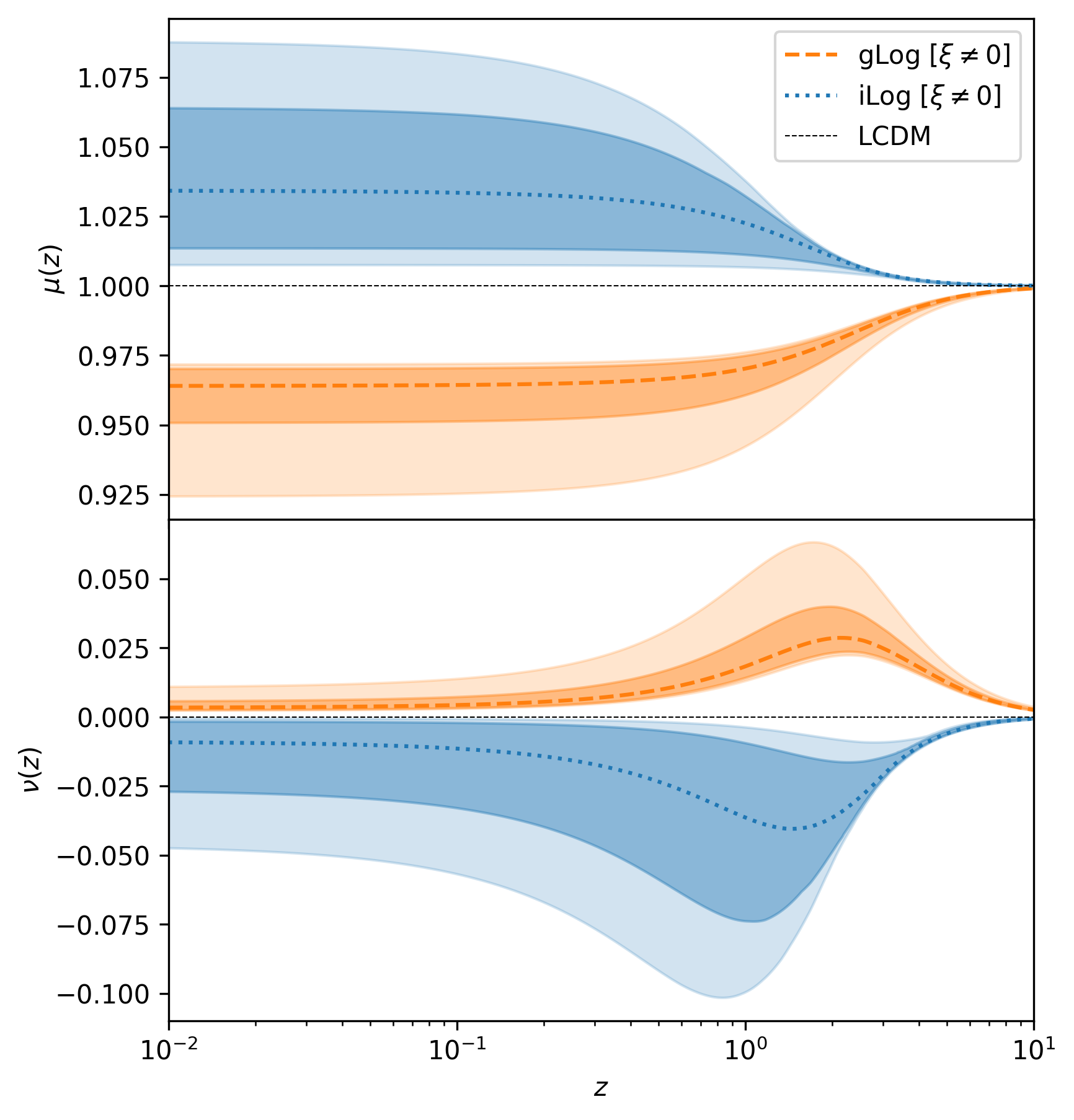}
    \caption{\textit{Left}: Evolution of the $\mu$ (\textit{top}) and $\nu$ (\textit{bottom}) parameters using DESI-DR2 \ac{BAO} + \Panp SNe + $\rd$ combination for the generalized Log (gLog) in orange and inverse Log (iLog) in blue. In purple we show the case when $\xi=0$ is fixed coinciding with the Log model in \cite{Najera:2023wcw}. \textit{Right}: Same as the left panel, but dataset combination used here is DESI-DR2 \ac{BAO} + \Panp SNe + $\Hrec$ prior. Note the limits on the parameter $\xi \in \{0, 10\}$.}
\label{fig:mu_nu_fQ}
\end{figure*}

Having assessed the background evolution of the $f(Q)$ models, we now discuss the implications for the perturbations and gravitational wave propagation, described by the $\mu$ (Eq.~\eqref{eq:Possion-mu}) and $\nu$ (eq.~\eqref{eq:GW-nu}) parameters. In the right panel of \Cref{fig:mu_nu_fQ}, we show the evolution of the $\mu$ and $\nu$ parameters as a function of redsfhit, for the gLog and iLog models. As expected both they converge to \ac{GR} based estimates at high-redshift  as, $\mu\to 1$ and $\nu\to 0$. 

In the \textit{left} panel of \Cref{fig:mu_nu_fQ}, we show the evolution of the $\mu$ and $\nu$ parameters for the two Log models alongside the case $\xi =0$ in the gLog model, labeled as Log. 
As one can infer in the Log model with $\xi =0$, both the $\mu(z\to0) \equiv 2/(2- \Omega_{de0})$ and $\nu(z\to0)$ are dependent solely on the dark energy density ($\Omega_{de0}$), which being constrained by the closure relation does not allow for additional modulation of their numerical values. 

In the \textit{right} panel of \cref{fig:mu_nu_fQ} we show the constraints in iLog (in blue) model indicates that the gravitational strength ($\GNeff = \mu(z) \GN $) increases at late-times slowly after $z\lesssim 3$. We find a $\sim3-4 \%$ increment in the gravitational strength, reaching up to $\sim 8\%$ at the $2\sigma$ level. The gLog model on the other hand, shows a decrement of a similar order of magnitude ($\sim 3-4 \%$) in the gravitational strength at late-times, also reaching upto $\sim 7\%$ at the $2\sigma$ level. Note that, we are primarily interested in the constraints away from $\LCDM$ limits ($\nu\to 0,\, \mu\to 1$).

It is interesting to note that for $\xi\to1$, the gLog model shows instability for the prediction of $\mu$ and $\nu$, providing singularities in each around $z\sim 0.3$. while for $\xi >1$ we find the reciprocal behavior (wrt Log and iLog models) of reduction in the gravitational strength $\mu<1$. For a value of $\xi < 1$ the gLog model behaves similar to the Log and iLog models to provide the increment of gravitational strength. However, as we have already seen earlier , the $\xi < 1$ case is not preferred by the data, as the gLog completely excludes this case at more than $\sim 3 \sigma$. Finally we conclude that the two models iLog and gLog while perform equivalently at the background level, provide complementary predictions for the perturbations and gravitational wave propagation and should be easily testable when contrasted against the relevant data, which we intend to examine in a future work. Note that the apparent limit on $\nu \neq 0$ in both the models gLog and iLog models is only an artifact of  numerical convergence as the models converge to $\LCDM$ limits ($\nu =0$) only in the $\xi \to \infty$, while we explore only to a limit of $\xi< 10, 100$. 

{We stress that the predicted variation in the gravitational strength is completely within the allowed range as anticipated in earlier analysis \cite{Nesseris:2017vor} (see also \cite{Anagnostopoulos:2021ydo}). Also, being a late-time effect $z\lesssim 10$, this variation does not affect the Big Bang Nucleosynthesis (BBN) and recombination physics, and is expected to be consistent with the \ac{CMB} based constraints \cite{Umezu:2005ee, Ballardini:2021evv, Wang:2020bjk}.}

\section{Conclusions}
\label{sec:conc}

In this work, we have explored the $f(Q)$ models constraining then using the DESI \ac{BAO} and SNe datasets, using the inverse distance ladder approach. We have constrained three models: Log, generalized Log (gLog), and inverse Log (iLog) $f(Q)$ models, with the $\xi$ parameter being a free parameter in the last two models.
We find that the $f(Q)$ models are well able to explain the \ac{BAO} and SNe data, being consistent with the $\Lambda$CDM based constraints, while the gLog and iLog models show a rich phenomenology with a multi-modal distribution in the $\xi$ parameter space. The inverse Log model is the most consistent with the $\Lambda$CDM based constraints using the DESI \ac{BAO} data and Planck based constraints for the $\Omega_m$ values, while the Log model with $\xi = 0$ showing the most deviation from the standard scenario, with larger values of $\Hzero$ and $\Omega_m$.

Imposing the \ac{CMB} based $\Hrec$ prior we find the gLog and iLog models can be consistent with the  pre-recombination physics while allowing for dynamics in the dark energy sector varying from quintessence-like to phantom-like behavior. The gLog model shows a phantom crossing at $z \sim 2$ and a phantom behavior at late-times, while the iLog model shows a quintessence-like behavior at late-times and a phantom behavior at high-redshift ($z \gtrsim 1.5$). The Log model with $\xi = 0$ shows a singular behavior, while the gLog and iLog models show a smooth evolution of the $\mu$ and $\nu$ parameters.

We also constrain the models using the RSD data, which allows us to assess the implications for the perturbations and gravitational wave propagation. We find that the inclusion of the RSD data leads to no major change of constraints on the $\Omega_m$ and $H_0$ parameters, while the constraints on the $\xi$ change mildly with clearly no consequences for the background evolution. The values of $\sigma_8$ are consistent with the $\Lambda$CDM ($\sigma_8 \sim 0.74$) model constraints using the same dataset combination. This in turn does not provide any additional advantage to address the $S_8$-tension \cite{CosmoVerse:2025txj}, however, requires an in depth assessment.

We have also discussed the implications for the perturbations and gravitational wave propagation, showing that the $\mu$ and $\nu$ parameters are well constrained by the data. These constraints show the predictions made for the perturbations and GW propagation in the $f(Q)$ models, based solely on the background data. We intend to extend the analysis to include the perturbations and gravitational wave propagation in the $f(Q)$ models in a future work, which will allow us to assess the consistency of the $f(Q)$ models with the current data on structure formation and gravitational wave propagation. In particular, the two models gLog and iLog provide complementary predictions for the perturbations and gravitational wave propagation and should be easily testable when contrasted against the relevant data.

%%%%%%%%%%%%%%%%%%%%%%%%%%%%%%%%%%%%%
%%%%%%%%%% Bibliography %%%%%%%%%%%%%
\section*{Acknowledgments}
We would like to thank K. Cannon, C. Chen, L. Järv, T. Koivisto, and M. Saal for useful discussions and comments on a preliminary version of this paper.  PK gratefully acknowledges the University of Tartu for hosting and for providing access to computing resources, and the hospitality of the University of Padova during the visiting period. PK and AN acknowledge support the JSPS Postdoctoral Fellowships for Research in Japan, Grant Number JP12345678. AN acknowledges support from JSPS KAKENHI Grant Nos. JP23K03408, JP23H00110, and JP23H04893. SH is supported by the INFN INDARK grant and acknowledges support from the COSMOS project of the Italian Space Agency (cosmosnet.it).

\bibliography{bibliography} % Produces the bibliography via BibTeX.

\appendix 

\section{gLog with $\xi = 0$}
We show the contour plots for the case of gLog model where $\xi =0$ is set apriori. 
 
\begin{figure}[t!]
	\centering 
    \includegraphics[width=1\linewidth]{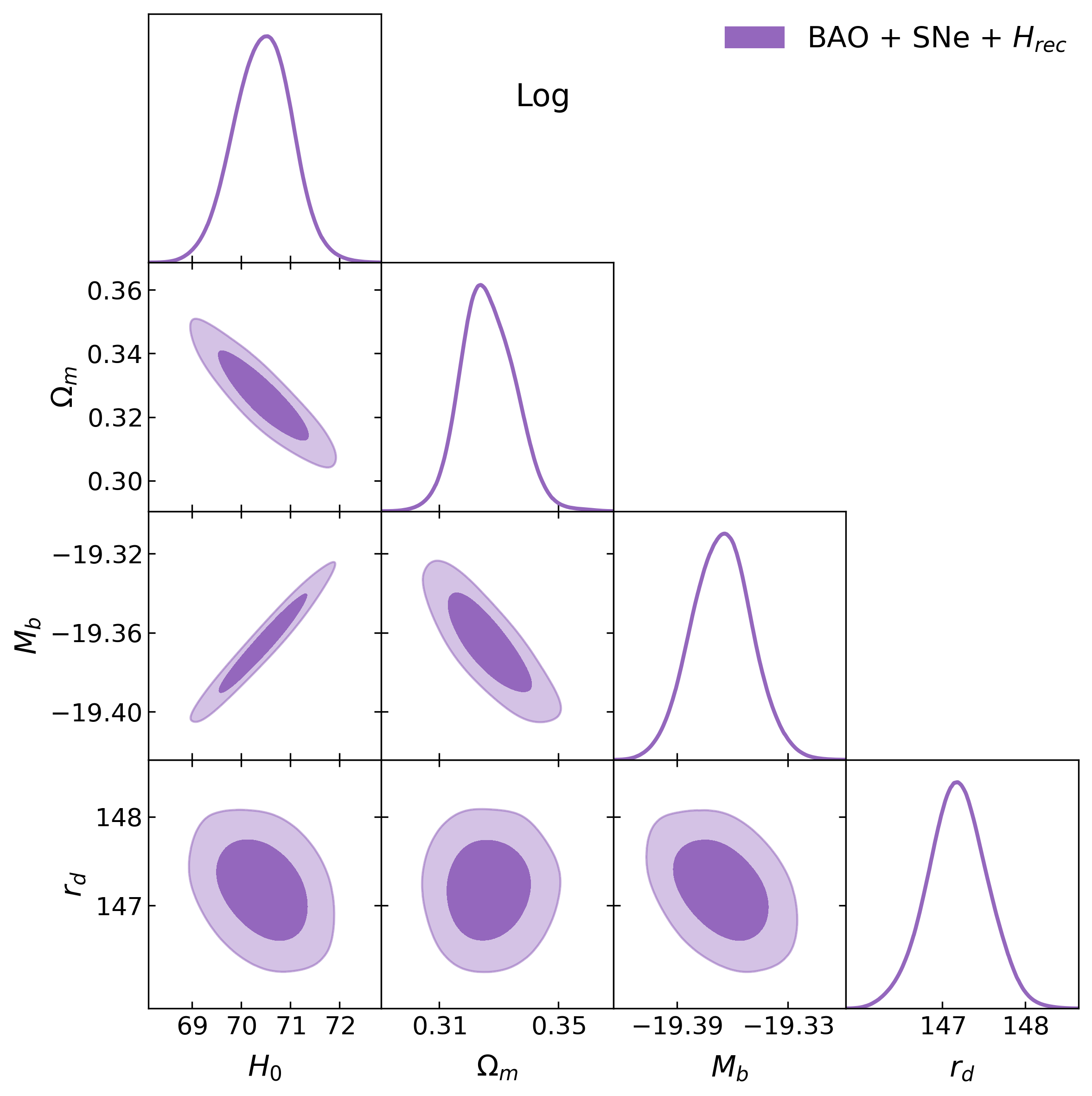}
    	\caption{Contours showing the $68\%$ and $95\%$ confidence levels for the Log model with $\xi=0$. The dataset combination used here is DESI-DR2 \ac{BAO} + \Panp SNe + $r_d$ prior. The parameter values shown on top of the 1D-posteriors correspond to the the $95\%$ C.L.}
\label{fig:contours_log}
\end{figure}
\end{document}

%% file: acros.tex
\DeclareAcronym{FLRW}{
  short = FLRW ,
  long = Friedmann-Lema{\^i}tre-Robertson-Walker
}
\DeclareAcronym{GR}{
  short = GR ,
  long = general relativity
}
\DeclareAcronym{BAO}{
  short = BAO ,
  long = Baryon acoustic oscillations
}
\DeclareAcronym{CMB}{
  short = CMB ,
  long = Cosmic microwave background
}
\DeclareAcronym{CMBR}{
  short = CMBR ,
  long = cosmic microwave background radiation
}

\DeclareAcronym{GW}{
  short = GW ,
  long = gravitational wave
}

\DeclareAcronym{RHS}{
  short = RHS ,
  long = right hand side
}

\DeclareAcronym{CDM}{
  short = CDM ,
  long = cold dark matter 
}

\DeclareAcronym{LCDM}{
  short = $\Lambda$CDM ,
  long = lambda cold dark matter 
}

\DeclareAcronym{STEGR}{
  short = STEGR ,
  long = Symmetric Teleparallel Equivalent of GR
}

\DeclareAcronym{TEGR}{
  short = TEGR ,
  long = Teleparallel Equivalent of GR
}

%% file: macross.tex
\definecolor{mypurple}{RGB}{135,10,190} % purple {135,10,110}
\definecolor{mygreen}{RGB}{0,130,0} 
\definecolor{softgreen}{RGB}{52, 168, 83} % Google Green
\definecolor{emerald}{RGB}{80,200,120}
\definecolor{forestgreen}{RGB}{34,139,34}  %ok
\definecolor{mediumseagreen}{RGB}{60,179,113} % Too faint
\definecolor{darkgreen}{RGB}{0, 128, 0}
\newcommand{\beq}{\begin{equation}}
\newcommand{\eeq}{\end{equation}}
\newcommand{\be}{\begin{equation}}
\newcommand{\ee}{\end{equation}}
\newcommand{\bea}{\begin{eqnarray}}
\newcommand{\eea}{\end{eqnarray}}
\newcommand{\eqeqref}{Eq. \eqref}

\newcommand{\LCDM}{\Lambda{\rm{CDM}}}

\newcommand{\GNeff}{\ensuremath{G_{\rm N}^{\rm eff}}}
\newcommand{\GN}{\ensuremath{G_{\rm N}}}
\newcommand{\Hzero}{H_{0}}
\newcommand{\Omzero}{\Omega_{\rm m0}}

\newcommand{\wdezero}{w_{\mathrm{de,0}}}
\newcommand{\Panp}{Pan$^{+}$}
\newcommand{\Hrec}{H_{\rm rec}}
\newcommand{\rd}{r_{\rm d}}
\newcommand{\Mb}{M_{\rm b}}

\newcommand{\diff}{{\rm d}}

%% file: fq_log_fsigma8.bbl
%merlin.mbs apsrev4-1.bst 2010-07-25 4.21a (PWD, AO, DPC) hacked
%Control: key (0)
%Control: author (8) initials jnrlst
%Control: editor formatted (1) identically to author
%Control: production of article title (-1) disabled
%Control: page (0) single
%Control: year (1) truncated
%Control: production of eprint (0) enabled
\begin{thebibliography}{106}%
\makeatletter
\providecommand \@ifxundefined [1]{%
 \@ifx{#1\undefined}
}%
\providecommand \@ifnum [1]{%
 \ifnum #1\expandafter \@firstoftwo
 \else \expandafter \@secondoftwo
 \fi
}%
\providecommand \@ifx [1]{%
 \ifx #1\expandafter \@firstoftwo
 \else \expandafter \@secondoftwo
 \fi
}%
\providecommand \natexlab [1]{#1}%
\providecommand \enquote  [1]{``#1''}%
\providecommand \bibnamefont  [1]{#1}%
\providecommand \bibfnamefont [1]{#1}%
\providecommand \citenamefont [1]{#1}%
\providecommand \href@noop [0]{\@secondoftwo}%
\providecommand \href [0]{\begingroup \@sanitize@url \@href}%
\providecommand \@href[1]{\@@startlink{#1}\@@href}%
\providecommand \@@href[1]{\endgroup#1\@@endlink}%
\providecommand \@sanitize@url [0]{\catcode `\\12\catcode `\$12\catcode `\&12\catcode `\#12\catcode `\^12\catcode `\_12\catcode `\%12\relax}%
\providecommand \@@startlink[1]{}%
\providecommand \@@endlink[0]{}%
\providecommand \url  [0]{\begingroup\@sanitize@url \@url }%
\providecommand \@url [1]{\endgroup\@href {#1}{\urlprefix }}%
\providecommand \urlprefix  [0]{URL }%
\providecommand \Eprint [0]{\href }%
\providecommand \doibase [0]{http://dx.doi.org/}%
\providecommand \selectlanguage [0]{\@gobble}%
\providecommand \bibinfo  [0]{\@secondoftwo}%
\providecommand \bibfield  [0]{\@secondoftwo}%
\providecommand \translation [1]{[#1]}%
\providecommand \BibitemOpen [0]{}%
\providecommand \bibitemStop [0]{}%
\providecommand \bibitemNoStop [0]{.\EOS\space}%
\providecommand \EOS [0]{\spacefactor3000\relax}%
\providecommand \BibitemShut  [1]{\csname bibitem#1\endcsname}%
\let\auto@bib@innerbib\@empty
%</preamble>
\bibitem [{\citenamefont {Aghanim}\ \emph {et~al.}(2020{\natexlab{a}})\citenamefont {Aghanim} \emph {et~al.}}]{Planck:2018nkj}%
  \BibitemOpen
  \bibfield  {author} {\bibinfo {author} {\bibfnamefont {N.}~\bibnamefont {Aghanim}} \emph {et~al.} (\bibinfo {collaboration} {Planck}),\ }\href {\doibase 10.1051/0004-6361/201833880} {\bibfield  {journal} {\bibinfo  {journal} {Astron. Astrophys.}\ }\textbf {\bibinfo {volume} {641}},\ \bibinfo {pages} {A1} (\bibinfo {year} {2020}{\natexlab{a}})},\ \Eprint {http://arxiv.org/abs/1807.06205} {arXiv:1807.06205 [astro-ph.CO]} \BibitemShut {NoStop}%
\bibitem [{\citenamefont {Ade}\ \emph {et~al.}(2016)\citenamefont {Ade} \emph {et~al.}}]{Planck:2015fie}%
  \BibitemOpen
  \bibfield  {author} {\bibinfo {author} {\bibfnamefont {P.~A.~R.}\ \bibnamefont {Ade}} \emph {et~al.} (\bibinfo {collaboration} {Planck}),\ }\href {\doibase 10.1051/0004-6361/201525830} {\bibfield  {journal} {\bibinfo  {journal} {Astron. Astrophys.}\ }\textbf {\bibinfo {volume} {594}},\ \bibinfo {pages} {A13} (\bibinfo {year} {2016})},\ \Eprint {http://arxiv.org/abs/1502.01589} {arXiv:1502.01589 [astro-ph.CO]} \BibitemShut {NoStop}%
\bibitem [{\citenamefont {Abdul~Karim}\ \emph {et~al.}(2025)\citenamefont {Abdul~Karim} \emph {et~al.}}]{DESI:2025zgx}%
  \BibitemOpen
  \bibfield  {author} {\bibinfo {author} {\bibfnamefont {M.}~\bibnamefont {Abdul~Karim}} \emph {et~al.} (\bibinfo {collaboration} {DESI}),\ }\href@noop {} {\  (\bibinfo {year} {2025})},\ \Eprint {http://arxiv.org/abs/2503.14738} {arXiv:2503.14738 [astro-ph.CO]} \BibitemShut {NoStop}%
\bibitem [{\citenamefont {Scolnic}\ \emph {et~al.}(2022)\citenamefont {Scolnic} \emph {et~al.}}]{Scolnic:2021amr}%
  \BibitemOpen
  \bibfield  {author} {\bibinfo {author} {\bibfnamefont {D.}~\bibnamefont {Scolnic}} \emph {et~al.},\ }\href {\doibase 10.3847/1538-4357/ac8b7a} {\bibfield  {journal} {\bibinfo  {journal} {Astrophys. J.}\ }\textbf {\bibinfo {volume} {938}},\ \bibinfo {pages} {113} (\bibinfo {year} {2022})},\ \Eprint {http://arxiv.org/abs/2112.03863} {arXiv:2112.03863 [astro-ph.CO]} \BibitemShut {NoStop}%
\bibitem [{\citenamefont {Alam}\ \emph {et~al.}(2017)\citenamefont {Alam} \emph {et~al.}}]{Alam:2016hwk}%
  \BibitemOpen
  \bibfield  {author} {\bibinfo {author} {\bibfnamefont {S.}~\bibnamefont {Alam}} \emph {et~al.} (\bibinfo {collaboration} {BOSS}),\ }\href {\doibase 10.1093/mnras/stx721} {\bibfield  {journal} {\bibinfo  {journal} {Mon. Not. Roy. Astron. Soc.}\ }\textbf {\bibinfo {volume} {470}},\ \bibinfo {pages} {2617} (\bibinfo {year} {2017})},\ \Eprint {http://arxiv.org/abs/1607.03155} {arXiv:1607.03155 [astro-ph.CO]} \BibitemShut {NoStop}%
\bibitem [{\citenamefont {Di~Valentino}\ \emph {et~al.}(2025)\citenamefont {Di~Valentino} \emph {et~al.}}]{CosmoVerse:2025txj}%
  \BibitemOpen
  \bibfield  {author} {\bibinfo {author} {\bibfnamefont {E.}~\bibnamefont {Di~Valentino}} \emph {et~al.} (\bibinfo {collaboration} {CosmoVerse Network}),\ }\href {\doibase 10.1016/j.dark.2025.101965} {\bibfield  {journal} {\bibinfo  {journal} {Phys. Dark Univ.}\ }\textbf {\bibinfo {volume} {49}},\ \bibinfo {pages} {101965} (\bibinfo {year} {2025})},\ \Eprint {http://arxiv.org/abs/2504.01669} {arXiv:2504.01669 [astro-ph.CO]} \BibitemShut {NoStop}%
\bibitem [{\citenamefont {Perivolaropoulos}(2024)}]{Perivolaropoulos:2024yxv}%
  \BibitemOpen
  \bibfield  {author} {\bibinfo {author} {\bibfnamefont {L.}~\bibnamefont {Perivolaropoulos}},\ }\href {\doibase 10.1103/PhysRevD.110.123518} {\bibfield  {journal} {\bibinfo  {journal} {Phys. Rev. D}\ }\textbf {\bibinfo {volume} {110}},\ \bibinfo {pages} {123518} (\bibinfo {year} {2024})},\ \Eprint {http://arxiv.org/abs/2408.11031} {arXiv:2408.11031 [astro-ph.CO]} \BibitemShut {NoStop}%
\bibitem [{\citenamefont {Perivolaropoulos}\ and\ \citenamefont {Skara}(2022)}]{Perivolaropoulos:2021jda}%
  \BibitemOpen
  \bibfield  {author} {\bibinfo {author} {\bibfnamefont {L.}~\bibnamefont {Perivolaropoulos}}\ and\ \bibinfo {author} {\bibfnamefont {F.}~\bibnamefont {Skara}},\ }\href {\doibase 10.1016/j.newar.2022.101659} {\bibfield  {journal} {\bibinfo  {journal} {New Astron. Rev.}\ }\textbf {\bibinfo {volume} {95}},\ \bibinfo {pages} {101659} (\bibinfo {year} {2022})},\ \Eprint {http://arxiv.org/abs/2105.05208} {arXiv:2105.05208 [astro-ph.CO]} \BibitemShut {NoStop}%
\bibitem [{\citenamefont {Beltr{\'a}n~Jim{\'e}nez}\ \emph {et~al.}(2019{\natexlab{a}})\citenamefont {Beltr{\'a}n~Jim{\'e}nez}, \citenamefont {Heisenberg},\ and\ \citenamefont {Koivisto}}]{BeltranJimenez:2019esp}%
  \BibitemOpen
  \bibfield  {author} {\bibinfo {author} {\bibfnamefont {J.}~\bibnamefont {Beltr{\'a}n~Jim{\'e}nez}}, \bibinfo {author} {\bibfnamefont {L.}~\bibnamefont {Heisenberg}}, \ and\ \bibinfo {author} {\bibfnamefont {T.~S.}\ \bibnamefont {Koivisto}},\ }\href {\doibase 10.3390/universe5070173} {\bibfield  {journal} {\bibinfo  {journal} {Universe}\ }\textbf {\bibinfo {volume} {5}},\ \bibinfo {pages} {173} (\bibinfo {year} {2019}{\natexlab{a}})},\ \Eprint {http://arxiv.org/abs/1903.06830} {arXiv:1903.06830 [hep-th]} \BibitemShut {NoStop}%
\bibitem [{\citenamefont {Beltr{\'a}n~Jim{\'e}nez}\ \emph {et~al.}(2019{\natexlab{b}})\citenamefont {Beltr{\'a}n~Jim{\'e}nez}, \citenamefont {Heisenberg},\ and\ \citenamefont {Koivisto}}]{BeltranJimenez:2019tjy}%
  \BibitemOpen
  \bibfield  {author} {\bibinfo {author} {\bibfnamefont {J.}~\bibnamefont {Beltr{\'a}n~Jim{\'e}nez}}, \bibinfo {author} {\bibfnamefont {L.}~\bibnamefont {Heisenberg}}, \ and\ \bibinfo {author} {\bibfnamefont {T.~S.}\ \bibnamefont {Koivisto}},\ }\href {\doibase 10.3390/universe5070173} {\bibfield  {journal} {\bibinfo  {journal} {Universe}\ }\textbf {\bibinfo {volume} {5}},\ \bibinfo {pages} {173} (\bibinfo {year} {2019}{\natexlab{b}})},\ \Eprint {http://arxiv.org/abs/1903.06830} {arXiv:1903.06830 [hep-th]} \BibitemShut {NoStop}%
\bibitem [{\citenamefont {Nester}\ and\ \citenamefont {Yo}(1999)}]{Nester:1998mp}%
  \BibitemOpen
  \bibfield  {author} {\bibinfo {author} {\bibfnamefont {J.~M.}\ \bibnamefont {Nester}}\ and\ \bibinfo {author} {\bibfnamefont {H.-J.}\ \bibnamefont {Yo}},\ }\href@noop {} {\bibfield  {journal} {\bibinfo  {journal} {Chin. J. Phys.}\ }\textbf {\bibinfo {volume} {37}},\ \bibinfo {pages} {113} (\bibinfo {year} {1999})},\ \Eprint {http://arxiv.org/abs/gr-qc/9809049} {arXiv:gr-qc/9809049} \BibitemShut {NoStop}%
\bibitem [{\citenamefont {Bahamonde}\ \emph {et~al.}(2023)\citenamefont {Bahamonde}, \citenamefont {Dialektopoulos}, \citenamefont {Escamilla-Rivera}, \citenamefont {Farrugia}, \citenamefont {Gakis}, \citenamefont {Hendry}, \citenamefont {Hohmann}, \citenamefont {Levi~Said}, \citenamefont {Mifsud},\ and\ \citenamefont {Di~Valentino}}]{Bahamonde:2021gfp}%
  \BibitemOpen
  \bibfield  {author} {\bibinfo {author} {\bibfnamefont {S.}~\bibnamefont {Bahamonde}}, \bibinfo {author} {\bibfnamefont {K.~F.}\ \bibnamefont {Dialektopoulos}}, \bibinfo {author} {\bibfnamefont {C.}~\bibnamefont {Escamilla-Rivera}}, \bibinfo {author} {\bibfnamefont {G.}~\bibnamefont {Farrugia}}, \bibinfo {author} {\bibfnamefont {V.}~\bibnamefont {Gakis}}, \bibinfo {author} {\bibfnamefont {M.}~\bibnamefont {Hendry}}, \bibinfo {author} {\bibfnamefont {M.}~\bibnamefont {Hohmann}}, \bibinfo {author} {\bibfnamefont {J.}~\bibnamefont {Levi~Said}}, \bibinfo {author} {\bibfnamefont {J.}~\bibnamefont {Mifsud}}, \ and\ \bibinfo {author} {\bibfnamefont {E.}~\bibnamefont {Di~Valentino}},\ }\href {\doibase 10.1088/1361-6633/ac9cef} {\bibfield  {journal} {\bibinfo  {journal} {Rept. Prog. Phys.}\ }\textbf {\bibinfo {volume} {86}},\ \bibinfo {pages} {026901} (\bibinfo {year} {2023})},\ \Eprint {http://arxiv.org/abs/2106.13793} {arXiv:2106.13793 [gr-qc]} \BibitemShut {NoStop}%
\bibitem [{\citenamefont {Beltr{\'a}n~Jim{\'e}nez}\ \emph {et~al.}(2020)\citenamefont {Beltr{\'a}n~Jim{\'e}nez}, \citenamefont {Heisenberg}, \citenamefont {Koivisto},\ and\ \citenamefont {Pekar}}]{BeltranJimenez:2019tme}%
  \BibitemOpen
  \bibfield  {author} {\bibinfo {author} {\bibfnamefont {J.}~\bibnamefont {Beltr{\'a}n~Jim{\'e}nez}}, \bibinfo {author} {\bibfnamefont {L.}~\bibnamefont {Heisenberg}}, \bibinfo {author} {\bibfnamefont {T.~S.}\ \bibnamefont {Koivisto}}, \ and\ \bibinfo {author} {\bibfnamefont {S.}~\bibnamefont {Pekar}},\ }\href {\doibase 10.1103/PhysRevD.101.103507} {\bibfield  {journal} {\bibinfo  {journal} {Phys. Rev. D}\ }\textbf {\bibinfo {volume} {101}},\ \bibinfo {pages} {103507} (\bibinfo {year} {2020})},\ \Eprint {http://arxiv.org/abs/1906.10027} {arXiv:1906.10027 [gr-qc]} \BibitemShut {NoStop}%
\bibitem [{\citenamefont {De~Felice}\ and\ \citenamefont {Tsujikawa}(2010)}]{DeFelice:2010aj}%
  \BibitemOpen
  \bibfield  {author} {\bibinfo {author} {\bibfnamefont {A.}~\bibnamefont {De~Felice}}\ and\ \bibinfo {author} {\bibfnamefont {S.}~\bibnamefont {Tsujikawa}},\ }\href {\doibase 10.12942/lrr-2010-3} {\bibfield  {journal} {\bibinfo  {journal} {Living Rev. Rel.}\ }\textbf {\bibinfo {volume} {13}},\ \bibinfo {pages} {3} (\bibinfo {year} {2010})},\ \Eprint {http://arxiv.org/abs/1002.4928} {arXiv:1002.4928 [gr-qc]} \BibitemShut {NoStop}%
\bibitem [{\citenamefont {Ferreira}\ \emph {et~al.}(2022)\citenamefont {Ferreira}, \citenamefont {Barreiro}, \citenamefont {Mimoso},\ and\ \citenamefont {Nunes}}]{Ferreira:2022jcd}%
  \BibitemOpen
  \bibfield  {author} {\bibinfo {author} {\bibfnamefont {J.}~\bibnamefont {Ferreira}}, \bibinfo {author} {\bibfnamefont {T.}~\bibnamefont {Barreiro}}, \bibinfo {author} {\bibfnamefont {J.}~\bibnamefont {Mimoso}}, \ and\ \bibinfo {author} {\bibfnamefont {N.~J.}\ \bibnamefont {Nunes}},\ }\href {\doibase 10.1103/PhysRevD.105.123531} {\bibfield  {journal} {\bibinfo  {journal} {Phys. Rev. D}\ }\textbf {\bibinfo {volume} {105}},\ \bibinfo {pages} {123531} (\bibinfo {year} {2022})},\ \Eprint {http://arxiv.org/abs/2203.13788} {arXiv:2203.13788 [astro-ph.CO]} \BibitemShut {NoStop}%
\bibitem [{\citenamefont {Ferreira}(2023)}]{Ferreira:2023tat}%
  \BibitemOpen
  \bibfield  {author} {\bibinfo {author} {\bibfnamefont {J.}~\bibnamefont {Ferreira}},\ }\href@noop {} {\  (\bibinfo {year} {2023})},\ \Eprint {http://arxiv.org/abs/2303.12674} {arXiv:2303.12674 [astro-ph.CO]} \BibitemShut {NoStop}%
\bibitem [{\citenamefont {Anagnostopoulos}\ \emph {et~al.}(2023)\citenamefont {Anagnostopoulos}, \citenamefont {Gakis}, \citenamefont {Saridakis},\ and\ \citenamefont {Basilakos}}]{Anagnostopoulos:2022gej}%
  \BibitemOpen
  \bibfield  {author} {\bibinfo {author} {\bibfnamefont {F.~K.}\ \bibnamefont {Anagnostopoulos}}, \bibinfo {author} {\bibfnamefont {V.}~\bibnamefont {Gakis}}, \bibinfo {author} {\bibfnamefont {E.~N.}\ \bibnamefont {Saridakis}}, \ and\ \bibinfo {author} {\bibfnamefont {S.}~\bibnamefont {Basilakos}},\ }\href {\doibase 10.1140/epjc/s10052-023-11190-x} {\bibfield  {journal} {\bibinfo  {journal} {Eur. Phys. J. C}\ }\textbf {\bibinfo {volume} {83}},\ \bibinfo {pages} {58} (\bibinfo {year} {2023})},\ \Eprint {http://arxiv.org/abs/2205.11445} {arXiv:2205.11445 [gr-qc]} \BibitemShut {NoStop}%
\bibitem [{\citenamefont {Yang}\ \emph {et~al.}(2024)\citenamefont {Yang}, \citenamefont {Ren}, \citenamefont {Wang}, \citenamefont {Cai},\ and\ \citenamefont {Saridakis}}]{Yang:2024tkw}%
  \BibitemOpen
  \bibfield  {author} {\bibinfo {author} {\bibfnamefont {Y.}~\bibnamefont {Yang}}, \bibinfo {author} {\bibfnamefont {X.}~\bibnamefont {Ren}}, \bibinfo {author} {\bibfnamefont {B.}~\bibnamefont {Wang}}, \bibinfo {author} {\bibfnamefont {Y.-F.}\ \bibnamefont {Cai}}, \ and\ \bibinfo {author} {\bibfnamefont {E.~N.}\ \bibnamefont {Saridakis}},\ }\href {\doibase 10.1093/mnras/stae1905} {\bibfield  {journal} {\bibinfo  {journal} {Mon. Not. Roy. Astron. Soc.}\ }\textbf {\bibinfo {volume} {533}},\ \bibinfo {pages} {2232} (\bibinfo {year} {2024})},\ \Eprint {http://arxiv.org/abs/2404.12140} {arXiv:2404.12140 [astro-ph.CO]} \BibitemShut {NoStop}%
\bibitem [{\citenamefont {Gadbail}\ and\ \citenamefont {Sahoo}(2024)}]{Gadbail:2024een}%
  \BibitemOpen
  \bibfield  {author} {\bibinfo {author} {\bibfnamefont {G.~N.}\ \bibnamefont {Gadbail}}\ and\ \bibinfo {author} {\bibfnamefont {P.~K.}\ \bibnamefont {Sahoo}},\ }\href {\doibase 10.1016/j.cjph.2024.04.037} {\bibfield  {journal} {\bibinfo  {journal} {Chin. J. Phys.}\ }\textbf {\bibinfo {volume} {89}},\ \bibinfo {pages} {1754} (\bibinfo {year} {2024})},\ \Eprint {http://arxiv.org/abs/2405.01594} {arXiv:2405.01594 [gr-qc]} \BibitemShut {NoStop}%
\bibitem [{\citenamefont {Su}\ \emph {et~al.}(2025)\citenamefont {Su}, \citenamefont {He},\ and\ \citenamefont {Zhang}}]{Su:2024avk}%
  \BibitemOpen
  \bibfield  {author} {\bibinfo {author} {\bibfnamefont {X.}~\bibnamefont {Su}}, \bibinfo {author} {\bibfnamefont {D.}~\bibnamefont {He}}, \ and\ \bibinfo {author} {\bibfnamefont {Y.}~\bibnamefont {Zhang}},\ }\href {\doibase 10.1140/epjc/s10052-025-14077-1} {\bibfield  {journal} {\bibinfo  {journal} {Eur. Phys. J. C}\ }\textbf {\bibinfo {volume} {85}},\ \bibinfo {pages} {358} (\bibinfo {year} {2025})},\ \Eprint {http://arxiv.org/abs/2408.03725} {arXiv:2408.03725 [gr-qc]} \BibitemShut {NoStop}%
\bibitem [{\citenamefont {Gadbail}\ \emph {et~al.}(2024)\citenamefont {Gadbail}, \citenamefont {Mandal},\ and\ \citenamefont {Sahoo}}]{Gadbail:2024rpp}%
  \BibitemOpen
  \bibfield  {author} {\bibinfo {author} {\bibfnamefont {G.~N.}\ \bibnamefont {Gadbail}}, \bibinfo {author} {\bibfnamefont {S.}~\bibnamefont {Mandal}}, \ and\ \bibinfo {author} {\bibfnamefont {P.~K.}\ \bibnamefont {Sahoo}},\ }\href {\doibase 10.3847/1538-4357/ad5cf4} {\bibfield  {journal} {\bibinfo  {journal} {Astrophys. J.}\ }\textbf {\bibinfo {volume} {972}},\ \bibinfo {pages} {174} (\bibinfo {year} {2024})},\ \Eprint {http://arxiv.org/abs/2404.13095} {arXiv:2404.13095 [gr-qc]} \BibitemShut {NoStop}%
\bibitem [{\citenamefont {Mhamdi}\ \emph {et~al.}(2024)\citenamefont {Mhamdi}, \citenamefont {Bargach}, \citenamefont {Dahmani}, \citenamefont {Bouali},\ and\ \citenamefont {Ouali}}]{Mhamdi:2024xqd}%
  \BibitemOpen
  \bibfield  {author} {\bibinfo {author} {\bibfnamefont {D.}~\bibnamefont {Mhamdi}}, \bibinfo {author} {\bibfnamefont {F.}~\bibnamefont {Bargach}}, \bibinfo {author} {\bibfnamefont {S.}~\bibnamefont {Dahmani}}, \bibinfo {author} {\bibfnamefont {A.}~\bibnamefont {Bouali}}, \ and\ \bibinfo {author} {\bibfnamefont {T.}~\bibnamefont {Ouali}},\ }\href {\doibase 10.1016/j.physletb.2024.139113} {\bibfield  {journal} {\bibinfo  {journal} {Phys. Lett. B}\ }\textbf {\bibinfo {volume} {859}},\ \bibinfo {pages} {139113} (\bibinfo {year} {2024})},\ \Eprint {http://arxiv.org/abs/2410.10480} {arXiv:2410.10480 [gr-qc]} \BibitemShut {NoStop}%
\bibitem [{\citenamefont {Vasquez}\ and\ \citenamefont {Oliveros}(2025)}]{Vasquez:2025hrz}%
  \BibitemOpen
  \bibfield  {author} {\bibinfo {author} {\bibfnamefont {I.~R.}\ \bibnamefont {Vasquez}}\ and\ \bibinfo {author} {\bibfnamefont {A.}~\bibnamefont {Oliveros}},\ }\href {\doibase 10.1007/s10714-025-03403-3} {\bibfield  {journal} {\bibinfo  {journal} {Gen. Rel. Grav.}\ }\textbf {\bibinfo {volume} {57}},\ \bibinfo {pages} {67} (\bibinfo {year} {2025})},\ \Eprint {http://arxiv.org/abs/2501.12585} {arXiv:2501.12585 [gr-qc]} \BibitemShut {NoStop}%
\bibitem [{\citenamefont {Ayuso}\ \emph {et~al.}(2025)\citenamefont {Ayuso}, \citenamefont {Bouhmadi-L{\'o}pez}, \citenamefont {Chen}, \citenamefont {Chew}, \citenamefont {Dialektopoulos},\ and\ \citenamefont {Ong}}]{Ayuso:2025vkc}%
  \BibitemOpen
  \bibfield  {author} {\bibinfo {author} {\bibfnamefont {I.}~\bibnamefont {Ayuso}}, \bibinfo {author} {\bibfnamefont {M.}~\bibnamefont {Bouhmadi-L{\'o}pez}}, \bibinfo {author} {\bibfnamefont {C.-Y.}\ \bibnamefont {Chen}}, \bibinfo {author} {\bibfnamefont {X.~Y.}\ \bibnamefont {Chew}}, \bibinfo {author} {\bibfnamefont {K.}~\bibnamefont {Dialektopoulos}}, \ and\ \bibinfo {author} {\bibfnamefont {Y.~C.}\ \bibnamefont {Ong}},\ }\href@noop {} {\  (\bibinfo {year} {2025})},\ \Eprint {http://arxiv.org/abs/2506.03506} {arXiv:2506.03506 [gr-qc]} \BibitemShut {NoStop}%
\bibitem [{\citenamefont {Roy}\ \emph {et~al.}(2025)\citenamefont {Roy}, \citenamefont {Beesham},\ and\ \citenamefont {Paul}}]{Roy:2025nde}%
  \BibitemOpen
  \bibfield  {author} {\bibinfo {author} {\bibfnamefont {B.~C.}\ \bibnamefont {Roy}}, \bibinfo {author} {\bibfnamefont {A.}~\bibnamefont {Beesham}}, \ and\ \bibinfo {author} {\bibfnamefont {B.~C.}\ \bibnamefont {Paul}},\ }\href@noop {} {\  (\bibinfo {year} {2025})},\ \Eprint {http://arxiv.org/abs/2504.15680} {arXiv:2504.15680 [gr-qc]} \BibitemShut {NoStop}%
\bibitem [{\citenamefont {Boiza}\ \emph {et~al.}(2025)\citenamefont {Boiza}, \citenamefont {Petronikolou}, \citenamefont {Bouhmadi-L{\'o}pez},\ and\ \citenamefont {Saridakis}}]{Boiza:2025xpn}%
  \BibitemOpen
  \bibfield  {author} {\bibinfo {author} {\bibfnamefont {C.~G.}\ \bibnamefont {Boiza}}, \bibinfo {author} {\bibfnamefont {M.}~\bibnamefont {Petronikolou}}, \bibinfo {author} {\bibfnamefont {M.}~\bibnamefont {Bouhmadi-L{\'o}pez}}, \ and\ \bibinfo {author} {\bibfnamefont {E.~N.}\ \bibnamefont {Saridakis}},\ }\href@noop {} {\  (\bibinfo {year} {2025})},\ \Eprint {http://arxiv.org/abs/2505.18264} {arXiv:2505.18264 [astro-ph.CO]} \BibitemShut {NoStop}%
\bibitem [{\citenamefont {Frusciante}(2021)}]{Frusciante:2021sio}%
  \BibitemOpen
  \bibfield  {author} {\bibinfo {author} {\bibfnamefont {N.}~\bibnamefont {Frusciante}},\ }\href {\doibase 10.1103/PhysRevD.103.044021} {\bibfield  {journal} {\bibinfo  {journal} {Phys. Rev. D}\ }\textbf {\bibinfo {volume} {103}},\ \bibinfo {pages} {044021} (\bibinfo {year} {2021})},\ \Eprint {http://arxiv.org/abs/2101.09242} {arXiv:2101.09242 [astro-ph.CO]} \BibitemShut {NoStop}%
\bibitem [{\citenamefont {Bhardwaj}\ \emph {et~al.}(2024)\citenamefont {Bhardwaj}, \citenamefont {Garg},\ and\ \citenamefont {Prakash}}]{Bhardwaj:2024mop}%
  \BibitemOpen
  \bibfield  {author} {\bibinfo {author} {\bibfnamefont {V.~K.}\ \bibnamefont {Bhardwaj}}, \bibinfo {author} {\bibfnamefont {P.}~\bibnamefont {Garg}}, \ and\ \bibinfo {author} {\bibfnamefont {S.}~\bibnamefont {Prakash}},\ }\href {\doibase 10.1007/s10509-024-04315-5} {\bibfield  {journal} {\bibinfo  {journal} {Astrophys. Space Sci.}\ }\textbf {\bibinfo {volume} {369}},\ \bibinfo {pages} {50} (\bibinfo {year} {2024})}\BibitemShut {NoStop}%
\bibitem [{\citenamefont {Sahlu}\ \emph {et~al.}(2025)\citenamefont {Sahlu}, \citenamefont {de~la Cruz-Dombriz},\ and\ \citenamefont {Abebe}}]{Sahlu:2024pxk}%
  \BibitemOpen
  \bibfield  {author} {\bibinfo {author} {\bibfnamefont {S.}~\bibnamefont {Sahlu}}, \bibinfo {author} {\bibfnamefont {{\'A}.}~\bibnamefont {de~la Cruz-Dombriz}}, \ and\ \bibinfo {author} {\bibfnamefont {A.}~\bibnamefont {Abebe}},\ }\href {\doibase 10.1093/mnras/staf439} {\bibfield  {journal} {\bibinfo  {journal} {Mon. Not. Roy. Astron. Soc.}\ }\textbf {\bibinfo {volume} {539}},\ \bibinfo {pages} {690} (\bibinfo {year} {2025})},\ \Eprint {http://arxiv.org/abs/2405.07361} {arXiv:2405.07361 [gr-qc]} \BibitemShut {NoStop}%
\bibitem [{\citenamefont {Enkhili}\ \emph {et~al.}(2024)\citenamefont {Enkhili}, \citenamefont {Dahmani}, \citenamefont {Mhamdi}, \citenamefont {Ouali},\ and\ \citenamefont {Errahmani}}]{Enkhili:2024dil}%
  \BibitemOpen
  \bibfield  {author} {\bibinfo {author} {\bibfnamefont {O.}~\bibnamefont {Enkhili}}, \bibinfo {author} {\bibfnamefont {S.}~\bibnamefont {Dahmani}}, \bibinfo {author} {\bibfnamefont {D.}~\bibnamefont {Mhamdi}}, \bibinfo {author} {\bibfnamefont {T.}~\bibnamefont {Ouali}}, \ and\ \bibinfo {author} {\bibfnamefont {A.}~\bibnamefont {Errahmani}},\ }\href {\doibase 10.1140/epjc/s10052-024-13143-4} {\bibfield  {journal} {\bibinfo  {journal} {Eur. Phys. J. C}\ }\textbf {\bibinfo {volume} {84}},\ \bibinfo {pages} {806} (\bibinfo {year} {2024})},\ \Eprint {http://arxiv.org/abs/2407.12236} {arXiv:2407.12236 [gr-qc]} \BibitemShut {NoStop}%
\bibitem [{\citenamefont {Kolhatkar}\ \emph {et~al.}(2024)\citenamefont {Kolhatkar}, \citenamefont {Mishra},\ and\ \citenamefont {Sahoo}}]{Kolhatkar:2024oyy}%
  \BibitemOpen
  \bibfield  {author} {\bibinfo {author} {\bibfnamefont {A.}~\bibnamefont {Kolhatkar}}, \bibinfo {author} {\bibfnamefont {S.~S.}\ \bibnamefont {Mishra}}, \ and\ \bibinfo {author} {\bibfnamefont {P.~K.}\ \bibnamefont {Sahoo}},\ }\href {\doibase 10.1140/epjc/s10052-024-13237-z} {\bibfield  {journal} {\bibinfo  {journal} {Eur. Phys. J. C}\ }\textbf {\bibinfo {volume} {84}},\ \bibinfo {pages} {888} (\bibinfo {year} {2024})},\ \Eprint {http://arxiv.org/abs/2409.01538} {arXiv:2409.01538 [gr-qc]} \BibitemShut {NoStop}%
\bibitem [{\citenamefont {Sakr}\ and\ \citenamefont {Schey}(2024)}]{Sakr:2024eee}%
  \BibitemOpen
  \bibfield  {author} {\bibinfo {author} {\bibfnamefont {Z.}~\bibnamefont {Sakr}}\ and\ \bibinfo {author} {\bibfnamefont {L.}~\bibnamefont {Schey}},\ }\href {\doibase 10.1088/1475-7516/2024/10/052} {\bibfield  {journal} {\bibinfo  {journal} {JCAP}\ }\textbf {\bibinfo {volume} {10}},\ \bibinfo {pages} {052} (\bibinfo {year} {2024})},\ \Eprint {http://arxiv.org/abs/2405.03627} {arXiv:2405.03627 [astro-ph.CO]} \BibitemShut {NoStop}%
\bibitem [{\citenamefont {Wang}\ \emph {et~al.}(2024)\citenamefont {Wang}, \citenamefont {Ren}, \citenamefont {Cai}, \citenamefont {Luo},\ and\ \citenamefont {Saridakis}}]{Wang:2024eai}%
  \BibitemOpen
  \bibfield  {author} {\bibinfo {author} {\bibfnamefont {Q.}~\bibnamefont {Wang}}, \bibinfo {author} {\bibfnamefont {X.}~\bibnamefont {Ren}}, \bibinfo {author} {\bibfnamefont {Y.-F.}\ \bibnamefont {Cai}}, \bibinfo {author} {\bibfnamefont {W.}~\bibnamefont {Luo}}, \ and\ \bibinfo {author} {\bibfnamefont {E.~N.}\ \bibnamefont {Saridakis}},\ }\href {\doibase 10.3847/1538-4357/ad6c4d} {\bibfield  {journal} {\bibinfo  {journal} {Astrophys. J.}\ }\textbf {\bibinfo {volume} {974}},\ \bibinfo {pages} {7} (\bibinfo {year} {2024})},\ \Eprint {http://arxiv.org/abs/2406.00242} {arXiv:2406.00242 [astro-ph.CO]} \BibitemShut {NoStop}%
\bibitem [{\citenamefont {Paliathanasis}(2025)}]{Paliathanasis:2025hjw}%
  \BibitemOpen
  \bibfield  {author} {\bibinfo {author} {\bibfnamefont {A.}~\bibnamefont {Paliathanasis}},\ }\href {\doibase 10.1016/j.dark.2025.101993} {\bibfield  {journal} {\bibinfo  {journal} {Phys. Dark Univ.}\ }\textbf {\bibinfo {volume} {49}},\ \bibinfo {pages} {101993} (\bibinfo {year} {2025})},\ \Eprint {http://arxiv.org/abs/2504.11132} {arXiv:2504.11132 [gr-qc]} \BibitemShut {NoStop}%
\bibitem [{\citenamefont {D'Agostino}\ and\ \citenamefont {Nunes}(2022)}]{DAgostino:2022tdk}%
  \BibitemOpen
  \bibfield  {author} {\bibinfo {author} {\bibfnamefont {R.}~\bibnamefont {D'Agostino}}\ and\ \bibinfo {author} {\bibfnamefont {R.~C.}\ \bibnamefont {Nunes}},\ }\href {\doibase 10.1103/PhysRevD.106.124053} {\bibfield  {journal} {\bibinfo  {journal} {Phys. Rev. D}\ }\textbf {\bibinfo {volume} {106}},\ \bibinfo {pages} {124053} (\bibinfo {year} {2022})},\ \Eprint {http://arxiv.org/abs/2210.11935} {arXiv:2210.11935 [gr-qc]} \BibitemShut {NoStop}%
\bibitem [{\citenamefont {N{\'a}jera}\ \emph {et~al.}(2023)\citenamefont {N{\'a}jera}, \citenamefont {Alvarado},\ and\ \citenamefont {Escamilla-Rivera}}]{Najera:2023wcw}%
  \BibitemOpen
  \bibfield  {author} {\bibinfo {author} {\bibfnamefont {J.~A.}\ \bibnamefont {N{\'a}jera}}, \bibinfo {author} {\bibfnamefont {C.~A.}\ \bibnamefont {Alvarado}}, \ and\ \bibinfo {author} {\bibfnamefont {C.}~\bibnamefont {Escamilla-Rivera}},\ }\href {\doibase 10.1093/mnras/stad2180} {\bibfield  {journal} {\bibinfo  {journal} {Mon. Not. Roy. Astron. Soc.}\ }\textbf {\bibinfo {volume} {524}},\ \bibinfo {pages} {5280} (\bibinfo {year} {2023})},\ \Eprint {http://arxiv.org/abs/2304.12601} {arXiv:2304.12601 [gr-qc]} \BibitemShut {NoStop}%
\bibitem [{\citenamefont {Anagnostopoulos}\ \emph {et~al.}(2021)\citenamefont {Anagnostopoulos}, \citenamefont {Basilakos},\ and\ \citenamefont {Saridakis}}]{Anagnostopoulos:2021ydo}%
  \BibitemOpen
  \bibfield  {author} {\bibinfo {author} {\bibfnamefont {F.~K.}\ \bibnamefont {Anagnostopoulos}}, \bibinfo {author} {\bibfnamefont {S.}~\bibnamefont {Basilakos}}, \ and\ \bibinfo {author} {\bibfnamefont {E.~N.}\ \bibnamefont {Saridakis}},\ }\href {\doibase 10.1016/j.physletb.2021.136634} {\bibfield  {journal} {\bibinfo  {journal} {Phys. Lett. B}\ }\textbf {\bibinfo {volume} {822}},\ \bibinfo {pages} {136634} (\bibinfo {year} {2021})},\ \Eprint {http://arxiv.org/abs/2104.15123} {arXiv:2104.15123 [gr-qc]} \BibitemShut {NoStop}%
\bibitem [{\citenamefont {Gon{\c{c}}alves}\ \emph {et~al.}(2024)\citenamefont {Gon{\c{c}}alves}, \citenamefont {Atayde},\ and\ \citenamefont {Frusciante}}]{Goncalves:2024sem}%
  \BibitemOpen
  \bibfield  {author} {\bibinfo {author} {\bibfnamefont {T.~B.}\ \bibnamefont {Gon{\c{c}}alves}}, \bibinfo {author} {\bibfnamefont {L.}~\bibnamefont {Atayde}}, \ and\ \bibinfo {author} {\bibfnamefont {N.}~\bibnamefont {Frusciante}},\ }\href {\doibase 10.1103/PhysRevD.109.084003} {\bibfield  {journal} {\bibinfo  {journal} {Phys. Rev. D}\ }\textbf {\bibinfo {volume} {109}},\ \bibinfo {pages} {084003} (\bibinfo {year} {2024})},\ \Eprint {http://arxiv.org/abs/2404.01742} {arXiv:2404.01742 [gr-qc]} \BibitemShut {NoStop}%
\bibitem [{\citenamefont {Jarv}\ and\ \citenamefont {Pati}(2024)}]{Jarv:2023sbp}%
  \BibitemOpen
  \bibfield  {author} {\bibinfo {author} {\bibfnamefont {L.}~\bibnamefont {Jarv}}\ and\ \bibinfo {author} {\bibfnamefont {L.}~\bibnamefont {Pati}},\ }\href {\doibase 10.1103/PhysRevD.109.064069} {\bibfield  {journal} {\bibinfo  {journal} {Phys. Rev. D}\ }\textbf {\bibinfo {volume} {109}},\ \bibinfo {pages} {064069} (\bibinfo {year} {2024})},\ \Eprint {http://arxiv.org/abs/2309.04262} {arXiv:2309.04262 [gr-qc]} \BibitemShut {NoStop}%
\bibitem [{\citenamefont {Bahamonde}\ and\ \citenamefont {J{\"a}rv}(2022)}]{Bahamonde:2022zgj}%
  \BibitemOpen
  \bibfield  {author} {\bibinfo {author} {\bibfnamefont {S.}~\bibnamefont {Bahamonde}}\ and\ \bibinfo {author} {\bibfnamefont {L.}~\bibnamefont {J{\"a}rv}},\ }\href {\doibase 10.1140/epjc/s10052-022-10922-9} {\bibfield  {journal} {\bibinfo  {journal} {Eur. Phys. J. C}\ }\textbf {\bibinfo {volume} {82}},\ \bibinfo {pages} {963} (\bibinfo {year} {2022})},\ \Eprint {http://arxiv.org/abs/2208.01872} {arXiv:2208.01872 [gr-qc]} \BibitemShut {NoStop}%
\bibitem [{\citenamefont {Heisenberg}(2024)}]{Heisenberg:2023lru}%
  \BibitemOpen
  \bibfield  {author} {\bibinfo {author} {\bibfnamefont {L.}~\bibnamefont {Heisenberg}},\ }\href {\doibase 10.1016/j.physrep.2024.02.001} {\bibfield  {journal} {\bibinfo  {journal} {Phys. Rept.}\ }\textbf {\bibinfo {volume} {1066}},\ \bibinfo {pages} {1} (\bibinfo {year} {2024})},\ \Eprint {http://arxiv.org/abs/2309.15958} {arXiv:2309.15958 [gr-qc]} \BibitemShut {NoStop}%
\bibitem [{\citenamefont {Aghanim}\ \emph {et~al.}(2020{\natexlab{b}})\citenamefont {Aghanim} \emph {et~al.}}]{Planck:2018vyg}%
  \BibitemOpen
  \bibfield  {author} {\bibinfo {author} {\bibfnamefont {N.}~\bibnamefont {Aghanim}} \emph {et~al.} (\bibinfo {collaboration} {Planck}),\ }\href {\doibase 10.1051/0004-6361/201833910} {\bibfield  {journal} {\bibinfo  {journal} {Astron. Astrophys.}\ }\textbf {\bibinfo {volume} {641}},\ \bibinfo {pages} {A6} (\bibinfo {year} {2020}{\natexlab{b}})},\ \bibinfo {note} {[Erratum: Astron.Astrophys. 652, C4 (2021)]},\ \Eprint {http://arxiv.org/abs/1807.06209} {arXiv:1807.06209 [astro-ph.CO]} \BibitemShut {NoStop}%
\bibitem [{\citenamefont {Nesseris}\ \emph {et~al.}(2017)\citenamefont {Nesseris}, \citenamefont {Pantazis},\ and\ \citenamefont {Perivolaropoulos}}]{Nesseris:2017vor}%
  \BibitemOpen
  \bibfield  {author} {\bibinfo {author} {\bibfnamefont {S.}~\bibnamefont {Nesseris}}, \bibinfo {author} {\bibfnamefont {G.}~\bibnamefont {Pantazis}}, \ and\ \bibinfo {author} {\bibfnamefont {L.}~\bibnamefont {Perivolaropoulos}},\ }\href {\doibase 10.1103/PhysRevD.96.023542} {\bibfield  {journal} {\bibinfo  {journal} {Phys. Rev. D}\ }\textbf {\bibinfo {volume} {96}},\ \bibinfo {pages} {023542} (\bibinfo {year} {2017})},\ \Eprint {http://arxiv.org/abs/1703.10538} {arXiv:1703.10538 [astro-ph.CO]} \BibitemShut {NoStop}%
\bibitem [{\citenamefont {Kar}\ \emph {et~al.}(2022)\citenamefont {Kar}, \citenamefont {Sadhukhan},\ and\ \citenamefont {Debnath}}]{Kar:2021juu}%
  \BibitemOpen
  \bibfield  {author} {\bibinfo {author} {\bibfnamefont {A.}~\bibnamefont {Kar}}, \bibinfo {author} {\bibfnamefont {S.}~\bibnamefont {Sadhukhan}}, \ and\ \bibinfo {author} {\bibfnamefont {U.}~\bibnamefont {Debnath}},\ }\href {\doibase 10.1142/S0217732322501838} {\bibfield  {journal} {\bibinfo  {journal} {Mod. Phys. Lett. A}\ }\textbf {\bibinfo {volume} {37}},\ \bibinfo {pages} {2250183} (\bibinfo {year} {2022})},\ \Eprint {http://arxiv.org/abs/2109.10906} {arXiv:2109.10906 [gr-qc]} \BibitemShut {NoStop}%
\bibitem [{\citenamefont {Capozziello}\ and\ \citenamefont {D'Agostino}(2022)}]{Capozziello:2022wgl}%
  \BibitemOpen
  \bibfield  {author} {\bibinfo {author} {\bibfnamefont {S.}~\bibnamefont {Capozziello}}\ and\ \bibinfo {author} {\bibfnamefont {R.}~\bibnamefont {D'Agostino}},\ }\href {\doibase 10.1016/j.physletb.2022.137229} {\bibfield  {journal} {\bibinfo  {journal} {Phys. Lett. B}\ }\textbf {\bibinfo {volume} {832}},\ \bibinfo {pages} {137229} (\bibinfo {year} {2022})},\ \Eprint {http://arxiv.org/abs/2204.01015} {arXiv:2204.01015 [gr-qc]} \BibitemShut {NoStop}%
\bibitem [{\citenamefont {Gadbail}\ \emph {et~al.}(2022)\citenamefont {Gadbail}, \citenamefont {Mandal},\ and\ \citenamefont {Sahoo}}]{Gadbail:2022jco}%
  \BibitemOpen
  \bibfield  {author} {\bibinfo {author} {\bibfnamefont {G.~N.}\ \bibnamefont {Gadbail}}, \bibinfo {author} {\bibfnamefont {S.}~\bibnamefont {Mandal}}, \ and\ \bibinfo {author} {\bibfnamefont {P.~K.}\ \bibnamefont {Sahoo}},\ }\href {\doibase 10.1016/j.physletb.2022.137509} {\bibfield  {journal} {\bibinfo  {journal} {Phys. Lett. B}\ }\textbf {\bibinfo {volume} {835}},\ \bibinfo {pages} {137509} (\bibinfo {year} {2022})},\ \Eprint {http://arxiv.org/abs/2210.09237} {arXiv:2210.09237 [gr-qc]} \BibitemShut {NoStop}%
\bibitem [{\citenamefont {Gadbail}\ \emph {et~al.}(2023{\natexlab{a}})\citenamefont {Gadbail}, \citenamefont {Arora},\ and\ \citenamefont {Sahoo}}]{Gadbail:2023klq}%
  \BibitemOpen
  \bibfield  {author} {\bibinfo {author} {\bibfnamefont {G.~N.}\ \bibnamefont {Gadbail}}, \bibinfo {author} {\bibfnamefont {S.}~\bibnamefont {Arora}}, \ and\ \bibinfo {author} {\bibfnamefont {P.~K.}\ \bibnamefont {Sahoo}},\ }\href {\doibase 10.1016/j.physletb.2023.137710} {\bibfield  {journal} {\bibinfo  {journal} {Phys. Lett. B}\ }\textbf {\bibinfo {volume} {838}},\ \bibinfo {pages} {137710} (\bibinfo {year} {2023}{\natexlab{a}})},\ \Eprint {http://arxiv.org/abs/2301.08876} {arXiv:2301.08876 [gr-qc]} \BibitemShut {NoStop}%
\bibitem [{\citenamefont {Naik}\ \emph {et~al.}(2023)\citenamefont {Naik}, \citenamefont {Kavya}, \citenamefont {Sudharani},\ and\ \citenamefont {Venkatesha}}]{Naik:2023ykt}%
  \BibitemOpen
  \bibfield  {author} {\bibinfo {author} {\bibfnamefont {D.~M.}\ \bibnamefont {Naik}}, \bibinfo {author} {\bibfnamefont {N.~S.}\ \bibnamefont {Kavya}}, \bibinfo {author} {\bibfnamefont {L.}~\bibnamefont {Sudharani}}, \ and\ \bibinfo {author} {\bibfnamefont {V.}~\bibnamefont {Venkatesha}},\ }\href {\doibase 10.1140/epjc/s10052-023-12029-1} {\bibfield  {journal} {\bibinfo  {journal} {Eur. Phys. J. C}\ }\textbf {\bibinfo {volume} {83}},\ \bibinfo {pages} {840} (\bibinfo {year} {2023})}\BibitemShut {NoStop}%
\bibitem [{\citenamefont {Mahmood}\ \emph {et~al.}(2024)\citenamefont {Mahmood}, \citenamefont {Sohail}, \citenamefont {Ditta}, \citenamefont {Shekh},\ and\ \citenamefont {Yadav}}]{Mahmood:2023mac}%
  \BibitemOpen
  \bibfield  {author} {\bibinfo {author} {\bibfnamefont {I.}~\bibnamefont {Mahmood}}, \bibinfo {author} {\bibfnamefont {H.}~\bibnamefont {Sohail}}, \bibinfo {author} {\bibfnamefont {A.}~\bibnamefont {Ditta}}, \bibinfo {author} {\bibfnamefont {S.~H.}\ \bibnamefont {Shekh}}, \ and\ \bibinfo {author} {\bibfnamefont {A.~K.}\ \bibnamefont {Yadav}},\ }\href {\doibase 10.1142/S0219887824502049} {\bibfield  {journal} {\bibinfo  {journal} {Int. J. Geom. Meth. Mod. Phys.}\ }\textbf {\bibinfo {volume} {21}},\ \bibinfo {pages} {2450204} (\bibinfo {year} {2024})},\ \Eprint {http://arxiv.org/abs/2311.16527} {arXiv:2311.16527 [gr-qc]} \BibitemShut {NoStop}%
\bibitem [{\citenamefont {Gadbail}\ \emph {et~al.}(2023{\natexlab{b}})\citenamefont {Gadbail}, \citenamefont {De},\ and\ \citenamefont {Sahoo}}]{Gadbail:2023mvu}%
  \BibitemOpen
  \bibfield  {author} {\bibinfo {author} {\bibfnamefont {G.~N.}\ \bibnamefont {Gadbail}}, \bibinfo {author} {\bibfnamefont {A.}~\bibnamefont {De}}, \ and\ \bibinfo {author} {\bibfnamefont {P.~K.}\ \bibnamefont {Sahoo}},\ }\href {\doibase 10.1140/epjc/s10052-023-12288-y} {\bibfield  {journal} {\bibinfo  {journal} {Eur. Phys. J. C}\ }\textbf {\bibinfo {volume} {83}},\ \bibinfo {pages} {1099} (\bibinfo {year} {2023}{\natexlab{b}})},\ \Eprint {http://arxiv.org/abs/2312.02492} {arXiv:2312.02492 [gr-qc]} \BibitemShut {NoStop}%
\bibitem [{\citenamefont {Kaczmarek}(2024)}]{Kaczmarek:2024yju}%
  \BibitemOpen
  \bibfield  {author} {\bibinfo {author} {\bibfnamefont {A.~Z.}\ \bibnamefont {Kaczmarek}},\ }\href {\doibase 10.1016/j.nuclphysb.2024.116677} {\bibfield  {journal} {\bibinfo  {journal} {Nucl. Phys. B}\ }\textbf {\bibinfo {volume} {1007}},\ \bibinfo {pages} {116677} (\bibinfo {year} {2024})},\ \Eprint {http://arxiv.org/abs/2401.04084} {arXiv:2401.04084 [gr-qc]} \BibitemShut {NoStop}%
\bibitem [{\citenamefont {Kaczmarek}\ \emph {et~al.}(2025)\citenamefont {Kaczmarek}, \citenamefont {Rosa},\ and\ \citenamefont {Szcz{\c{e}}{\'s}niak}}]{Kaczmarek:2024quk}%
  \BibitemOpen
  \bibfield  {author} {\bibinfo {author} {\bibfnamefont {A.~Z.}\ \bibnamefont {Kaczmarek}}, \bibinfo {author} {\bibfnamefont {J.~L.}\ \bibnamefont {Rosa}}, \ and\ \bibinfo {author} {\bibfnamefont {D.}~\bibnamefont {Szcz{\c{e}}{\'s}niak}},\ }\href {\doibase 10.1140/epjc/s10052-025-13919-2} {\bibfield  {journal} {\bibinfo  {journal} {Eur. Phys. J. C}\ }\textbf {\bibinfo {volume} {85}},\ \bibinfo {pages} {203} (\bibinfo {year} {2025})},\ \Eprint {http://arxiv.org/abs/2410.00707} {arXiv:2410.00707 [gr-qc]} \BibitemShut {NoStop}%
\bibitem [{\citenamefont {Gadbail}\ \emph {et~al.}(2025)\citenamefont {Gadbail}, \citenamefont {Mandal}, \citenamefont {Sahoo},\ and\ \citenamefont {Bamba}}]{Gadbail:2024lek}%
  \BibitemOpen
  \bibfield  {author} {\bibinfo {author} {\bibfnamefont {G.~N.}\ \bibnamefont {Gadbail}}, \bibinfo {author} {\bibfnamefont {S.}~\bibnamefont {Mandal}}, \bibinfo {author} {\bibfnamefont {P.~K.}\ \bibnamefont {Sahoo}}, \ and\ \bibinfo {author} {\bibfnamefont {K.}~\bibnamefont {Bamba}},\ }\href {\doibase 10.1016/j.physletb.2024.139232} {\bibfield  {journal} {\bibinfo  {journal} {Phys. Lett. B}\ }\textbf {\bibinfo {volume} {860}},\ \bibinfo {pages} {139232} (\bibinfo {year} {2025})},\ \Eprint {http://arxiv.org/abs/2411.00051} {arXiv:2411.00051 [gr-qc]} \BibitemShut {NoStop}%
\bibitem [{\citenamefont {El~Ouardi}\ \emph {et~al.}(2025)\citenamefont {El~Ouardi}, \citenamefont {Bouali}, \citenamefont {Dahmani}, \citenamefont {Errahmani},\ and\ \citenamefont {Ouali}}]{ElOuardi:2025okl}%
  \BibitemOpen
  \bibfield  {author} {\bibinfo {author} {\bibfnamefont {R.}~\bibnamefont {El~Ouardi}}, \bibinfo {author} {\bibfnamefont {A.}~\bibnamefont {Bouali}}, \bibinfo {author} {\bibfnamefont {S.}~\bibnamefont {Dahmani}}, \bibinfo {author} {\bibfnamefont {A.}~\bibnamefont {Errahmani}}, \ and\ \bibinfo {author} {\bibfnamefont {T.}~\bibnamefont {Ouali}},\ }\href {\doibase 10.1016/j.physletb.2025.139374} {\bibfield  {journal} {\bibinfo  {journal} {Phys. Lett. B}\ }\textbf {\bibinfo {volume} {863}},\ \bibinfo {pages} {139374} (\bibinfo {year} {2025})}\BibitemShut {NoStop}%
\bibitem [{\citenamefont {Yadav}\ \emph {et~al.}(2024)\citenamefont {Yadav}, \citenamefont {Bhoyar}, \citenamefont {Dhabe}, \citenamefont {Shekh},\ and\ \citenamefont {Ahmad}}]{Yadav:2024vmt}%
  \BibitemOpen
  \bibfield  {author} {\bibinfo {author} {\bibfnamefont {A.~K.}\ \bibnamefont {Yadav}}, \bibinfo {author} {\bibfnamefont {S.~R.}\ \bibnamefont {Bhoyar}}, \bibinfo {author} {\bibfnamefont {M.~C.}\ \bibnamefont {Dhabe}}, \bibinfo {author} {\bibfnamefont {S.~H.}\ \bibnamefont {Shekh}}, \ and\ \bibinfo {author} {\bibfnamefont {N.}~\bibnamefont {Ahmad}},\ }\href {\doibase 10.1016/j.jheap.2024.06.012} {\bibfield  {journal} {\bibinfo  {journal} {JHEAp}\ }\textbf {\bibinfo {volume} {43}},\ \bibinfo {pages} {114} (\bibinfo {year} {2024})}\BibitemShut {NoStop}%
\bibitem [{\citenamefont {Saha}\ \emph {et~al.}(2024)\citenamefont {Saha}, \citenamefont {Chakrabortty},\ and\ \citenamefont {Sanyal}}]{Saha:2024gmk}%
  \BibitemOpen
  \bibfield  {author} {\bibinfo {author} {\bibfnamefont {D.}~\bibnamefont {Saha}}, \bibinfo {author} {\bibfnamefont {M.}~\bibnamefont {Chakrabortty}}, \ and\ \bibinfo {author} {\bibfnamefont {A.~K.}\ \bibnamefont {Sanyal}},\ }\href {\doibase 10.3390/universe10010044} {\bibfield  {journal} {\bibinfo  {journal} {Universe}\ }\textbf {\bibinfo {volume} {10}},\ \bibinfo {pages} {44} (\bibinfo {year} {2024})},\ \Eprint {http://arxiv.org/abs/2401.11226} {arXiv:2401.11226 [gr-qc]} \BibitemShut {NoStop}%
\bibitem [{\citenamefont {Esposito}\ \emph {et~al.}(2022)\citenamefont {Esposito}, \citenamefont {Carloni}, \citenamefont {Cianci},\ and\ \citenamefont {Vignolo}}]{Esposito:2021ect}%
  \BibitemOpen
  \bibfield  {author} {\bibinfo {author} {\bibfnamefont {F.}~\bibnamefont {Esposito}}, \bibinfo {author} {\bibfnamefont {S.}~\bibnamefont {Carloni}}, \bibinfo {author} {\bibfnamefont {R.}~\bibnamefont {Cianci}}, \ and\ \bibinfo {author} {\bibfnamefont {S.}~\bibnamefont {Vignolo}},\ }\href {\doibase 10.1103/PhysRevD.105.084061} {\bibfield  {journal} {\bibinfo  {journal} {Phys. Rev. D}\ }\textbf {\bibinfo {volume} {105}},\ \bibinfo {pages} {084061} (\bibinfo {year} {2022})},\ \Eprint {http://arxiv.org/abs/2107.14522} {arXiv:2107.14522 [gr-qc]} \BibitemShut {NoStop}%
\bibitem [{\citenamefont {Kang}(2021)}]{Kang:2021osc}%
  \BibitemOpen
  \bibfield  {author} {\bibinfo {author} {\bibfnamefont {J.}~\bibnamefont {Kang}},\ }\href {\doibase 10.1016/j.dark.2021.100784} {\bibfield  {journal} {\bibinfo  {journal} {Phys. Dark Univ.}\ }\textbf {\bibinfo {volume} {31}},\ \bibinfo {pages} {100784} (\bibinfo {year} {2021})},\ \Eprint {http://arxiv.org/abs/2102.04232} {arXiv:2102.04232 [astro-ph.CO]} \BibitemShut {NoStop}%
\bibitem [{\citenamefont {Albuquerque}\ and\ \citenamefont {Frusciante}(2022)}]{Albuquerque:2022eac}%
  \BibitemOpen
  \bibfield  {author} {\bibinfo {author} {\bibfnamefont {I.~S.}\ \bibnamefont {Albuquerque}}\ and\ \bibinfo {author} {\bibfnamefont {N.}~\bibnamefont {Frusciante}},\ }\href {\doibase 10.1016/j.dark.2022.100980} {\bibfield  {journal} {\bibinfo  {journal} {Phys. Dark Univ.}\ }\textbf {\bibinfo {volume} {35}},\ \bibinfo {pages} {100980} (\bibinfo {year} {2022})},\ \Eprint {http://arxiv.org/abs/2202.04637} {arXiv:2202.04637 [astro-ph.CO]} \BibitemShut {NoStop}%
\bibitem [{\citenamefont {Lodha}\ \emph {et~al.}(2025)\citenamefont {Lodha} \emph {et~al.}}]{DESI:2025fii}%
  \BibitemOpen
  \bibfield  {author} {\bibinfo {author} {\bibfnamefont {K.}~\bibnamefont {Lodha}} \emph {et~al.} (\bibinfo {collaboration} {DESI}),\ }\href@noop {} {\  (\bibinfo {year} {2025})},\ \Eprint {http://arxiv.org/abs/2503.14743} {arXiv:2503.14743 [astro-ph.CO]} \BibitemShut {NoStop}%
\bibitem [{\citenamefont {Basilakos}\ \emph {et~al.}(2025)\citenamefont {Basilakos}, \citenamefont {Paliathanasis},\ and\ \citenamefont {Saridakis}}]{Basilakos:2025olm}%
  \BibitemOpen
  \bibfield  {author} {\bibinfo {author} {\bibfnamefont {S.}~\bibnamefont {Basilakos}}, \bibinfo {author} {\bibfnamefont {A.}~\bibnamefont {Paliathanasis}}, \ and\ \bibinfo {author} {\bibfnamefont {E.~N.}\ \bibnamefont {Saridakis}},\ }\href {\doibase 10.1016/j.physletb.2025.139658} {\bibfield  {journal} {\bibinfo  {journal} {Phys. Lett. B}\ }\textbf {\bibinfo {volume} {868}},\ \bibinfo {pages} {139658} (\bibinfo {year} {2025})},\ \Eprint {http://arxiv.org/abs/2503.19864} {arXiv:2503.19864 [gr-qc]} \BibitemShut {NoStop}%
\bibitem [{\citenamefont {Lymperis}(2022)}]{Lymperis:2022oyo}%
  \BibitemOpen
  \bibfield  {author} {\bibinfo {author} {\bibfnamefont {A.}~\bibnamefont {Lymperis}},\ }\href {\doibase 10.1088/1475-7516/2022/11/018} {\bibfield  {journal} {\bibinfo  {journal} {JCAP}\ }\textbf {\bibinfo {volume} {11}},\ \bibinfo {pages} {018} (\bibinfo {year} {2022})},\ \Eprint {http://arxiv.org/abs/2207.10997} {arXiv:2207.10997 [gr-qc]} \BibitemShut {NoStop}%
\bibitem [{\citenamefont {Narawade}\ and\ \citenamefont {Mishra}(2023)}]{Narawade:2022cgb}%
  \BibitemOpen
  \bibfield  {author} {\bibinfo {author} {\bibfnamefont {S.~A.}\ \bibnamefont {Narawade}}\ and\ \bibinfo {author} {\bibfnamefont {B.}~\bibnamefont {Mishra}},\ }\href {\doibase 10.1002/andp.202200626} {\bibfield  {journal} {\bibinfo  {journal} {Annalen Phys.}\ }\textbf {\bibinfo {volume} {535}},\ \bibinfo {pages} {2200626} (\bibinfo {year} {2023})},\ \Eprint {http://arxiv.org/abs/2211.09701} {arXiv:2211.09701 [gr-qc]} \BibitemShut {NoStop}%
\bibitem [{\citenamefont {Arora}\ and\ \citenamefont {Sahoo}(2022)}]{Arora:2022mlo}%
  \BibitemOpen
  \bibfield  {author} {\bibinfo {author} {\bibfnamefont {S.}~\bibnamefont {Arora}}\ and\ \bibinfo {author} {\bibfnamefont {P.~K.}\ \bibnamefont {Sahoo}},\ }\href {\doibase 10.1002/andp.202200233} {\bibfield  {journal} {\bibinfo  {journal} {Annalen Phys.}\ }\textbf {\bibinfo {volume} {534}},\ \bibinfo {pages} {2200233} (\bibinfo {year} {2022})},\ \Eprint {http://arxiv.org/abs/2206.05110} {arXiv:2206.05110 [gr-qc]} \BibitemShut {NoStop}%
\bibitem [{\citenamefont {Hu}\ \emph {et~al.}(2022)\citenamefont {Hu}, \citenamefont {Katsuragawa},\ and\ \citenamefont {Qiu}}]{Hu:2022anq}%
  \BibitemOpen
  \bibfield  {author} {\bibinfo {author} {\bibfnamefont {K.}~\bibnamefont {Hu}}, \bibinfo {author} {\bibfnamefont {T.}~\bibnamefont {Katsuragawa}}, \ and\ \bibinfo {author} {\bibfnamefont {T.}~\bibnamefont {Qiu}},\ }\href {\doibase 10.1103/PhysRevD.106.044025} {\bibfield  {journal} {\bibinfo  {journal} {Phys. Rev. D}\ }\textbf {\bibinfo {volume} {106}},\ \bibinfo {pages} {044025} (\bibinfo {year} {2022})},\ \Eprint {http://arxiv.org/abs/2204.12826} {arXiv:2204.12826 [gr-qc]} \BibitemShut {NoStop}%
\bibitem [{\citenamefont {D'Ambrosio}\ \emph {et~al.}(2023)\citenamefont {D'Ambrosio}, \citenamefont {Heisenberg},\ and\ \citenamefont {Zentarra}}]{DAmbrosio:2023asf}%
  \BibitemOpen
  \bibfield  {author} {\bibinfo {author} {\bibfnamefont {F.}~\bibnamefont {D'Ambrosio}}, \bibinfo {author} {\bibfnamefont {L.}~\bibnamefont {Heisenberg}}, \ and\ \bibinfo {author} {\bibfnamefont {S.}~\bibnamefont {Zentarra}},\ }\href {\doibase 10.1002/prop.202300185} {\bibfield  {journal} {\bibinfo  {journal} {Fortsch. Phys.}\ }\textbf {\bibinfo {volume} {71}},\ \bibinfo {pages} {2300185} (\bibinfo {year} {2023})},\ \Eprint {http://arxiv.org/abs/2308.02250} {arXiv:2308.02250 [gr-qc]} \BibitemShut {NoStop}%
\bibitem [{\citenamefont {Tomonari}\ and\ \citenamefont {Bahamonde}(2024)}]{Tomonari:2023wcs}%
  \BibitemOpen
  \bibfield  {author} {\bibinfo {author} {\bibfnamefont {K.}~\bibnamefont {Tomonari}}\ and\ \bibinfo {author} {\bibfnamefont {S.}~\bibnamefont {Bahamonde}},\ }\href {\doibase 10.1140/epjc/s10052-024-12677-x} {\bibfield  {journal} {\bibinfo  {journal} {Eur. Phys. J. C}\ }\textbf {\bibinfo {volume} {84}},\ \bibinfo {pages} {349} (\bibinfo {year} {2024})},\ \bibinfo {note} {[Erratum: Eur.Phys.J.C 84, 508 (2024)]},\ \Eprint {http://arxiv.org/abs/2308.06469} {arXiv:2308.06469 [gr-qc]} \BibitemShut {NoStop}%
\bibitem [{\citenamefont {Guzman}(2023)}]{Guzman:2023oyl}%
  \BibitemOpen
  \bibfield  {author} {\bibinfo {author} {\bibfnamefont {M.-J.}\ \bibnamefont {Guzman}},\ }\href@noop {} {\  (\bibinfo {year} {2023})},\ \Eprint {http://arxiv.org/abs/2311.01424} {arXiv:2311.01424 [gr-qc]} \BibitemShut {NoStop}%
\bibitem [{\citenamefont {Gomes}\ \emph {et~al.}(2024)\citenamefont {Gomes}, \citenamefont {Beltr{\'a}n~Jim{\'e}nez}, \citenamefont {Cano},\ and\ \citenamefont {Koivisto}}]{Gomes:2023tur}%
  \BibitemOpen
  \bibfield  {author} {\bibinfo {author} {\bibfnamefont {D.~A.}\ \bibnamefont {Gomes}}, \bibinfo {author} {\bibfnamefont {J.}~\bibnamefont {Beltr{\'a}n~Jim{\'e}nez}}, \bibinfo {author} {\bibfnamefont {A.~J.}\ \bibnamefont {Cano}}, \ and\ \bibinfo {author} {\bibfnamefont {T.~S.}\ \bibnamefont {Koivisto}},\ }\href {\doibase 10.1103/PhysRevLett.132.141401} {\bibfield  {journal} {\bibinfo  {journal} {Phys. Rev. Lett.}\ }\textbf {\bibinfo {volume} {132}},\ \bibinfo {pages} {141401} (\bibinfo {year} {2024})},\ \Eprint {http://arxiv.org/abs/2311.04201} {arXiv:2311.04201 [gr-qc]} \BibitemShut {NoStop}%
\bibitem [{\citenamefont {Guzm{\'a}n}\ \emph {et~al.}(2024)\citenamefont {Guzm{\'a}n}, \citenamefont {J{\"a}rv},\ and\ \citenamefont {Pati}}]{Guzman:2024cwa}%
  \BibitemOpen
  \bibfield  {author} {\bibinfo {author} {\bibfnamefont {M.-J.}\ \bibnamefont {Guzm{\'a}n}}, \bibinfo {author} {\bibfnamefont {L.}~\bibnamefont {J{\"a}rv}}, \ and\ \bibinfo {author} {\bibfnamefont {L.}~\bibnamefont {Pati}},\ }\href {\doibase 10.1103/PhysRevD.110.124013} {\bibfield  {journal} {\bibinfo  {journal} {Phys. Rev. D}\ }\textbf {\bibinfo {volume} {110}},\ \bibinfo {pages} {124013} (\bibinfo {year} {2024})},\ \Eprint {http://arxiv.org/abs/2406.11621} {arXiv:2406.11621 [gr-qc]} \BibitemShut {NoStop}%
\bibitem [{\citenamefont {Bello-Morales}\ \emph {et~al.}(2024)\citenamefont {Bello-Morales}, \citenamefont {Beltr{\'a}n~Jim{\'e}nez}, \citenamefont {Jim{\'e}nez~Cano}, \citenamefont {Koivisto},\ and\ \citenamefont {Maroto}}]{Bello-Morales:2024vqk}%
  \BibitemOpen
  \bibfield  {author} {\bibinfo {author} {\bibfnamefont {A.~G.}\ \bibnamefont {Bello-Morales}}, \bibinfo {author} {\bibfnamefont {J.}~\bibnamefont {Beltr{\'a}n~Jim{\'e}nez}}, \bibinfo {author} {\bibfnamefont {A.}~\bibnamefont {Jim{\'e}nez~Cano}}, \bibinfo {author} {\bibfnamefont {T.~S.}\ \bibnamefont {Koivisto}}, \ and\ \bibinfo {author} {\bibfnamefont {A.~L.}\ \bibnamefont {Maroto}},\ }\href {\doibase 10.1007/JHEP12(2024)146} {\bibfield  {journal} {\bibinfo  {journal} {JHEP}\ }\textbf {\bibinfo {volume} {12}},\ \bibinfo {pages} {146} (\bibinfo {year} {2024})},\ \Eprint {http://arxiv.org/abs/2406.19355} {arXiv:2406.19355 [gr-qc]} \BibitemShut {NoStop}%
\bibitem [{\citenamefont {Dutta}\ \emph {et~al.}(2025)\citenamefont {Dutta}, \citenamefont {Khyllep}, \citenamefont {Chakraborty}, \citenamefont {Gregoris},\ and\ \citenamefont {Karwan}}]{Dutta:2025fqw}%
  \BibitemOpen
  \bibfield  {author} {\bibinfo {author} {\bibfnamefont {J.}~\bibnamefont {Dutta}}, \bibinfo {author} {\bibfnamefont {W.}~\bibnamefont {Khyllep}}, \bibinfo {author} {\bibfnamefont {S.}~\bibnamefont {Chakraborty}}, \bibinfo {author} {\bibfnamefont {D.}~\bibnamefont {Gregoris}}, \ and\ \bibinfo {author} {\bibfnamefont {K.}~\bibnamefont {Karwan}},\ }\href@noop {} {\  (\bibinfo {year} {2025})},\ \Eprint {http://arxiv.org/abs/2508.09530} {arXiv:2508.09530 [gr-qc]} \BibitemShut {NoStop}%
\bibitem [{\citenamefont {Khyllep}\ \emph {et~al.}(2023)\citenamefont {Khyllep}, \citenamefont {Dutta}, \citenamefont {Saridakis},\ and\ \citenamefont {Yesmakhanova}}]{Khyllep:2022spx}%
  \BibitemOpen
  \bibfield  {author} {\bibinfo {author} {\bibfnamefont {W.}~\bibnamefont {Khyllep}}, \bibinfo {author} {\bibfnamefont {J.}~\bibnamefont {Dutta}}, \bibinfo {author} {\bibfnamefont {E.~N.}\ \bibnamefont {Saridakis}}, \ and\ \bibinfo {author} {\bibfnamefont {K.}~\bibnamefont {Yesmakhanova}},\ }\href {\doibase 10.1103/PhysRevD.107.044022} {\bibfield  {journal} {\bibinfo  {journal} {Phys. Rev. D}\ }\textbf {\bibinfo {volume} {107}},\ \bibinfo {pages} {044022} (\bibinfo {year} {2023})},\ \Eprint {http://arxiv.org/abs/2207.02610} {arXiv:2207.02610 [gr-qc]} \BibitemShut {NoStop}%
\bibitem [{\citenamefont {Narawade}\ \emph {et~al.}(2022)\citenamefont {Narawade}, \citenamefont {Pati}, \citenamefont {Mishra},\ and\ \citenamefont {Tripathy}}]{Narawade:2022jeg}%
  \BibitemOpen
  \bibfield  {author} {\bibinfo {author} {\bibfnamefont {S.~A.}\ \bibnamefont {Narawade}}, \bibinfo {author} {\bibfnamefont {L.}~\bibnamefont {Pati}}, \bibinfo {author} {\bibfnamefont {B.}~\bibnamefont {Mishra}}, \ and\ \bibinfo {author} {\bibfnamefont {S.~K.}\ \bibnamefont {Tripathy}},\ }\href {\doibase 10.1016/j.dark.2022.101020} {\bibfield  {journal} {\bibinfo  {journal} {Phys. Dark Univ.}\ }\textbf {\bibinfo {volume} {36}},\ \bibinfo {pages} {101020} (\bibinfo {year} {2022})},\ \Eprint {http://arxiv.org/abs/2203.14121} {arXiv:2203.14121 [gr-qc]} \BibitemShut {NoStop}%
\bibitem [{\citenamefont {Tomonari}\ \emph {et~al.}(2025)\citenamefont {Tomonari}, \citenamefont {Katsuragawa},\ and\ \citenamefont {Nojiri}}]{Tomonari:2025axd}%
  \BibitemOpen
  \bibfield  {author} {\bibinfo {author} {\bibfnamefont {K.}~\bibnamefont {Tomonari}}, \bibinfo {author} {\bibfnamefont {T.}~\bibnamefont {Katsuragawa}}, \ and\ \bibinfo {author} {\bibfnamefont {S.}~\bibnamefont {Nojiri}},\ }\href@noop {} {\  (\bibinfo {year} {2025})},\ \Eprint {http://arxiv.org/abs/2506.22158} {arXiv:2506.22158 [gr-qc]} \BibitemShut {NoStop}%
\bibitem [{\citenamefont {Green}\ and\ \citenamefont {Wald}(2014)}]{Green:2014aga}%
  \BibitemOpen
  \bibfield  {author} {\bibinfo {author} {\bibfnamefont {S.~R.}\ \bibnamefont {Green}}\ and\ \bibinfo {author} {\bibfnamefont {R.~M.}\ \bibnamefont {Wald}},\ }\href {\doibase 10.1088/0264-9381/31/23/234003} {\bibfield  {journal} {\bibinfo  {journal} {Class. Quant. Grav.}\ }\textbf {\bibinfo {volume} {31}},\ \bibinfo {pages} {234003} (\bibinfo {year} {2014})},\ \Eprint {http://arxiv.org/abs/1407.8084} {arXiv:1407.8084 [gr-qc]} \BibitemShut {NoStop}%
\bibitem [{Note1()}]{Note1}%
  \BibitemOpen
  \bibinfo {note} {We assume that the matter, i.e., baryons and \ac {CDM}is pressureless ($p_m = 0$) dust.}\BibitemShut {Stop}%
\bibitem [{\citenamefont {Beltr{\'a}n~Jim{\'e}nez}\ \emph {et~al.}(2018)\citenamefont {Beltr{\'a}n~Jim{\'e}nez}, \citenamefont {Heisenberg},\ and\ \citenamefont {Koivisto}}]{BeltranJimenez:2017tkd}%
  \BibitemOpen
  \bibfield  {author} {\bibinfo {author} {\bibfnamefont {J.}~\bibnamefont {Beltr{\'a}n~Jim{\'e}nez}}, \bibinfo {author} {\bibfnamefont {L.}~\bibnamefont {Heisenberg}}, \ and\ \bibinfo {author} {\bibfnamefont {T.}~\bibnamefont {Koivisto}},\ }\href {\doibase 10.1103/PhysRevD.98.044048} {\bibfield  {journal} {\bibinfo  {journal} {Phys. Rev. D}\ }\textbf {\bibinfo {volume} {98}},\ \bibinfo {pages} {044048} (\bibinfo {year} {2018})},\ \Eprint {http://arxiv.org/abs/1710.03116} {arXiv:1710.03116 [gr-qc]} \BibitemShut {NoStop}%
\bibitem [{\citenamefont {Amendola}\ \emph {et~al.}(2008)\citenamefont {Amendola}, \citenamefont {Kunz},\ and\ \citenamefont {Sapone}}]{Amendola:2007rr}%
  \BibitemOpen
  \bibfield  {author} {\bibinfo {author} {\bibfnamefont {L.}~\bibnamefont {Amendola}}, \bibinfo {author} {\bibfnamefont {M.}~\bibnamefont {Kunz}}, \ and\ \bibinfo {author} {\bibfnamefont {D.}~\bibnamefont {Sapone}},\ }\href {\doibase 10.1088/1475-7516/2008/04/013} {\bibfield  {journal} {\bibinfo  {journal} {JCAP}\ }\textbf {\bibinfo {volume} {04}},\ \bibinfo {pages} {013} (\bibinfo {year} {2008})},\ \Eprint {http://arxiv.org/abs/0704.2421} {arXiv:0704.2421 [astro-ph]} \BibitemShut {NoStop}%
\bibitem [{\citenamefont {Nishizawa}(2018)}]{Nishizawa:2017nef}%
  \BibitemOpen
  \bibfield  {author} {\bibinfo {author} {\bibfnamefont {A.}~\bibnamefont {Nishizawa}},\ }\href {\doibase 10.1103/PhysRevD.97.104037} {\bibfield  {journal} {\bibinfo  {journal} {Phys. Rev. D}\ }\textbf {\bibinfo {volume} {97}},\ \bibinfo {pages} {104037} (\bibinfo {year} {2018})},\ \Eprint {http://arxiv.org/abs/1710.04825} {arXiv:1710.04825 [gr-qc]} \BibitemShut {NoStop}%
\bibitem [{\citenamefont {Flathmann}\ and\ \citenamefont {Hohmann}(2021)}]{Flathmann:2020zyj}%
  \BibitemOpen
  \bibfield  {author} {\bibinfo {author} {\bibfnamefont {K.}~\bibnamefont {Flathmann}}\ and\ \bibinfo {author} {\bibfnamefont {M.}~\bibnamefont {Hohmann}},\ }\href {\doibase 10.1103/PhysRevD.103.044030} {\bibfield  {journal} {\bibinfo  {journal} {Phys. Rev. D}\ }\textbf {\bibinfo {volume} {103}},\ \bibinfo {pages} {044030} (\bibinfo {year} {2021})},\ \Eprint {http://arxiv.org/abs/2012.12875} {arXiv:2012.12875 [gr-qc]} \BibitemShut {NoStop}%
\bibitem [{\citenamefont {Adame}\ \emph {et~al.}(2025)\citenamefont {Adame} \emph {et~al.}}]{DESI:2024mwx}%
  \BibitemOpen
  \bibfield  {author} {\bibinfo {author} {\bibfnamefont {A.~G.}\ \bibnamefont {Adame}} \emph {et~al.} (\bibinfo {collaboration} {DESI}),\ }\href {\doibase 10.1088/1475-7516/2025/02/021} {\bibfield  {journal} {\bibinfo  {journal} {JCAP}\ }\textbf {\bibinfo {volume} {02}},\ \bibinfo {pages} {021} (\bibinfo {year} {2025})},\ \Eprint {http://arxiv.org/abs/2404.03002} {arXiv:2404.03002 [astro-ph.CO]} \BibitemShut {NoStop}%
\bibitem [{\citenamefont {M{\"o}ller}\ \emph {et~al.}(2024)\citenamefont {M{\"o}ller} \emph {et~al.}}]{DES:2024haq}%
  \BibitemOpen
  \bibfield  {author} {\bibinfo {author} {\bibfnamefont {A.}~\bibnamefont {M{\"o}ller}} \emph {et~al.} (\bibinfo {collaboration} {DES}),\ }\href {\doibase 10.1093/mnras/stae1953} {\bibfield  {journal} {\bibinfo  {journal} {Mon. Not. Roy. Astron. Soc.}\ }\textbf {\bibinfo {volume} {533}},\ \bibinfo {pages} {2073} (\bibinfo {year} {2024})},\ \Eprint {http://arxiv.org/abs/2402.18690} {arXiv:2402.18690 [astro-ph.CO]} \BibitemShut {NoStop}%
\bibitem [{\citenamefont {Addison}\ \emph {et~al.}(2018)\citenamefont {Addison}, \citenamefont {Watts}, \citenamefont {Bennett}, \citenamefont {Halpern}, \citenamefont {Hinshaw},\ and\ \citenamefont {Weiland}}]{Addison:2017fdm}%
  \BibitemOpen
  \bibfield  {author} {\bibinfo {author} {\bibfnamefont {G.~E.}\ \bibnamefont {Addison}}, \bibinfo {author} {\bibfnamefont {D.~J.}\ \bibnamefont {Watts}}, \bibinfo {author} {\bibfnamefont {C.~L.}\ \bibnamefont {Bennett}}, \bibinfo {author} {\bibfnamefont {M.}~\bibnamefont {Halpern}}, \bibinfo {author} {\bibfnamefont {G.}~\bibnamefont {Hinshaw}}, \ and\ \bibinfo {author} {\bibfnamefont {J.~L.}\ \bibnamefont {Weiland}},\ }\href {\doibase 10.3847/1538-4357/aaa1ed} {\bibfield  {journal} {\bibinfo  {journal} {Astrophys. J.}\ }\textbf {\bibinfo {volume} {853}},\ \bibinfo {pages} {119} (\bibinfo {year} {2018})},\ \Eprint {http://arxiv.org/abs/1707.06547} {arXiv:1707.06547 [astro-ph.CO]} \BibitemShut {NoStop}%
\bibitem [{\citenamefont {Aubourg}\ \emph {et~al.}(2015)\citenamefont {Aubourg} \emph {et~al.}}]{BOSS:2014hhw}%
  \BibitemOpen
  \bibfield  {author} {\bibinfo {author} {\bibfnamefont {{\'E}.}~\bibnamefont {Aubourg}} \emph {et~al.} (\bibinfo {collaboration} {BOSS}),\ }\href {\doibase 10.1103/PhysRevD.92.123516} {\bibfield  {journal} {\bibinfo  {journal} {Phys. Rev. D}\ }\textbf {\bibinfo {volume} {92}},\ \bibinfo {pages} {123516} (\bibinfo {year} {2015})},\ \Eprint {http://arxiv.org/abs/1411.1074} {arXiv:1411.1074 [astro-ph.CO]} \BibitemShut {NoStop}%
\bibitem [{\citenamefont {Haridasu}\ \emph {et~al.}(2018)\citenamefont {Haridasu}, \citenamefont {Lukovi{\'c}}, \citenamefont {Moresco},\ and\ \citenamefont {Vittorio}}]{Haridasu:2018gqm}%
  \BibitemOpen
  \bibfield  {author} {\bibinfo {author} {\bibfnamefont {B.~S.}\ \bibnamefont {Haridasu}}, \bibinfo {author} {\bibfnamefont {V.~V.}\ \bibnamefont {Lukovi{\'c}}}, \bibinfo {author} {\bibfnamefont {M.}~\bibnamefont {Moresco}}, \ and\ \bibinfo {author} {\bibfnamefont {N.}~\bibnamefont {Vittorio}},\ }\href {\doibase 10.1088/1475-7516/2018/10/015} {\bibfield  {journal} {\bibinfo  {journal} {JCAP}\ }\textbf {\bibinfo {volume} {10}},\ \bibinfo {pages} {015} (\bibinfo {year} {2018})},\ \Eprint {http://arxiv.org/abs/1805.03595} {arXiv:1805.03595 [astro-ph.CO]} \BibitemShut {NoStop}%
\bibitem [{\citenamefont {Lapi}\ \emph {et~al.}(2025)\citenamefont {Lapi}, \citenamefont {Haridasu}, \citenamefont {Boco}, \citenamefont {Cueli}, \citenamefont {Baccigalupi},\ and\ \citenamefont {Danese}}]{Lapi:2025mgr}%
  \BibitemOpen
  \bibfield  {author} {\bibinfo {author} {\bibfnamefont {A.}~\bibnamefont {Lapi}}, \bibinfo {author} {\bibfnamefont {B.~S.}\ \bibnamefont {Haridasu}}, \bibinfo {author} {\bibfnamefont {L.}~\bibnamefont {Boco}}, \bibinfo {author} {\bibfnamefont {M.~M.}\ \bibnamefont {Cueli}}, \bibinfo {author} {\bibfnamefont {C.}~\bibnamefont {Baccigalupi}}, \ and\ \bibinfo {author} {\bibfnamefont {L.}~\bibnamefont {Danese}},\ }\href {\doibase 10.1088/1475-7516/2025/04/015} {\bibfield  {journal} {\bibinfo  {journal} {JCAP}\ }\textbf {\bibinfo {volume} {04}},\ \bibinfo {pages} {015} (\bibinfo {year} {2025})},\ \Eprint {http://arxiv.org/abs/2502.05823} {arXiv:2502.05823 [astro-ph.CO]} \BibitemShut {NoStop}%
\bibitem [{\citenamefont {Barros}\ \emph {et~al.}(2020)\citenamefont {Barros}, \citenamefont {Barreiro}, \citenamefont {Koivisto},\ and\ \citenamefont {Nunes}}]{Barros:2020bgg}%
  \BibitemOpen
  \bibfield  {author} {\bibinfo {author} {\bibfnamefont {B.~J.}\ \bibnamefont {Barros}}, \bibinfo {author} {\bibfnamefont {T.}~\bibnamefont {Barreiro}}, \bibinfo {author} {\bibfnamefont {T.}~\bibnamefont {Koivisto}}, \ and\ \bibinfo {author} {\bibfnamefont {N.~J.}\ \bibnamefont {Nunes}},\ }\href {\doibase 10.1016/j.dark.2020.100616} {\bibfield  {journal} {\bibinfo  {journal} {Phys. Dark Univ.}\ }\textbf {\bibinfo {volume} {30}},\ \bibinfo {pages} {100616} (\bibinfo {year} {2020})},\ \Eprint {http://arxiv.org/abs/2004.07867} {arXiv:2004.07867 [gr-qc]} \BibitemShut {NoStop}%
\bibitem [{\citenamefont {Alcock}\ and\ \citenamefont {Paczynski}(1979)}]{Alcock:1979mp}%
  \BibitemOpen
  \bibfield  {author} {\bibinfo {author} {\bibfnamefont {C.}~\bibnamefont {Alcock}}\ and\ \bibinfo {author} {\bibfnamefont {B.}~\bibnamefont {Paczynski}},\ }\href {\doibase 10.1038/281358a0} {\bibfield  {journal} {\bibinfo  {journal} {Nature}\ }\textbf {\bibinfo {volume} {281}},\ \bibinfo {pages} {358} (\bibinfo {year} {1979})}\BibitemShut {NoStop}%
\bibitem [{\citenamefont {Alam}\ \emph {et~al.}(2016)\citenamefont {Alam}, \citenamefont {Ho},\ and\ \citenamefont {Silvestri}}]{Alam:2015rsa}%
  \BibitemOpen
  \bibfield  {author} {\bibinfo {author} {\bibfnamefont {S.}~\bibnamefont {Alam}}, \bibinfo {author} {\bibfnamefont {S.}~\bibnamefont {Ho}}, \ and\ \bibinfo {author} {\bibfnamefont {A.}~\bibnamefont {Silvestri}},\ }\href {\doibase 10.1093/mnras/stv2935} {\bibfield  {journal} {\bibinfo  {journal} {Mon. Not. Roy. Astron. Soc.}\ }\textbf {\bibinfo {volume} {456}},\ \bibinfo {pages} {3743} (\bibinfo {year} {2016})},\ \Eprint {http://arxiv.org/abs/1509.05034} {arXiv:1509.05034 [astro-ph.CO]} \BibitemShut {NoStop}%
\bibitem [{\citenamefont {Haridasu}\ \emph {et~al.}(2021)\citenamefont {Haridasu}, \citenamefont {Viel},\ and\ \citenamefont {Vittorio}}]{Haridasu:2020pms}%
  \BibitemOpen
  \bibfield  {author} {\bibinfo {author} {\bibfnamefont {B.~S.}\ \bibnamefont {Haridasu}}, \bibinfo {author} {\bibfnamefont {M.}~\bibnamefont {Viel}}, \ and\ \bibinfo {author} {\bibfnamefont {N.}~\bibnamefont {Vittorio}},\ }\href {\doibase 10.1103/PhysRevD.103.063539} {\bibfield  {journal} {\bibinfo  {journal} {Phys. Rev. D}\ }\textbf {\bibinfo {volume} {103}},\ \bibinfo {pages} {063539} (\bibinfo {year} {2021})},\ \Eprint {http://arxiv.org/abs/2012.10324} {arXiv:2012.10324 [astro-ph.CO]} \BibitemShut {NoStop}%
\bibitem [{\citenamefont {Verde}\ \emph {et~al.}(2017)\citenamefont {Verde}, \citenamefont {Bellini}, \citenamefont {Pigozzo}, \citenamefont {Heavens},\ and\ \citenamefont {Jimenez}}]{Verde:2016wmz}%
  \BibitemOpen
  \bibfield  {author} {\bibinfo {author} {\bibfnamefont {L.}~\bibnamefont {Verde}}, \bibinfo {author} {\bibfnamefont {E.}~\bibnamefont {Bellini}}, \bibinfo {author} {\bibfnamefont {C.}~\bibnamefont {Pigozzo}}, \bibinfo {author} {\bibfnamefont {A.~F.}\ \bibnamefont {Heavens}}, \ and\ \bibinfo {author} {\bibfnamefont {R.}~\bibnamefont {Jimenez}},\ }\href {\doibase 10.1088/1475-7516/2017/04/023} {\bibfield  {journal} {\bibinfo  {journal} {JCAP}\ }\textbf {\bibinfo {volume} {04}},\ \bibinfo {pages} {023} (\bibinfo {year} {2017})},\ \Eprint {http://arxiv.org/abs/1611.00376} {arXiv:1611.00376 [astro-ph.CO]} \BibitemShut {NoStop}%
\bibitem [{\citenamefont {Lemos}\ and\ \citenamefont {Lewis}(2023)}]{Lemos:2023xhs}%
  \BibitemOpen
  \bibfield  {author} {\bibinfo {author} {\bibfnamefont {P.}~\bibnamefont {Lemos}}\ and\ \bibinfo {author} {\bibfnamefont {A.}~\bibnamefont {Lewis}},\ }\href {\doibase 10.1103/PhysRevD.107.103505} {\bibfield  {journal} {\bibinfo  {journal} {Phys. Rev. D}\ }\textbf {\bibinfo {volume} {107}},\ \bibinfo {pages} {103505} (\bibinfo {year} {2023})},\ \Eprint {http://arxiv.org/abs/2302.12911} {arXiv:2302.12911 [astro-ph.CO]} \BibitemShut {NoStop}%
\bibitem [{\citenamefont {Lynch}\ and\ \citenamefont {Knox}(2025)}]{Lynch:2025ine}%
  \BibitemOpen
  \bibfield  {author} {\bibinfo {author} {\bibfnamefont {G.~P.}\ \bibnamefont {Lynch}}\ and\ \bibinfo {author} {\bibfnamefont {L.}~\bibnamefont {Knox}},\ }\href@noop {} {\  (\bibinfo {year} {2025})},\ \Eprint {http://arxiv.org/abs/2503.14470} {arXiv:2503.14470 [astro-ph.CO]} \BibitemShut {NoStop}%
\bibitem [{\citenamefont {Foreman-Mackey}\ \emph {et~al.}(2013)\citenamefont {Foreman-Mackey}, \citenamefont {Hogg}, \citenamefont {Lang},\ and\ \citenamefont {Goodman}}]{Foreman-Mackey:2012any}%
  \BibitemOpen
  \bibfield  {author} {\bibinfo {author} {\bibfnamefont {D.}~\bibnamefont {Foreman-Mackey}}, \bibinfo {author} {\bibfnamefont {D.~W.}\ \bibnamefont {Hogg}}, \bibinfo {author} {\bibfnamefont {D.}~\bibnamefont {Lang}}, \ and\ \bibinfo {author} {\bibfnamefont {J.}~\bibnamefont {Goodman}},\ }\href {\doibase 10.1086/670067} {\bibfield  {journal} {\bibinfo  {journal} {Publ. Astron. Soc. Pac.}\ }\textbf {\bibinfo {volume} {125}},\ \bibinfo {pages} {306} (\bibinfo {year} {2013})},\ \Eprint {http://arxiv.org/abs/1202.3665} {arXiv:1202.3665 [astro-ph.IM]} \BibitemShut {NoStop}%
\bibitem [{\citenamefont {Lange}(2023)}]{Lange:2023ydq}%
  \BibitemOpen
  \bibfield  {author} {\bibinfo {author} {\bibfnamefont {J.~U.}\ \bibnamefont {Lange}},\ }\href {\doibase 10.1093/mnras/stad2441} {\bibfield  {journal} {\bibinfo  {journal} {Mon. Not. Roy. Astron. Soc.}\ }\textbf {\bibinfo {volume} {525}},\ \bibinfo {pages} {3181} (\bibinfo {year} {2023})},\ \Eprint {http://arxiv.org/abs/2306.16923} {arXiv:2306.16923 [astro-ph.IM]} \BibitemShut {NoStop}%
\bibitem [{Note2()}]{Note2}%
  \BibitemOpen
  \bibinfo {note} {\protect \url {https://corner.readthedocs.io/en/latest/}}\BibitemShut {NoStop}%
\bibitem [{Note3()}]{Note3}%
  \BibitemOpen
  \bibinfo {note} {\protect \url {https://getdist.readthedocs.io/}}\BibitemShut {NoStop}%
\bibitem [{\citenamefont {Lewis}(2025)}]{Lewis:2019xzd}%
  \BibitemOpen
  \bibfield  {author} {\bibinfo {author} {\bibfnamefont {A.}~\bibnamefont {Lewis}},\ }\href {\doibase 10.1088/1475-7516/2025/08/025} {\bibfield  {journal} {\bibinfo  {journal} {JCAP}\ }\textbf {\bibinfo {volume} {08}},\ \bibinfo {pages} {025} (\bibinfo {year} {2025})},\ \Eprint {http://arxiv.org/abs/1910.13970} {arXiv:1910.13970 [astro-ph.IM]} \BibitemShut {NoStop}%
\bibitem [{Note4()}]{Note4}%
  \BibitemOpen
  \bibinfo {note} {Within the context of model-independent reconstructions the singularity in the dark energy EoS is accompanied by transition of dark energy density from positive to negative.}\BibitemShut {Stop}%
\bibitem [{\citenamefont {Chevallier}\ and\ \citenamefont {Polarski}(2001)}]{Chevallier:2000qy}%
  \BibitemOpen
  \bibfield  {author} {\bibinfo {author} {\bibfnamefont {M.}~\bibnamefont {Chevallier}}\ and\ \bibinfo {author} {\bibfnamefont {D.}~\bibnamefont {Polarski}},\ }\href {\doibase 10.1142/S0218271801000822} {\bibfield  {journal} {\bibinfo  {journal} {Int. J. Mod. Phys. D}\ }\textbf {\bibinfo {volume} {10}},\ \bibinfo {pages} {213} (\bibinfo {year} {2001})},\ \Eprint {http://arxiv.org/abs/gr-qc/0009008} {arXiv:gr-qc/0009008} \BibitemShut {NoStop}%
\bibitem [{\citenamefont {Linder}(2003)}]{Linder:2002et}%
  \BibitemOpen
  \bibfield  {author} {\bibinfo {author} {\bibfnamefont {E.~V.}\ \bibnamefont {Linder}},\ }\href {\doibase 10.1103/PhysRevLett.90.091301} {\bibfield  {journal} {\bibinfo  {journal} {Phys. Rev. Lett.}\ }\textbf {\bibinfo {volume} {90}},\ \bibinfo {pages} {091301} (\bibinfo {year} {2003})},\ \Eprint {http://arxiv.org/abs/astro-ph/0208512} {arXiv:astro-ph/0208512} \BibitemShut {NoStop}%
\bibitem [{Note5()}]{Note5}%
  \BibitemOpen
  \bibinfo {note} {Dark energy EoS in the CPL model is parametrized as $w(a) = w_0 +w_{\protect \rm a} (1-a)$.}\BibitemShut {Stop}%
\bibitem [{\citenamefont {Umezu}\ \emph {et~al.}(2005)\citenamefont {Umezu}, \citenamefont {Ichiki},\ and\ \citenamefont {Yahiro}}]{Umezu:2005ee}%
  \BibitemOpen
  \bibfield  {author} {\bibinfo {author} {\bibfnamefont {K.-i.}\ \bibnamefont {Umezu}}, \bibinfo {author} {\bibfnamefont {K.}~\bibnamefont {Ichiki}}, \ and\ \bibinfo {author} {\bibfnamefont {M.}~\bibnamefont {Yahiro}},\ }\href {\doibase 10.1103/PhysRevD.72.044010} {\bibfield  {journal} {\bibinfo  {journal} {Phys. Rev. D}\ }\textbf {\bibinfo {volume} {72}},\ \bibinfo {pages} {044010} (\bibinfo {year} {2005})},\ \Eprint {http://arxiv.org/abs/astro-ph/0503578} {arXiv:astro-ph/0503578} \BibitemShut {NoStop}%
\bibitem [{\citenamefont {Ballardini}\ \emph {et~al.}(2022)\citenamefont {Ballardini}, \citenamefont {Finelli},\ and\ \citenamefont {Sapone}}]{Ballardini:2021evv}%
  \BibitemOpen
  \bibfield  {author} {\bibinfo {author} {\bibfnamefont {M.}~\bibnamefont {Ballardini}}, \bibinfo {author} {\bibfnamefont {F.}~\bibnamefont {Finelli}}, \ and\ \bibinfo {author} {\bibfnamefont {D.}~\bibnamefont {Sapone}},\ }\href {\doibase 10.1088/1475-7516/2022/06/004} {\bibfield  {journal} {\bibinfo  {journal} {JCAP}\ }\textbf {\bibinfo {volume} {06}},\ \bibinfo {pages} {004} (\bibinfo {year} {2022})},\ \Eprint {http://arxiv.org/abs/2111.09168} {arXiv:2111.09168 [astro-ph.CO]} \BibitemShut {NoStop}%
\bibitem [{\citenamefont {Wang}\ and\ \citenamefont {Chen}(2020)}]{Wang:2020bjk}%
  \BibitemOpen
  \bibfield  {author} {\bibinfo {author} {\bibfnamefont {K.}~\bibnamefont {Wang}}\ and\ \bibinfo {author} {\bibfnamefont {L.}~\bibnamefont {Chen}},\ }\href {\doibase 10.1140/epjc/s10052-020-8137-x} {\bibfield  {journal} {\bibinfo  {journal} {Eur. Phys. J. C}\ }\textbf {\bibinfo {volume} {80}},\ \bibinfo {pages} {570} (\bibinfo {year} {2020})},\ \Eprint {http://arxiv.org/abs/2004.13976} {arXiv:2004.13976 [astro-ph.CO]} \BibitemShut {NoStop}%
\end{thebibliography}%
